\documentclass[aps,prd,reprint,a4paper,preprintnumbers,superscriptaddress,longbibliography,amsmath,amssymb,floatfix,nofootinbib]{revtex4-2}
\pdfoutput=1
\usepackage{graphicx}
\usepackage{scalefnt}
\usepackage[acronym]{glossaries}
\usepackage[export]{adjustbox}
\usepackage[colorlinks=true, linkcolor=blue, filecolor=blue, urlcolor=blue, citecolor=blue, pdftex, plainpages=false]{hyperref}

\glsdisablehyper
\newcommand{\abbrev}[1]{{\scalefont{.9}#1}}
\let\oldglsentryshort\glsentryshort
\renewcommand{\glsentryshort}[1]{\abbrev{\oldglsentryshort{#1}}}

\usepackage[capitalize]{cleveref}

\newcommand{\myacrodef}[3]{\newacronym{#2}{#2}{#3}\newcommand{#1}{\gls{#2}}}
\myacrodef{\dwf}{DWF}{domain-wall fermion}
\myacrodef{\qcd}{QCD}{Quantum Chromodynamics}
\myacrodef{\lo}{LO}{leading order}
\myacrodef{\nlo}{NLO}{next-to-leading order}
\myacrodef{\nnlo}{NNLO}{next-to-next-to-leading order}
\myacrodef{\sftx}{SFTX}{short-flow-time expansion}
\myacrodef{\sm}{SM}{Standard Model}
\myacrodef{\pt}{PT}{perturbation theory}
\myacrodef{\gf}{GF}{gradient flow}
\myacrodef{\rg}{RG}{renormalization group}
\myacrodef{\pcac}{PCAC}{partially conserved axial-vector current}
\myacrodef{\uv}{UV}{ultra-violet}

\newcommand{\MSbar}{\ensuremath{\overline{\text{\abbrev{MS}}}}}

\begin{document}

\title{Renormalized quark masses using gradient flow}


\author {Matthew Black}
\affiliation{Higgs Centre for Theoretical Physics, School of Physics and Astronomy, University of Edinburgh, Edinburgh EH9 3JZ, UK}
\affiliation{Theoretische Physik 1, Center for Particle Physics Siegen, Naturwissenschaftlich-Technische Fakult\"at, Universit\"at Siegen, 57068 Siegen, Germany}

\author{Robert V. Harlander}
\affiliation{Institute for Theoretical Particle Physics and Cosmology, RWTH Aachen University, 52056 Aachen, Germany}

\author{Anna Hasenfratz}
\email[Contact author: ] {anna.hasenfratz@colorado.edu}
\affiliation{Department of Physics, University of Colorado, 
Boulder, Colorado 80309, USA}

\author{Antonio Rago}
\affiliation{IMADA and Quantum Theory Center, University of Southern Denmark, Odense, Denmark}

\author{Oliver Witzel}
\email[Contact author: ]{oliver.witzel@uni-siegen.de}
\affiliation{Theoretische Physik 1, Center for Particle Physics Siegen, Naturwissenschaftlich-Technische Fakult\"at, Universit\"at Siegen, 57068 Siegen, Germany}

\date{\today}
\preprint{P3H-25-044,~~SI-HEP-2025-15,~~TTK-25-16}

\begin{abstract}
We propose a new and simple method for determining the renormalized quark masses from
lattice simulations. Renormalized quark masses are an important input to many
phenomenological applications, including searching and modeling physics beyond the
Standard Model.  The non-perturbative renormalization is performed using
gradient flow combined with the short-flow-time expansion that is improved by
\rg{} running to match to the \MSbar\ scheme.  
Implementing the \rg{} running perturbatively, we demonstrate this method works reliably at least up to the charm-quark mass and exhibits an easily-attainable ``windowing condition''.
Using \abbrev{RBC/UKQCD}'s (2+1)-flavor Shamir domain-wall
fermion ensembles with Iwasaki gauge action, we find $m_s^\text{\MSbar}(\mu=2
\text{ GeV}) = 89(3)$ MeV and $m_c^\text{\MSbar}(\mu=3 \text{ GeV}) = 972(16)$
MeV.  These results predict the scale-independent ratio $m_c/m_s= 12.1(4)$.
Generalization to other observables is possible, providing an efficient
approach to determine non-perturbatively renormalized fermionic observables
like form factors or bag parameters from lattice simulations.
\end{abstract}

\maketitle
\glsresetall
\section{Introduction}
Theoretical predictions for \sm\ processes require that non-perturbative
effects due to the strong force are accounted for.  These are typically
expressed in terms of decay constants, form factors, bag parameters, etc.
An ab initio way to determine such non-perturbative quantities is
the framework of lattice \qcd\ which allows us to improve the associated
uncertainties systematically.  Lattice quantities need to be renormalized and
matched to a continuum scheme. This matching requires a renormalization scheme
that can be implemented on the lattice as well as in \pt.
Different methods are employed, each with distinct pros and cons: The
Schr\"odinger
Functional~\cite{Luscher:1992an,Luscher:1993gh,Sint:1993un,Capitani:1998mq}
scheme uses gauge-invariant correlation functions, but the boundary conditions
make perturbative calculations challenging.  The regularization-independent
momentum subtraction (\abbrev{RI/MOM}) schemes~\cite{Martinelli:1994ty} break
gauge invariance and have a ``window problem'' arising from the need to choose a matching scale $\mu$
that is large enough to control the perturbative expansion, $\mu\gg
\Lambda_\text{\qcd}$, but small enough to avoid large cutoff effects, $\mu \ll
1/a$. Position-space renormalization schemes  \cite{Martinelli:1997zc,Gimenez:2004me}  use gauge
invariant correlation functions, where source-sink separation has to satisfy
$a\ll t \ll \Lambda_\text{\qcd}^{-1}$, creating a window problem.

In this work we develop a method to calculate
non-perturbatively renormalized quark masses using lattice \qcd{} simulations
and match them to the
\MSbar\ scheme~\cite{Suzuki:2013gza,Black:2023vju,Black:2024iwb} using the
\sftx~\cite{Luscher:2011bx,Luscher:2013vga,Suzuki:2013gza}.  The procedure
combines lattice and perturbative
\gf~\cite{Narayanan:2006rf,Luscher:2010iy,Luscher:2011bx,Luscher:2013cpa},
improved by \rg{} running.  Our approach is
gauge invariant, numerically precise, avoids most of the window problems, and
is straightforward to implement.\footnote{An alternative approach based on
flowed vacuum-expectation values has been recently presented in
Ref.\,\cite{Takaura:2025pao}.}

We demonstrate our method by predicting renormalized strange-
and charm-quark masses. Precise values for the quark masses are an important
input to many phenomenological applications, helping us to explore potential
physics beyond the \sm{}.  We extract the renormalized quark-mass values in
the \gf{} scheme from the two-point
correlation functions of flowed and regular pseudo-scalar and axial-vector
currents using the \pcac{} relation.  The lattice calculation is performed on
a set of (2+1)-flavor gauge-field configurations, and we use tuned
strange- and charm-quark masses in the valence sector.  After extrapolating to
the continuum limit, we match the resulting \gf{} renormalized quantities to
the \MSbar\ scheme by multiplying with the corresponding \sftx{} coefficients
and taking the limit to zero flow time ($\tau\to 0$).

We aid this matching by \rg{} running the flow time from the low-energy
lattice scale to values where \pt\ becomes reliable. Currently, this step is
based on the perturbatively evaluated \gf\ mass anomalous
dimension~\cite{Harlander:2020duo}, but we envision replacing this with a
non-perturbative running in the future~\cite{Hasenfratz:2022wll}.  This \rg{}
running improves the $\tau\to 0$ limit, making it better controlled and more
stable compared to the unimproved extrapolation.

\section{Notation and definitions} 

\subsection{Renormalized quark mass in the \texorpdfstring{\gf{}}{GF} scheme}
The gradient-flow equation for the gauge field reads
\begin{align}
\partial_\tau B_\mu = D_\nu G_{\nu\mu} 
\,,
\label{Eq.gaugeflow}
\end{align}
where the flowed gauge field is related to the regular gluon field by
$B_\mu(\tau=0) = A_\mu$, 
and
\begin{align}
 G_{\mu\nu} = \partial_\mu B_\nu - \partial_\nu B_\mu + [B_\mu, B_\nu]\, ,
 \,\, D_\mu = \partial_\mu + [B_\mu, \cdot] \,. \nonumber
\end{align}
The evolution of the quark fields is given on the gauge background:
\begin{align}
  \partial_\tau \chi &= \Delta \chi 
 \, , & \partial_\tau \bar{\chi} &= \bar{\chi} \overleftarrow{\Delta},
& \Delta &= D_\mu D_\mu.  
\label{Eq.fermionflow}
\end{align}
The flow regularizes \uv\ divergences in composite operators of
flowed gauge fields. Composite operators which also involve flowed fermions
are finite only after a multiplicative fermion wave function renormalization
is taken into account, $\chi_R =
Z_\chi^{1/2}\chi$~\cite{Luscher:2013cpa,Carosso:2018bmz}.  The perturbative approach to
the flow equations has been discussed in
Refs.\,\cite{Luscher:2010iy,Luscher:2011bx,Luscher:2013cpa,Artz:2019bpr}. 

On the lattice, the gradient flow is an invertible smoothing transformation. 
Its numerical evaluation on the gauge
fields is straightforward~\cite{Luscher:2009eq}. Fermionic quantities are always expressed as 
expectation values of quark propagators, requiring the inversion of the Dirac
operator on some source. While flowing both ``ends'' of the propagator is
numerically challenging, flowing only one end is again straightforward~\cite{Luscher:2013cpa}. In the
following, we consider current correlators where only the sinks are flowed.
 
Specifically, we are interested in ratios
\begin{align}
 R^\mathcal{O}_{rs}(t;\tau) = - \frac{\langle A_0(t;\tau) \mathcal{O}(t=0;\tau=0)\rangle_{rs}}{\langle
   P(t;\tau) \mathcal{O}(t=0;\tau=0)\rangle_{rs}},
 \label{eq:R}
\end{align}
where $t$ denotes Euclidean time, $A_0^{rs}(t;\tau)$ and $P^{rs}(t;\tau)$ are
the zero-momentum axial and pseudo-scalar charge operators of (flowed) quark
fields with flavors $r$ and $s$, and $\mathcal{O}$ could be either $P^{rs}$ or
$A_0^{rs}$. In this paper, we only consider flavor-nonsinglet (connected)
correlators. Since $Z_\chi$ as
well as the divergences stemming from $\mathcal{O}(t=0;\tau=0)$ cancel in the
ratio, $R^{\mathcal{O}}_{rs}(t;\tau)$ is UV-finite.

Both correlators in \cref{eq:R} couple to the same pseudo-scalar state of
  mass $M_\text{PS}$.  Thus, at (sufficiently) large source-sink separation
  and assuming $t$ is also much larger than the footprint $\sqrt{8\tau}$ of the flow, the ground
  state dominates and the ratio $R^{\mathcal{O}}_{rs}(t;\tau)$ is independent
  of $t$. We define
\begin{align}
\bar R^{\mathcal{O}} (\tau) &= \lim_{t\to\infty} R^{\mathcal{O}}(t;\tau)\quad \text{with}\quad \sqrt{8\tau} \ll t.
\label{eq:Rbar}
\end{align}
The renormalized \pcac\ relation (Ward identity) then implies~\cite{Endo:2015iea}
\begin{align}\label{eq:mGF1}
\left(m^{(r)}_\text{\gf}(\tau) +m^{(s)}_\text{\gf}(\tau)\right) = 
M^{(rs)}_\text{PS} \bar R^{\mathcal{O}}_{rs}(\tau)\, ,
\end{align}
where $m^{(r,s)}_\text{\gf}(\tau)$ is the renormalized quark mass of flavor
$r$ or $s$ at scale $\tau$ in the \gf\ scheme, and $M^{(rs)}_\text{PS}$ is the
pseudo-scalar mass in the channel.

If the time extent of the lattice configuration is finite, the mixed
  correlators $\langle A_0(t;\tau) P(0;0)\rangle_{rs}$ and $\langle P(t;\tau)
  A_0(0;0)\rangle_{rs}$ depend on $\text{sinh}(M_\text{PS} ( T/2-t))$, while the
  $\langle A_0(t;\tau) A_0(0;0)\rangle_{rs}$ and $\langle P(t;\tau)
  P(0;0)\rangle_{rs}$ correlators show a $\text{cosh}(M_\text{PS} ( T/2-t))$ time
  dependence. Hence the time dependence in the ratios defined in
  Eq.~\eqref{eq:R} does not cancel for lattices with final extent $T/a$ and
  the $R^{\mathcal{O}}(t;\tau)$ in Eq.~\eqref{eq:Rbar} should be replaced by
  the ratio of the corresponding amplitudes. This can be achieved
by multiplying/dividing the ratio by $\text{cosh}(M_\text{PS} ( T/2-t))/\text{sinh}(M_\text{PS} ( T/2-t))$, since $aM_\text{PS}$ can be determined with high precision.
Alternatively, we can avoid this difficulty by considering the product of $\bar R^P_{rs}(\tau)$ and $\bar R^{A_0}_{rs}(\tau)$ 
\begin{align}\label{eq:mGF2}
\left(m^{(r)}_\text{\gf}(\tau) +m^{(s)}_\text{\gf}(\tau)\right) = 
M^{(rs)}_\text{PS} \sqrt{\bar R^{A_0}_{rs}(\tau) \bar R^{P}_{rs}(\tau)}\, .
\end{align}
We have analyzed our data using both Eq.~\eqref{eq:mGF1} as well as
Eq.~\eqref{eq:mGF2}. For the range of flow times $\tau$ of interest, both ratios are consistent within our statistical uncertainties. To simplify the discussion later on we introduce
\begin{align}
  \bar R^{AP}(\tau) \equiv \sqrt{\bar R^{A_0}_{rs}(\tau) \bar R^{P}_{rs}(\tau)},
  \label{eq.RAP}
\end{align}
and also refer to $\bar R^{AP}$ as ratio.

\subsection{Matching to the \texorpdfstring{\MSbar{}}{MSbar} scheme}
\begin{figure*}[tp]
\includegraphics[width=\columnwidth]{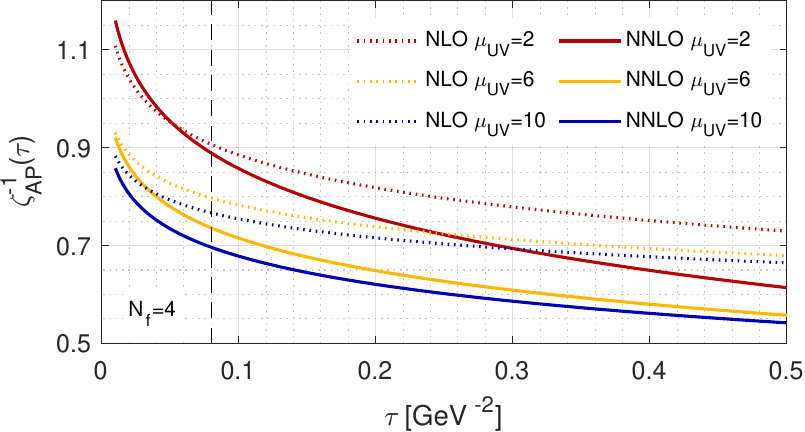}\hfill
\includegraphics[width=\columnwidth]{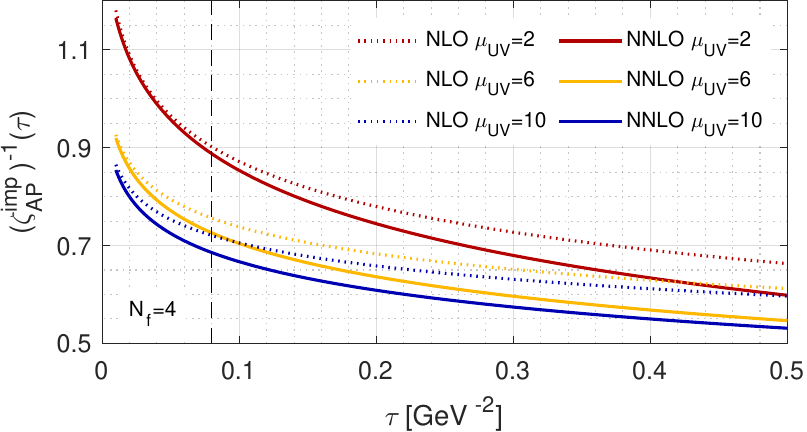}
\caption{The matching factor  $\zeta_{AP}^{-1}$ as the function of the flow time $\tau$ for different values of the matching scale $\mu$. The left panel shows the \nlo{} (dotted lines) and \nnlo{} (solid lines) predictions without,  the right panel with \rg\ improvement. The vertical dashed lines indicate $\tau_{\text{min}}$ used in our analysis. For consistency we use the three-loop $\gamma^\text{\gf}_m$ function when evaluating \nnlo, but only the two-loop function for \nlo. Using three-loop $\gamma^\text{GF}_m$ for both \nnlo\ and \nlo, the differences would mostly disappear. Recalling that $\zeta_{AP}^{-1}\big|_\text{\lo}=1$, these plots demonstrate the perturbative validity of the approach for our choice of parameters.}
\label{Fig.zeta}
\end{figure*}

Finally, we need to connect \gf{} renormalized quark masses $m_\text{\gf}^{(r,s)}(\tau)$ to the renormalized
quark mass in the \MSbar\ scheme at some energy scale
$\mu_\text{\uv}$. This matching can be performed in the \sftx\
\begin{align}
m_\text{\MSbar}(\mu_\text{\uv}) = \lim_{\tau\to 0} \,\, \zeta^{-1}_{AP}(\mu_\text{\uv},\tau)
m_\text{\gf}(\tau) \, ,
\label{eq:conversion}
\end{align}
where $\zeta_{AP}(\mu_\text{\uv},\tau) = \zeta_{A}(\mu_\text{\uv},\tau) \zeta^{-1}_P(\mu_\text{\uv},\tau)$. The $\tau\to 0$ limit implements the transition from the massive \gf{} scheme to the massless \MSbar{} scheme.
 $\zeta_{A}$ and  $\zeta_{P}$ are the
\sftx\ matching coefficients of the axial-vector and the pseudo-scalar
currents.   They are known perturbatively to \nnlo{}
level~\cite{Borgulat:2023xml}, and depend on $\tau$ and $\mu_\text{\uv}$
  via $\alpha_s(\mu_\text{\uv})$ and $\ln \tau \mu^2_\text{\uv}$.  For the
  strong coupling $\alpha_s(\mu_\text{\uv})$ to be in the perturbative regime,
  we need to choose $\mu_\text{\uv}$ sufficiently large. At the same time, the
  flow time $\tau$ should not be too different from $1/\mu^2_\text{\uv}$, so
  that the logarithms do not spoil the perturbative series. On the other hand,
  the smearing radius $\sqrt{8\tau}$ induced by the gradient flow should be
  larger than the lattice spacing $a$ in order to minimize cutoff effects. 
  These conditions can be fulfilled on typical lattice
  ensembles, including the ones analyzed in this work.  To illustrate this, the left panel of Fig.~\ref{Fig.zeta}
  compares the \nlo{} and \nnlo{} values of $\zeta^{-1}_{PA}$ as functions of
  the flow time $\tau$ at different values of $\mu_\text{\uv}$. For $\tau\to
  0$, $\zeta_{AP}^{-1}$ diverges logarithmically as required to
    compensate the singularity from $m_\text{\gf}(\tau)$ determined on the
    lattice. The gap between \nlo{} and \nnlo{} increases with $\tau$,
    but the difference between the \lo{}   ($\zeta_{AP}^{-1}=1$), \nlo{}, and \nnlo{} values progressively decreases in the entire $\tau$ range relevant for this analysis.

Nevertheless, we can further improve the matching in \cref{eq:conversion} by
resumming the logarithmic terms with the help of the \rg\ equation in the
\gf\ scheme,
\begin{equation}\label{eq::brix}
  \begin{aligned}
    \tau\partial_\tau\mathcal{O}(\tau) = \gamma^\text{\gf}_{\mathcal{O}}
    \mathcal{O}(\tau)\,,
  \end{aligned}
\end{equation}
where, in our case, $\mathcal{O} \in\{ A_0,P\}$.  As shown in
Refs.\,\cite{Artz:2019bpr,Borgulat:2025gys}, the perturbative flowed anomalous
dimension can be expressed in terms of the matching coefficients as
$\gamma^\text{\gf}_\mathcal{O} =
\tau (\partial_\tau\zeta_\mathcal{O})\zeta^{-1}_\mathcal{O}$, and --- under certain
conditions --- knowledge of $\zeta_\mathcal{O}$ to order $\alpha_s^n$ allows to
obtain $\gamma^\text{\gf}_\mathcal{O}$ to order $\alpha_s^{n+1}$.  Integrating
\cref{eq::brix} leads to
\begin{align}\label{eq:mimp}
m_\text{\MSbar}(\mu_\text{\uv}) &= \lim_{\tau\to 0}
(\zeta_{AP}^\text{imp}(\mu_\text{\uv},\tau_\mu))^{-1}m_\text{\gf}(\tau)
\,,
\end{align}
with
\begin{align}
(\zeta_{AP}^\text{imp}(\mu_\text{\uv},\tau_\mu))^{-1} &=
\zeta_{AP}^{-1}(\mu_\text{\uv},\tau_\mu) \nonumber\\&\times \text{exp}\left( -
\int_{ \tau_\mu}^{\tau} d\tau^\prime
\frac{\gamma^\text{\gf}_m(\tau^\prime)}{\tau^\prime}\right)
\,,
\label{eq:impconversion}
\end{align}
where it follows from the previous discussion that $\gamma^\text{\gf}_m =
\gamma^\text{\gf}_A-\gamma^\text{\gf}_P$.  The upper limit of the
  integral in Eq.~\eqref{eq:impconversion} is the lattice flow time, $\tau \ge
  \tau_\text{min} \approx 0.08$ in our setup, while the lower limit is
  $\tau_\mu =
  e^{-\gamma_\text{E}}/(2\mu_\text{\uv}^2)$~\cite{Harlander:2018zpi}. As
  $\mu_\text{UV}$ increases from 2 to 6 to 10 GeV, $\tau_\mu$ decreases from
  $\approx 0.07$ to $\approx 0.008$ to $\approx 0.003$. The corresponding
  increase of the integration range requires increasingly precise
  determination of $\gamma^\text{\gf}_m$.

Ideally, the non-perturbative expression for $\gamma^\text{\gf}_m(\tau)$
should be used in \cref{eq:impconversion} and a way to obtain it is
demonstrated in
Refs.\,\cite{Hasenfratz:2022wll,Hasenfratz:2023sqa,Hasenfratz:2023wbr}.  For
two-flavor \qcd{} $\gamma^\text{\gf}_m$ was found to be close to the
perturbative prediction~\cite{Hasenfratz:2022wll}. In our analysis, we
use the perturbative expression, which is known to
$\alpha_s^3$~\cite{Borgulat:2023xml}. In this case, the integral in
Eq.~\eqref{eq:impconversion} can be obtained analytically.

The right panel of \cref{Fig.zeta} compares the \nlo{} and \nnlo{}
  values of $\zeta^{-1}_{AP}$ after the logarithmic resummation according
    to \cref{eq:impconversion}. We indeed observe an improved perturbative
    convergence for all values of $\mu_\text{\uv}$ under consideration.
    Although the final result should be independent of the choice of
    $\mu_\text{\uv}$ in \cref{eq:conversion,eq:mimp}, the $\tau\to 0$
    extrapolation in \cref{eq:conversion,eq:impconversion} might not be. 
    To account for systematic uncertainties arising from the integral in
  Eq.~\eqref{eq:impconversion}, while still ensuring convergence of the
  \sftx{}, we consider $\mu_\text{UV}\in(2,6)$~GeV in the
  following. At the end, we convert the
  result for the masses to the commonly preferred reference scales
    $\mu=2$ or 3~GeV value by regular four-loop running in the \MSbar{}
  scheme. 

\section{Numerical setup, analysis, and results}

\begin{table*}[tp]
    \caption{RBC/UKQCD's $N_f=2+1$ Shamir \dwf\ ensembles with Iwasaki gauge action~\cite{Allton:2008pn,Aoki:2010dy,Blum:2014tka,Boyle:2017jwu,Boyle:2018knm} specified by the inverse lattice spacing ($a^{-1}$), unitary pion mass ($M_\pi$), light and strange sea quark masses ($am_\ell^\text{sea}$ and $am_s^\text{sea}$), valence strange ($am_s^\text{val}$) and charm ($am_c$) quark masses. 
    For the coarse (C1, C2), medium (M1, M2, M3), and fine ensemble (F1S) we use $N_\text{conf}$ number of configurations and place $N_\text{src}$ $Z_2$ wall-sources per configuration. }
    \label{Tab.Ensembles}
    \centering
    \begin{tabular}{l@{~~~}c@{~~~}c@{~~~}c@{~~~}c@{~~~}c@{~~~}cccccc}
        \hline\hline & L/a & T/a & $a^{-1}[\text{GeV}]$ & $M_\pi[\text{MeV}]$ & $am_\ell^\text{sea}$ & $am_s^\text{sea}$ & $am_s^\text{val}$ & $am_c$ 
                      & $N_\text{src}\times\text{N}_{\text{conf}}$ \\\hline\hline
        C1 & 24 & 64 & 1.7848(50) & 339.8(1.2) & 0.005\phantom{144} & 0.04\phantom{144} & 0.03224 & 0.64& $32\times101$ \\
        C2 & 24 & 64 & 1.7848(50) & 430.6(1.4) & 0.010\phantom{144} & 0.04\phantom{144} & 0.03224 & 0.64 & $32\times101$ \\
        M1 & 32 & 64 & 2.3833(86) & 303.6(1.4) & 0.004\phantom{144} & 0.03\phantom{144} & 0.02477 & 0.45 & $32\times\phantom{0}79$  \\
        M2 & 32 & 64 & 2.3833(86) & 360.7(1.6) & 0.006\phantom{144} & 0.03\phantom{144} & 0.02477 & 0.45 & $32\times\phantom{0}89$\\
        M3 & 32 & 64 & 2.3833(86) & 410.8(1.7) & 0.008\phantom{144} & 0.03\phantom{144} & 0.02477 & 0.45 & $32\times\phantom{0}68$\\
        F1S & 48 & 96 & 2.785(11) & 267.6(1.3) & 0.002144 & 0.02144 & 0.02167 & 0.37 & $24\times\phantom{0}98$ 
        \\\hline\hline 
    \end{tabular}
\end{table*}

\begin{table*}[tp]
\caption{Values of the pseudo-scalar meson masses $M_{\eta_s}$, $M_{D_s}$, and $M_{\eta_c}$ obtained for our six ensembles. First we list each mass in lattice units with statistical uncertainty only. Next we convert the value to MeV using $a^{-1}$ quoted in \cref{Tab.Ensembles} and the statistical uncertainty as well as the uncertainty propagated from $a^{-1}$.}
\label{Tab.M_PS}
\begin{tabular}{c@{~~~}ll@{~~~}ll@{~~~}ll}
\hline\hline
    & $aM_{\eta_s}$ & $M_{\eta_s}$ [MeV] & $aM_{D_s}$ & $M_{D_s}$ [MeV] & $aM_{\eta_c}$ & $M_{\eta_c}$ [MeV] \\ \hline
C1  & 0.38961(51) & 695.38(0.9)(1.9) & 1.10198(66) & 1966.8(1.2)(5.5) &1.64713(21)  & 2939.81(0.4)(8.2)  \\
C2  & 0.39204(44) & 699.71(0.8)(2.0) &  1.10408(63) & 1970.6(1.1)(5.5) &1.64812(17)  & 2941.57(0.3)(8.2)  \\
M1  & 0.29175(35) & 695.33(0.8)(2.5) & 0.82889(48) & 1975.5(1.1)(7.1) &1.24895(16)  & 2976.61(0.4)(10.7) \\
M2  & 0.29214(41) & 696.26(1.0)(2.5) & 0.82933(49) & 1976.5(1.2)(7.1) &1.24923(16)  & 2977.28(0.4)(10.7) \\
M3  & 0.29281(36) & 697.86(0.9)(2.5) & 0.82995(57) & 1978.0(1.4)(7.1) &1.24964(20)  & 2978.26(0.5)(10.7) \\
F1S & 0.25512(19) & 710.50(0.5)(2.8) &  0.70717(37) & 1969.5(1.0)(7.8) &1.065287(87) & 2966.82(0.2)(11.7) \\
\hline\hline
\end{tabular}
\end{table*}

\glsreset{DWF}
We determine the strange and charm quark masses $m_s$ and $m_c$, and obtain
their ratio $m_c/m_s$ using \abbrev{RBC/UKQCD}'s set of (2+1)-flavor Shamir \dwf{} and
Iwasaki gauge field ensembles~\cite{Allton:2008pn,Aoki:2010dy,Blum:2014tka,Boyle:2017jwu,Boyle:2018knm}. These
ensembles feature three different bare gauge couplings, corresponding to
inverse lattice spacings $a^{-1}$ ranging from 1.78 to 2.78~GeV, as well
as different unitary pion masses (due to different light sea-quark
masses). Details of the ensembles are provided in \cref{Tab.Ensembles}.
Bare physical strange-quark masses and inverse lattice spacings are taken from Refs.\,\cite{Blum:2014tka,Boyle:2018knm}, and bare physical charm-quark masses are tuned using the $D_s$ meson. 
Our lattice calculations are based on \texttt{Grid}~\cite{Boyle:2015tjk} and \texttt{Hadrons}~\cite{Hadrons,Hadrons22}, and we implemented fermionic gradient flow~\cite{Black:2023vju,Black:2024iwb,Lifetimes:2025} in \texttt{Hadrons}.\footnote{Cf.~\url{https:
//github.com/aportelli/Hadrons/pull/137} and examples given at \url{https://github.com/mbr-phys/HeavyMesonLifetimes}.}  

We calculate the correlation functions of \cref{eq:R} at $\tau=0$ by placing
$Z_2$ wall sources~\cite{McNeile:2006bz,Boyle:2008rh} on several time slices  
per configuration and individually invert the Dirac operator to obtain quark
propagators. For the strange quark, we apply Gaussian source smearing~\cite{Gusken:1989ad} before inverting Shamir
\dwf~\cite{Kaplan:1992bt,Shamir:1993zy,Furman:1994ky}, whereas for the charm
quark we directly invert stout-smeared M\"obius
\dwf~\cite{Morningstar:2003gk,Brower:2012vk,Cho:2015ffa}.  Our calculation
proceeds by evolving the gauge-field configuration together with the strange-
and charm-quark propagators with  \gf{} as specified by
\cref{Eq.gaugeflow,Eq.fermionflow}. We choose a step size
$\epsilon=0.01$ for the \gf{} evolution and contract the quark propagators
imposing a point sink every 10\textsuperscript{th} (for larger flow times
every 40\textsuperscript{th}) step to obtain the corresponding two-point
correlators as functions of the \gf{} time $\tau$. Sources and sinks are
projected to zero momentum.

\begin{figure*}[tp]
  \includegraphics[width=0.95\columnwidth]{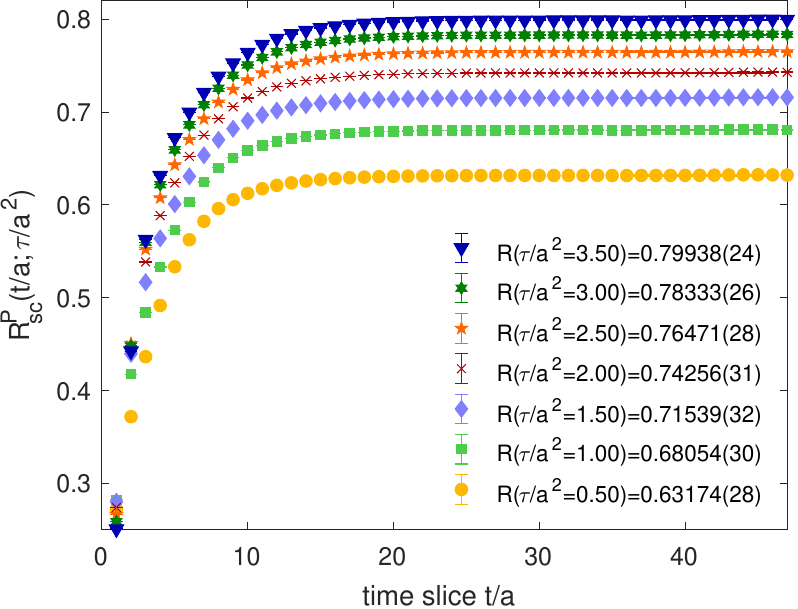}
\includegraphics[width=0.95\columnwidth]{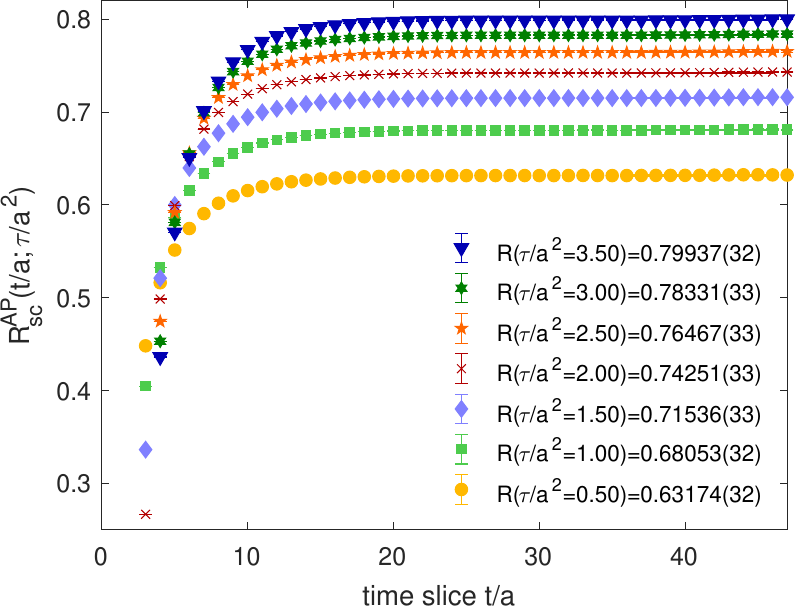}  
\caption{On the left we present the ratios $R^P(t/a;\tau/a^2)$ defined in \cref{eq:R} and on the right
$R^{AP}(t/a;\tau/a^2)$ introduced in Eq.~\eqref{eq.RAP} for selected flow times
$\tau/a^2$ for the $D_s$ meson on the F1S ensemble. Statistical errors are within the symbol sizes. 
\newline
Due to the finite extent $T/a$ of our gauge-field configurations and the use of
anti-periodic boundary conditions for fermions in the time direction, we need
to explicitly account for the ``around-the-world effect'' in  $R^P(t/a;\tau/a^2)$ which enters with a different
sign for $\langle AP\rangle$ than for the $\langle PP\rangle$ correlators. We
therefore obtain $\bar R$ in \cref{eq:Rbar} from the combination $R(t/a;\tau/a^2) \text{cosh}(M_\text{PS}(T/2-t))/\text{sinh}(M_\text{PS}(T/2-t))$. Since the pseudo-scalar mass $M_\text{PS}$ can be measured very precisely (see \cref{Tab.M_PS}), this combination shows a long, flat plateau. The alternative combination  $R^{AP}(t/a;\tau/a^2)$  avoids the problem of different ``around-the-world'' contributions and directly shows long, flat plateaus, as illustrated on the right panel.   For both quantities we perform correlated fits to the plateaus from time slice 36 to 46, extracting the values of $\bar R(\tau/a^2)$ shown in the legend. Magnified plateau regions are shown in Fig.~\ref{Fig.zoom}.}
\label{Fig.Ratios}
\end{figure*}

We average correlators calculated with different
source positions on the same gauge field and afterwards use jackknife
resampling to propagate statistical uncertainties. We extract
the pseudo-scalar meson mass $M_\text{PS}$ and verify that it is indeed
independent of flow time $\tau$. The corresponding values are listed in
\cref{Tab.M_PS}. Next we extract $R^P(t/a;\tau/a^2)$ as defined in
\cref{eq:R} with ${\cal O}=P^{rs}$ for a range of lattice flow times $0\le
\tau/a^2 \le~3.5$.
We repeat this analysis using $R^{AP}(t/a;\tau/a^2)=\sqrt{R^{A_0}(t/a;\tau/a^2)R^P(t/a;\tau/a^2)}$ defined in \cref{eq.RAP}.
\begin{figure}[tb]
\includegraphics[width=0.95\columnwidth]{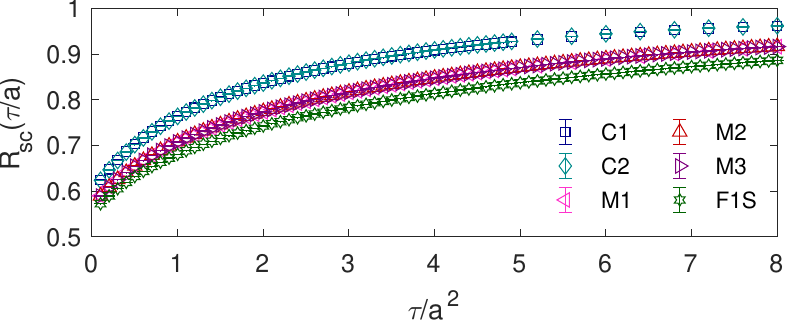}
\includegraphics[width=0.95\columnwidth]{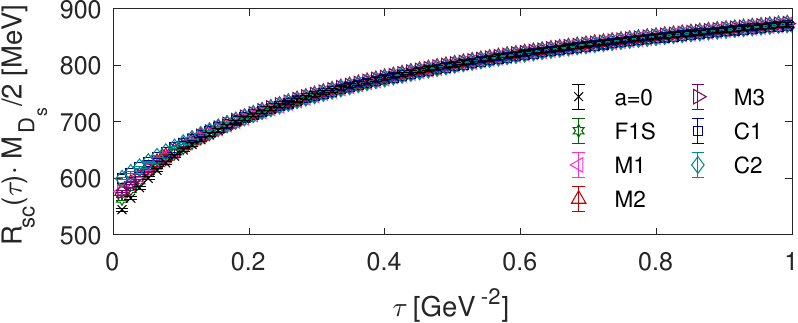}
\caption{Extracting $\bar R$-ratios (see \cref{eq:Rbar}) for $D_s$ meson correlators. Top: dependence on the flow time $\tau/a$ in lattice units for our six ensembles. Bottom: converting to physical units and adding the $a\to 0$ continuum limit predictions.}
\label{Fig.GFratios}
\end{figure}

\begin{figure}[tb]
  \includegraphics[width=0.95\columnwidth]{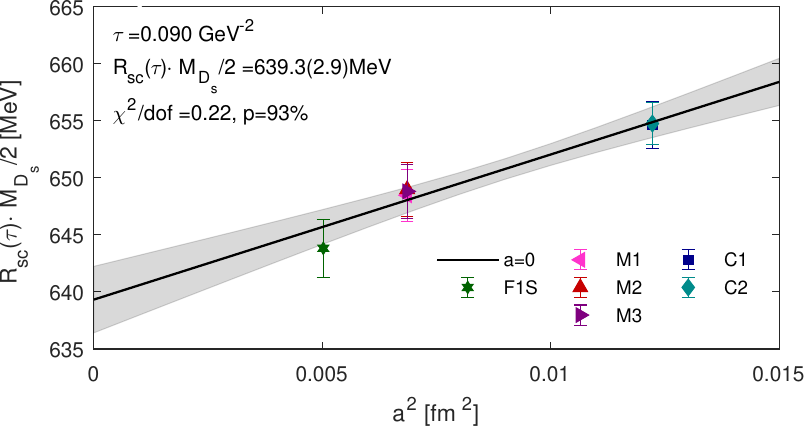}
  \includegraphics[width=0.95\columnwidth]{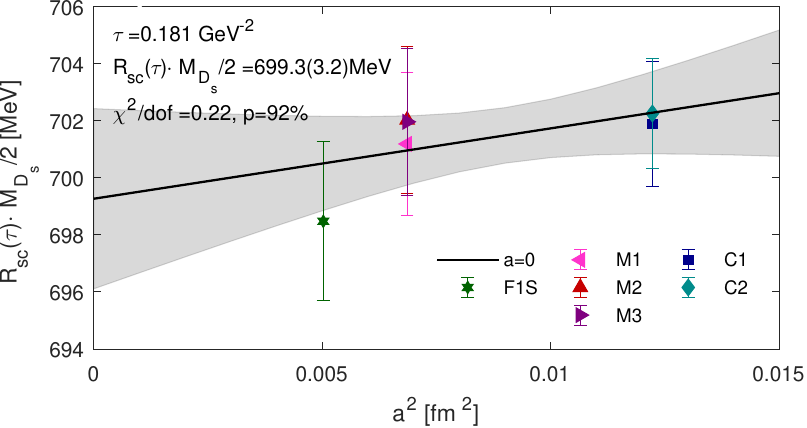}
  \includegraphics[width=0.95\columnwidth]{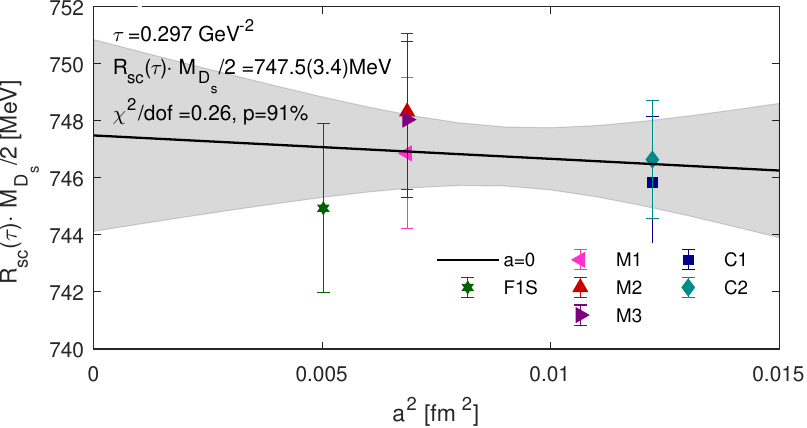}  
  \caption{Details of the continuum-limit extrapolation at selected flow times $\tau=0.090,\, 0.181,\, 0.297\, \text{GeV}^2$ for the $D_s$ meson. Since the data points for M1, M2, M3 as well as C1 and C2 sit on top of each other, sea quark mass effects are not resolved. Hence we obtain the $a\to 0$ limit simply by fitting an ansatz linear in $a^2$.}
  \label{Fig.ContinuumLimitDs}
\end{figure}

 As an example, we show both resulting correlator ratios for the $D_s$ meson on
 the F1S ensemble for a selection of \gf{} times in \cref{Fig.Ratios}.
 The values of the ratios are extracted by
 performing correlated fits to the plateaus and we report the results for the relevant flow times $\tau$ in Tab.~\ref{Tab.RP} for $\bar R^P(\tau/a^2)$ and in Tab.~\ref{Tab.RAP} for  $\bar R^{AP}(\tau/a^2)$. Both ratios lead to consistent results within our statistical uncertainties. We proceed with our analysis using $\bar R^{AP}(\tau/a^2)$ because we observe better $p$-values in the extraction of $\bar R^{AP}(\tau/a^2)$ than for $\bar R^{P}(\tau/a^2)$.
   After repeating this step for all
six ensembles, we show $\bar R$ as a function of the lattice flow time
$\tau/a^2$ in the upper panel of \cref{Fig.GFratios}.  Next, we multiply our
ratios with the corresponding lattice value of $aM_\text{PS}/2$ and convert both
$aM_\text{PS}$ and the lattice flow time $\tau/a^2$ to physical units using the inverse lattice spacings listed in
\cref{Tab.Ensembles}. The result is shown in the lower panel of
\cref{Fig.GFratios}.  In the upper panel, data corresponding to the different
coarse and medium ensembles sit on top of each other, indicating that
sea-quark effects are not resolved in the $\bar R$ ratios. In the lower panel,
the six ratios form a unique curve, showing deviation due to cutoff effects
only in the small-$\tau$ region.  We interpolate the flow times in the coarse
and medium ensembles to match the values on the F1S ensemble and take the
continuum limit $a\to 0$ for the different flow times $\tau$.  Since all
actions used in this calculation are ${\cal O}(a)$-improved and no sea-quark
effects are resolved, we simply perform a linear extrapolation in $a^2$ to
arrive at the continuum-limit results shown by the black crosses in the lower
panel of \cref{Fig.GFratios}. Details of the $a\to 0$ continuum limit extrapolation are shown for selected flow times in Fig.~\ref{Fig.ContinuumLimitDs} and in Tab.~\ref{Tab.ContinuumExtra}.

\begin{figure}[tb]
\includegraphics[width=0.95\columnwidth]{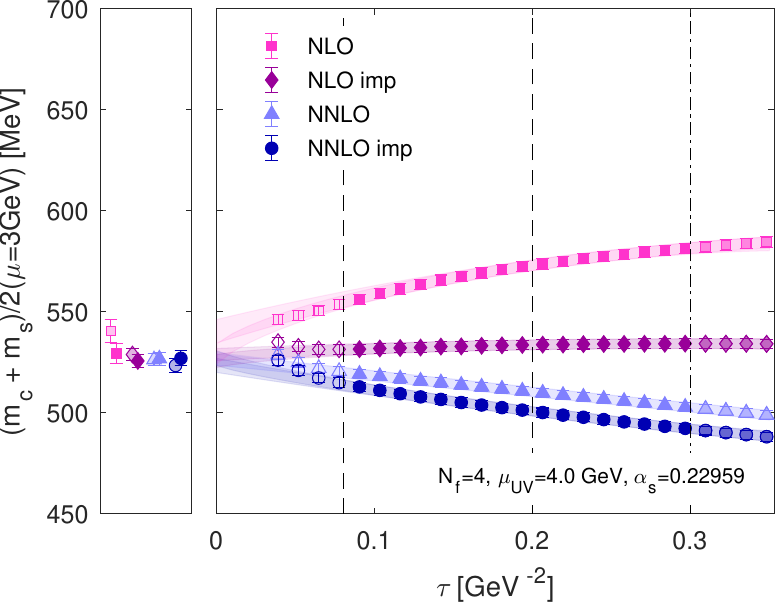}
\caption{The large panel illustrates extracting $(m_c+m_s)/2$ at the renormalization scale $\mu_\text{\uv}=4 \text{ GeV}$, converted to $\mu=3 \text{ GeV}$ using 4-loop $N_f=4$ \MSbar{} running~\cite{Chetyrkin:2000yt,Herren:2017osy}. 
Pink squares and light blue triangles denote to the \nlo{}
and the \nnlo{} values, respectively.
 Purple diamonds and dark blue circles show the corresponding \rg{} improved \nlo{}
and \nnlo{} values.
The $\tau\to 0$ limit is taken in a $(\tau_\text{min},\tau_\text{max})$ range as explained in the text.   The shaded error bands show the maximal spread of quadratic or linear-log fits using the central values plus/minus their uncertainty.  Only filled symbols enter the $\tau\to 0$ extrapolation.  The small panel compares the final predictions using solid filled symbols for linear-log and shaded fillings for quadratic fits.}
\label{Fig.Ds}
\end{figure}

The continuum limit results for $m_\text{\gf}(\tau)$ are renormalized in the
\gf{}\ scheme.  We have to take the $\tau\to 0$ limit to connect to the
 mass-independent \MSbar{}
scheme following \cref{eq:conversion}.
We can improve the matching using \rg{} running according to
\cref{eq:impconversion}. To
maintain consistency of our results in the \rg\ running, we use $\gamma_m^\text{\gf}$ at order
$\alpha_s^2$ for the \nlo\ improvement, and at order $\alpha_s^3$ for the
\nnlo\ improvement. The larger right panel of \cref{Fig.Ds} shows $\zeta^{-1}_{AP}(\tau,\mu)m_\text{\gf}(\tau)$ and the corresponding \rg\ improved values for the $D_s$ system as a function of the flow time $\tau$ both at \nlo{} and \nnlo{} level.  We use $N_f=4$  for the $\zeta_{AP}(\tau,\mu)$  factor and the \rg{} running. 

\begin{figure}[tb]
\includegraphics[width=0.95\columnwidth]{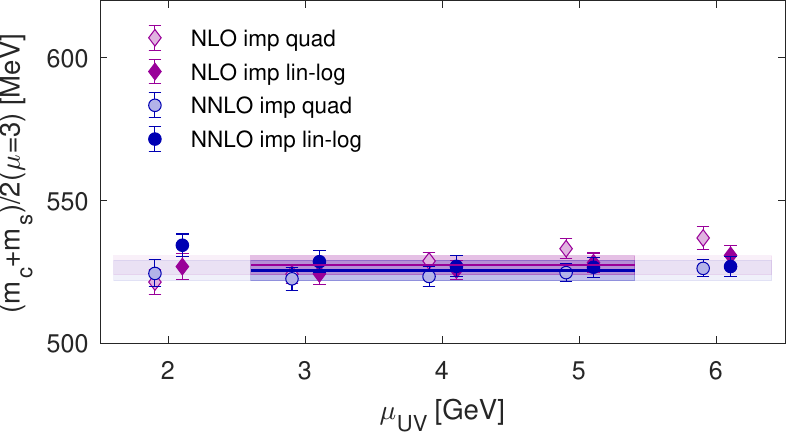}
\caption{Dependence of $(m_c+m_s)/2$ obtained from $D_s$ correlators on $\mu_\text{\uv}$. Our final results are obtained by calculating the correlated average for \nlo\ and \nnlo\ using the values at $\mu_\text{UV}=3$, 4, and 5 GeV.}
\label{Fig.muDs}
\end{figure}

At small flow time $\sqrt{8\tau}\lesssim a$, the lattice data are contaminated
by lattice artifacts.  On the coarsest ensembles, this corresponds to $\tau <
0.039 \text{ GeV}^{-2}$.  To obtain the $\tau\to0$ limit we consider two fit
ans\"atze:\footnote{Alternative approaches are discussed in
Ref.\,\cite{Suzuki:2021tlr}.}  quadratic $f_2(\tau) = \tau^2 c_2 + \tau c_1 +
c_0$ and linear-log $f_l(\tau) = \tau (c_l \log(\tau\mu^2) + c_1) + c_0$.
Moreover, we vary the fit range and always account for the maximal variation
using $\tau_\text{min} \in (0.08,0.2)\text{ GeV}^{-2}$ with
$\tau_\text{max}=0.3\text{ GeV}^{-2}$, and $\tau_\text{max}\in
(0.25,0.35)\text{ GeV}^{-2}$ with $\tau_\text{min} = 0.14\text{ GeV}^{-2}$ to
obtain the color-shaded bands in \cref{Fig.Ds}.  Because successive data in
\gf{} time tend to be highly correlated, the full
variance-covariance matrix is nearly singular and cannot reliably be inverted.
For one choice of $(\tau_\text{min},\,\tau_\text{max})$, we therefore take the $\tau\to 0$ limit by performing two uncorrelated fits where 
all central values are coherently shifted by plus/minus their uncertainties. 
The full spread across both fits feeds into our extrapolated result which is then taken as the symmetric uncertainty over the interval spanned by all variations of the fit range.
For better visibility, we show the two final
fit values for the four different approximations (\nlo{}, \nnlo{}, with and
without \rg{} improvement) in the small left panel of \cref{Fig.Ds}.  While
there is only little difference for the \nnlo{} results with and without \rg{}
running, the \nlo{} result improves substantially. The overall approach to
zero becomes flatter, and the spread between our different fit ansätze
reduces.  In \cref{Fig.muDs} we show the dependence on the $\mu_\text{\uv}$
scale. Our improved results show very litte variation in the $3\, \text{GeV} 
< \mu_\text{\uv} \le 6\, \text{GeV}$ range. We therefore calculate correlated 
averages~\cite{Schmelling:1994pz} for our \nlo{} and \nnlo{} determinations at 
$\mu_\text{UV} = 3$, 4, and 5 GeV and find
\begin{align}
  \begin{aligned}
  &\left(\frac{m_c+m_s}{2}\right)^\text{NLO} = 527.4(3.3)_\text{GF}~\text{MeV}, \\
  &\left(\frac{m_c+m_s}{2}\right)^\text{NNLO} = 525.5(3.6)_\text{GF}~\text{MeV},
  \end{aligned}
\label{Eq.ResultDs}
\end{align}  
using the renormalization scale $\mu=3\,\text{GeV}$.
The uncertainty labeled ``GF'' covers both statistical and systematic effects 
from the $\tau\to 0$ extrapolation. As part of our analysis, we have
propagated the uncertainty of $a^{-1}$ and the tuning of the bare charm mass,
whereas effects due to choosing the fit range to obtain $\bar R(\tau/a)$ or
finite volume effects are negligible. Additional discretization errors may arise 
for the heavy charm quark that we generate in our partially-quenched set-up using 
stout-smeared M\"obius \dwf. This smeared action has been explored and used by 
\abbrev{JLQCD} and \abbrev{RBC/UKQCD} for other calculations involving charm and 
bottom quarks (see, e.g., Refs.\,\cite{Boyle:2018knm,Aoki:2023qpa,Colquhoun:2022atw}). 
To check for discretization effects for \dwf, we determine the residual mass for
 $am_c$. As can be seen in Appendix \ref{Mres}, we find small values and long
 plateaus for $am_\text{res}$ on all ensembles for the bare quark-mass values
 considered. This indicates that charm discretization errors are
 negligible.
 
In order to estimate further systematic effects, we repeat our analysis 
discarding the two coarse ensembles (C1 and C2). On the one hand, this is a sensitive test
of charm quark discretization errors, because the largest bare charm quark mass is
reduced from $0.64$ to $0.45$; on the other hand, we substantially alter the $a\to 0$ 
continuum limit. For the remaining data at two lattice spacings, we 
perform a linear fit in $a^2$ as well as a fit to a constant to take the continuum limit. For the
$D_s$ correlators, the fit to a constant results in the larger shift of the central value, amounting to
13.3 MeV at \nnlo. We assign half of this difference as a systematic uncertainty to our final result, labeled by ``CL''.

Similarly, we use half the difference between the central values of the \nlo\ and the \nnlo\ results in \cref{Eq.ResultDs} in order to estimate the systematic uncertainty due to the truncation of the perturbative series (labeled ``PT'' in the final result). Thus, we obtain
\begin{align}
  \left(\frac{m_c+m_s}{2}\right) = 526(4)_\text{GF}(7)_\text{CL}(1)_\text{PT}~\text{MeV},
\label{Eq.FinalResultDs}
\end{align}
At present, the \abbrev{CL}-error estimate dominates 
our uncertainty, which reflects the fact that our data set is currently quite limited. We expect a significant improvement from including a larger number of gauge field ensembles in the future.

\begin{figure*}[tp]
  \begin{minipage}{\columnwidth}
    \includegraphics[width=0.95\columnwidth]{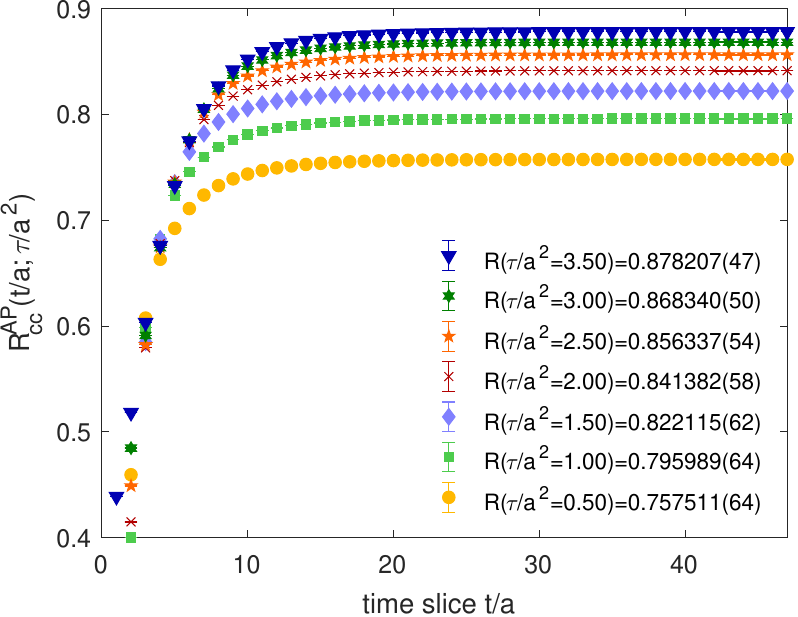} \\
    (a) Extracting the ratios $R^{AP}(t/a;\tau/a^2)$ using $\eta_c$ correlators on the F1S ensemble.\\[3mm]
    \includegraphics[width=0.95\columnwidth]{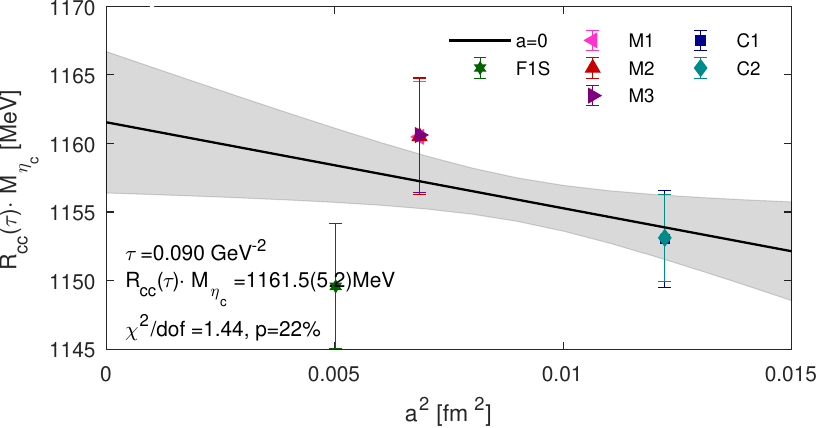}\\
    \includegraphics[width=0.95\columnwidth]{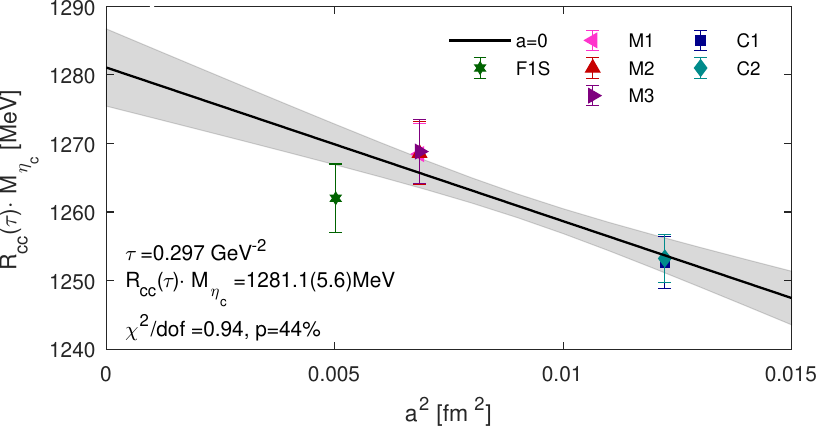}    \\
    (b) Continuum limit extrapolations at $\tau=0.090$ and 0.297 $\text{GeV}^{-2}$ using an ansatz linear in $a^2$.
  \end{minipage}
    \begin{minipage}{\columnwidth}
  \includegraphics[width=0.95\columnwidth]{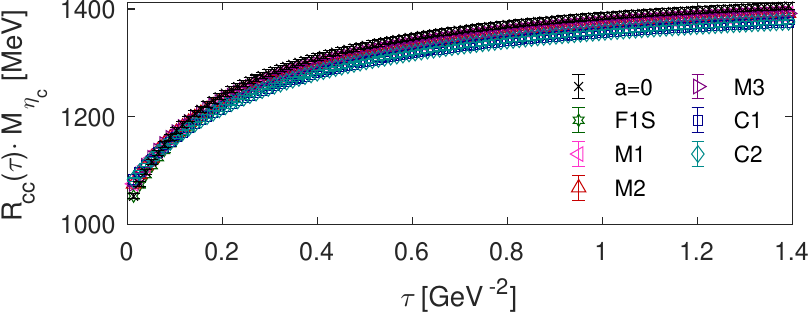}\\
  (c) Ratios $\bar R^{AP}(\tau)$ for the $\eta_c$ meson converted to physical units and including $a\to 0$ continuum limits. \\[3mm]
  \includegraphics[width=0.95\columnwidth]{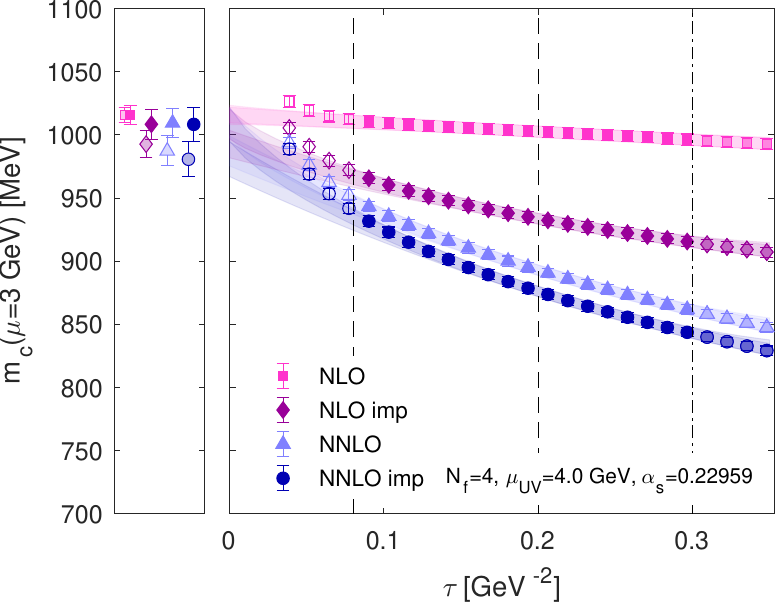} \\
  (d) Extracting the charm quark mass $m_c$ by performing $\tau\to0$ limit extrapolations.\\[3mm]
  \includegraphics[width=0.95\columnwidth]{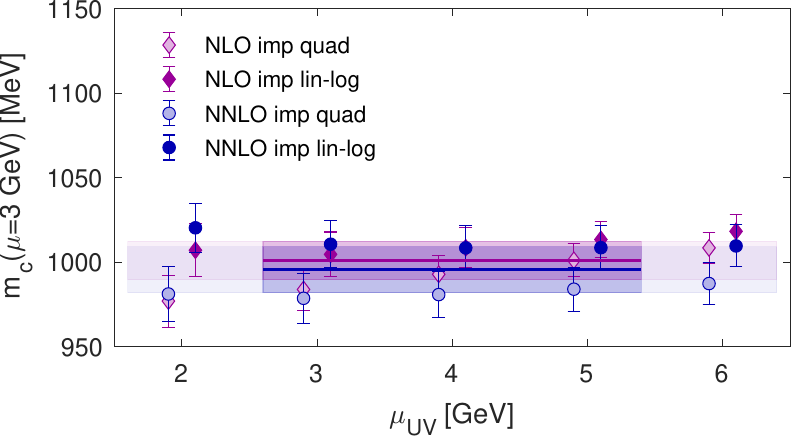} 
  (e) Dependence of $m_c$ obtained from the $\eta_c$ correlators on $\mu_\text{UV}$.
  \end{minipage}
\caption{Details of our charm-quark mass determination using $\eta_c$ correlators. Figure (a) shows the extraction of the ratios $R^{AP}(t/a;\tau/a^2)$ by performing correlated fits using time slices 42 to 46 for selected flow times $\tau/a^2$ on the F1S ensemble. For magnified plateau regions see Fig.~\ref{Fig.zoom}. Interpolating results on the coarse and medium ensembles to flow times matching values on F1S, we perform $a\to 0$ continuum limit extrapolations as shown in panels of (b). Subsequently we present $\bar R^{AP}(\tau)$ in physical units in panel (c). The large panel in Figure (d) illustrates extracting $m_c$  using $\eta_c$ correlators at the renormalization scale $\mu_\text{\uv}=4 \text{ GeV}$, converted to $\mu=3 \text{ GeV}$ using 4-loop $N_f=4$ \MSbar{} running~\cite{Chetyrkin:2000yt,Herren:2017osy}. 
Pink squares and light blue triangles denote the \nlo{} 
and \nnlo{} values, respectively.
 Purple diamonds and dark blue circles show the corresponding \rg{} improved \nlo{}
and \nnlo{} values.
The $\tau\to 0$ limit is taken in a $(\tau_\text{min},\tau_\text{max})$ range as explained in the text.   The shaded error bands show the maximal spread of quadratic or linear-log fits using the central values plus/minus their uncertainty.  Only filled symbols enter the $\tau\to 0$ extrapolation.  The small panel compares the final predictions using solid filled symbols for linear-log and shaded fillings for quadratic fits. Figure (e) demonstrates the dependence of the result for $m_c$ on the scale $\mu_\text{\uv}$. We obtain our final values by calculating correlated averages for \nlo{} and \nnlo{} using values at $\mu_{\text{UV}}=3,\, 4,\, 5\, \text{GeV}$.}
\label{Fig.etac}
\end{figure*}

\begin{figure*}[tp]
  \begin{minipage}{\columnwidth}
    \includegraphics[width=0.9\columnwidth]{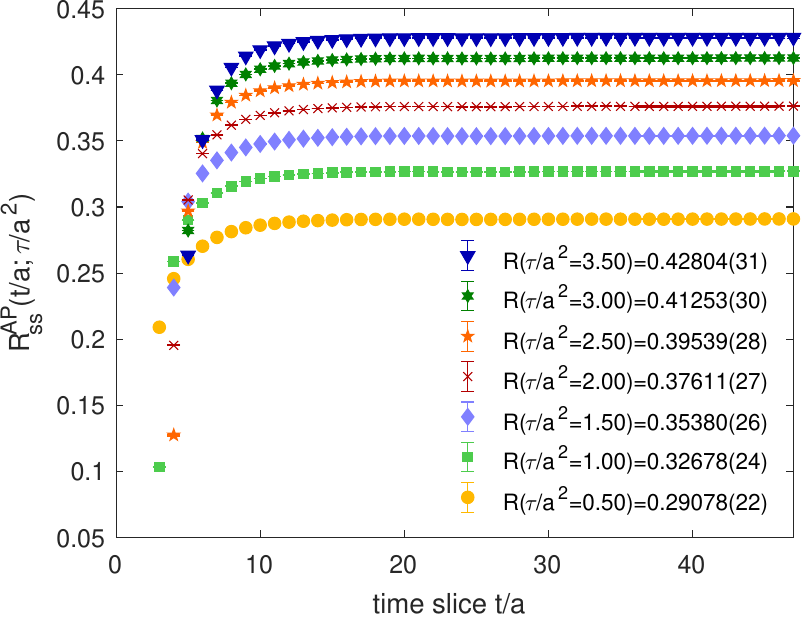} \\
    (a) Extracting the ratios $R^{AP}(t/a;\tau/a^2)$ using $\eta_s$ correlators on the F1S ensemble.\\[2mm]
    \includegraphics[width=0.92\columnwidth]{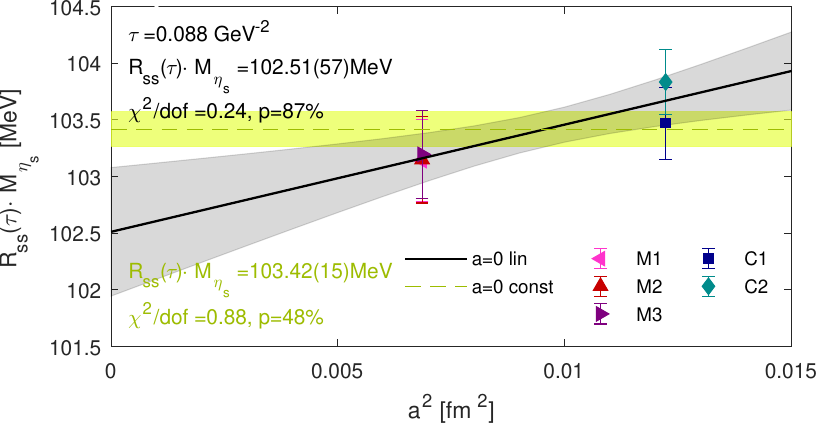}\\
    \includegraphics[width=0.92\columnwidth]{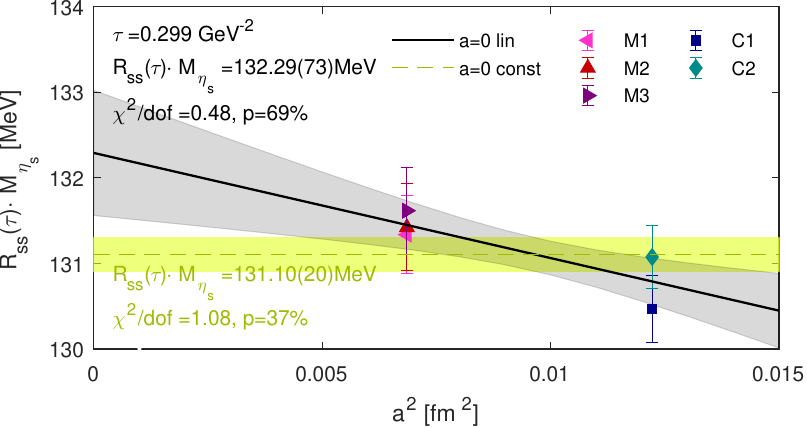} \\
    (b) Continuum limits at $\tau=0.088$ and $0.299 \text{ GeV}^{-2}$ based on an ansatz linear in $a^2$ and fitting a constant.\\[2mm]
    \includegraphics[width=0.9\columnwidth]{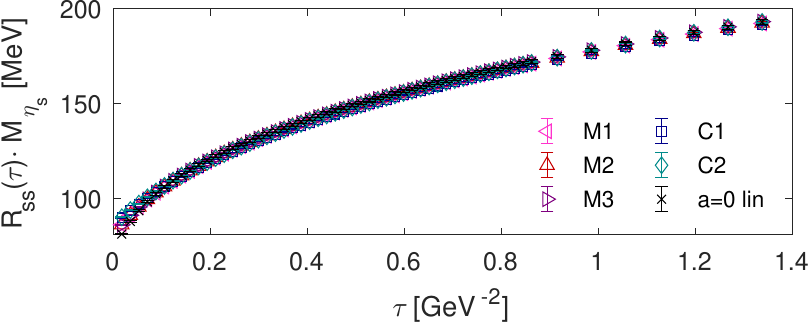}\\    
    (c) Ratios $\bar R^{AP}(\tau)$ for the $\eta_s$ meson converted to physical units and including $a\to 0$ continuum limits.   
  \end{minipage}
  \begin{minipage}{\columnwidth}
    \includegraphics[width=0.95\columnwidth]{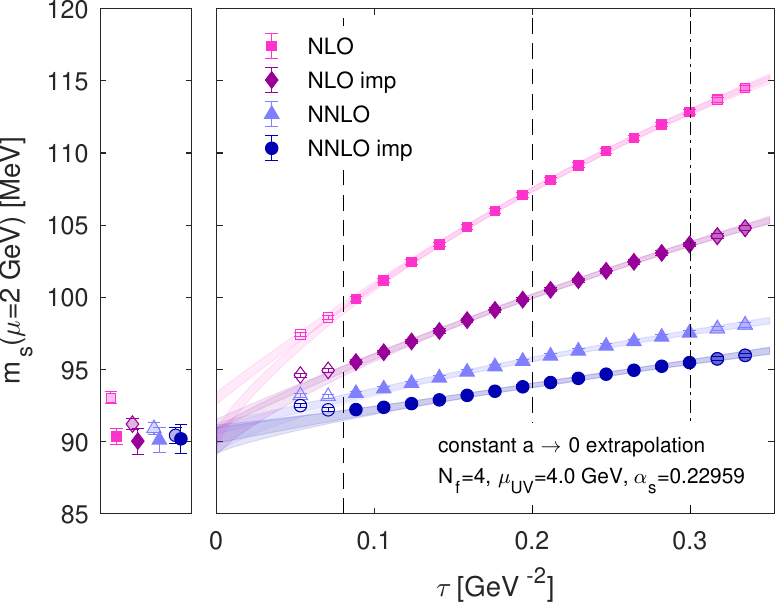}\\[-8mm]
    \includegraphics[width=0.95\columnwidth]{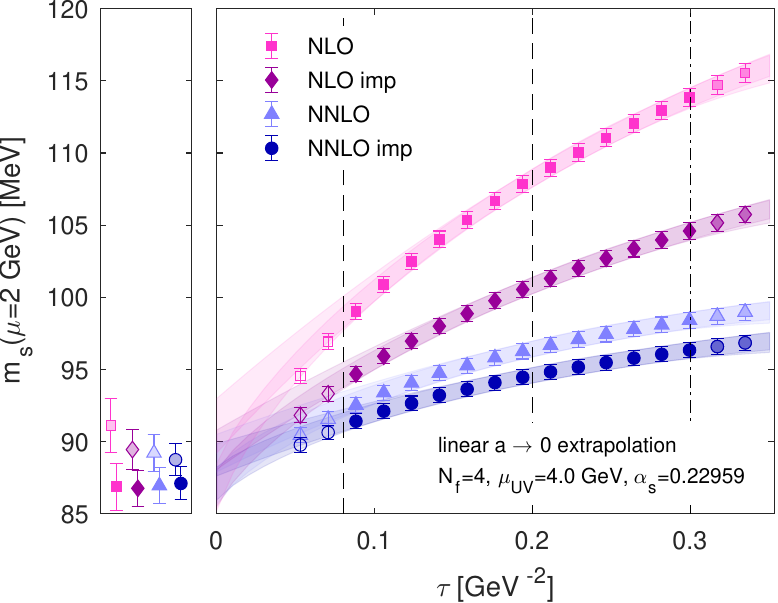}\\    
    (d) Extracting the strange-quark mass $m_s$ from $\tau\to0$  extrapolations for both choices of the $a\to 0$ continuum limit.\\[3mm]
    \includegraphics[width=0.95\columnwidth]{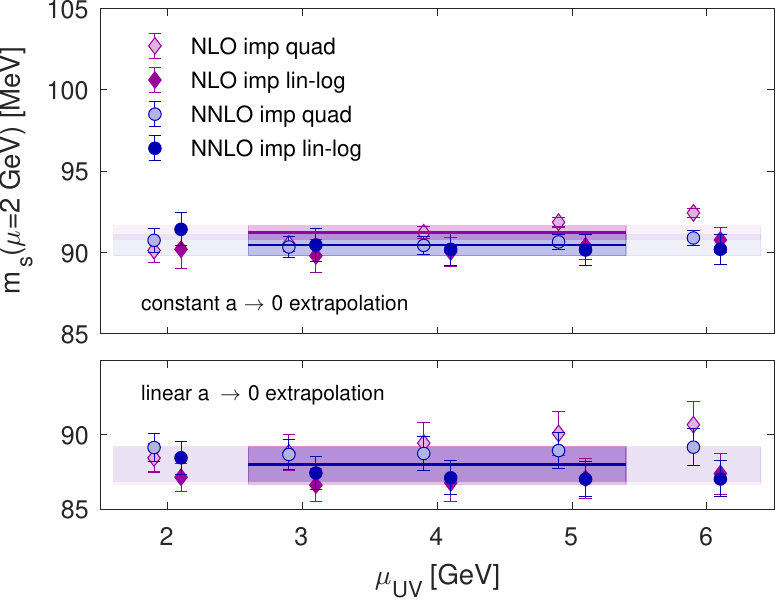}\\
    (e) Dependence of $m_s$ obtained from the $\eta_s$ correlators on $\mu_\text{UV}$ using constant or linear continuum limit extrapolation.
  \end{minipage}    
\caption{Details of our strange quark mass determination using $\eta_s$ correlators. Figure (a) shows the extraction of the ratios $R^{AP}(t/a;\tau/a^2)$ by performing correlated fits using time slices 36 to 46 for selected flow times $\tau/a^2$ on the F1S ensemble. For magnified plateau regions see Fig.~\ref{Fig.zoom}.  As pointed out in the text, we observe a mistuning of the bare strange quark mass on F1S. We therefore perform $a\to 0$ continuum limit extrapolations as shown in panels (b) using only coarse and medium data but consider both an ansatz linear in $a^2$ as well as a fit to a constant.  Next we obtain $\bar R^{AP}(\tau)$ in physical units as shown in panel (c). The large panel in Figure (d) illustrates extracting $m_s$  using $\eta_s$ correlators at the renormalization scale $\mu_\text{\uv}=4 \text{ GeV}$, converted to $\mu=2 \text{ GeV}$ using 4-loop $N_f=4$ \MSbar{} running~\cite{Chetyrkin:2000yt,Herren:2017osy}. 
Pink squares and light blue triangles denote the \nlo{} 
and \nnlo{} values, respectively.
 Purple diamonds and dark blue circles show the corresponding \rg{} improved \nlo{}
and \nnlo{} values.
The $\tau\to 0$ limit is taken in a $(\tau_\text{min},\tau_\text{max})$ range as explained in the text.   The shaded error bands show the maximal spread of quadratic or linear-log fits using the central values plus/minus their uncertainty.  Only filled symbols enter the $\tau\to 0$ extrapolation.  The small panel compares the final predictions using solid filled symbols for linear-log and shaded fillings for quadratic fits. Figure (e) demonstrates the dependence of the result for $m_s$ on the scale $\mu_\text{\uv}$. We obtain our final values by calculating correlated averages for \nlo{} and \nnlo{} using values at $\mu_{\text{UV}}=3,\, 4,\, 5\, \text{GeV}$ and then accounting for the full spread due to different choices for the $a\to 0$ extrapolation.}
\label{Fig.etas}
\end{figure*}

Repeating the steps outlined above, we analyze data that describe the charmonium state $\eta_c$ as well as the (unphysical) $(s\bar s)$-bound state, which we refer to as $\eta_s$. Details for $\eta_c$ are presented in \cref{Fig.etac} and in Tables \ref{Tab.RP}, \ref{Tab.RAP}, and \ref{Tab.ContinuumExtra}. 
After the $\tau\to 0$ extrapolation we find
\begin{align}
\begin{aligned}
&m_c^{\eta_c,\text{\nlo}}(\mu=3 \text{ GeV}) = 1001(11)_\text{GF}~\text{MeV}\,,\\
&m_c^{\eta_c,\text{\nnlo}}(\mu=3 \text{ GeV}) = 996(14)_\text{GF}~\text{MeV}\,.
\end{aligned}
\end{align}
Repeating the $\eta_c$ analysis dropping the coarse ensembles, we observe that 
the linear $a\to 0$ continuum limit using only fine and medium ensembles results 
in a shift of the central \nnlo\ value of 44~MeV, 
whereas using a fit to a constant barely changes the central values. Again we quote as
final value our \nnlo{} result and assign half of the shift of the central values as 
additional systematic uncertainties:
\begin{align}
m_c^{\eta_c}(\mu=3 \text{ GeV}) &= 996(14)_\text{GF}(22)_\text{CL}(3)_\text{PT}~\text{MeV}\,.
\end{align}

\begin{figure}[tb]
\includegraphics[width=0.95\columnwidth,valign=t]{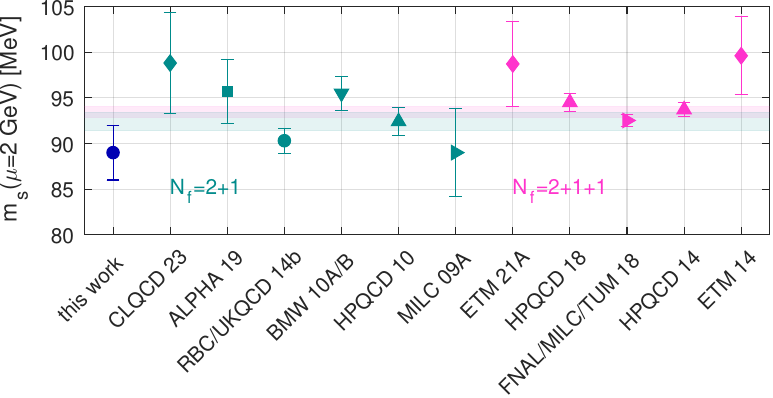}\\
\includegraphics[width=0.95\columnwidth,valign=t]{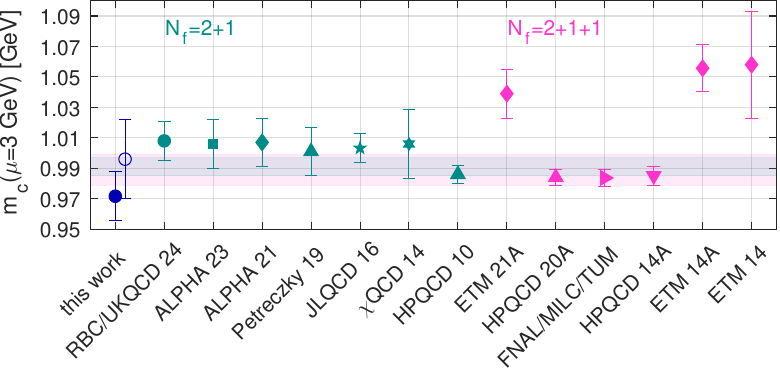}
\caption{Comparison of our strange and charm quark masses (blue circles) to the \abbrev{FLAG} averages~\cite{FlavourLatticeAveragingGroupFLAG:2024oxs} (shaded bands) for 2+1 and 2+1+1 flavors. We show the direct prediction for $m_c$ using the $\eta_c$ correlators with an open circle, and the determination from the $D_s$ meson with a filled circle. Also shown are results entering \abbrev{FLAG} averages: strange 2+1~\cite{CLQCD:2023sdb,Bruno:2019vup, RBC:2014ntl, McNeile:2010ji, BMW:2010ucx,BMW:2010skj, MILC:2009ltw}, strange 2+1+1~\cite{ExtendedTwistedMass:2021gbo,Lytle:2018evc,FermilabLattice:2018est,Chakraborty:2014aca,EuropeanTwistedMass:2014osg}, charm 2+1~\cite{Bussone:2023kag,Heitger:2021apz,Petreczky:2019ozv,Nakayama:2016atf,Yang:2014sea,McNeile:2010ji}, charm 2+1+1~\cite{ExtendedTwistedMass:2021gbo,Hatton:2020qhk,FermilabLattice:2018est,EuropeanTwistedMass:2014osg,Chakraborty:2014aca,Alexandrou:2014sha}, as well as a new result not part of the average~\cite{DelDebbio:2024hca}.}
\label{Fig.Comparison}
\end{figure}

In the case of $\eta_s$, we observe a tension in $M_{\eta_s}$ on the F1S
ensemble w.r.t.~our determinations on the coarse and medium ensembles
(cf.~\cref{Tab.M_PS}) which we attribute to a slight misdetermination of the
bare strange quark mass.
Hence we discard the F1S data for this analysis, interpolate the flow times on
the coarse ensembles to the values matching the medium ensembles, and show our
final result based on data from the medium and coarse ensembles in
Fig.~\ref{Fig.etas}. Since with only two lattice spacings a linear ansatz for
the $a\to 0$ limit is poorly constrained, we consider in addition fitting a
constant. Both $a\to 0$ extrapolations have good $p$-values
(cf.~Tab.~\ref{Tab.ContinuumExtraEtas}), so we repeat the remaining steps of
analysis by once using the linear in $a^2$ continuum limit (lin),
and once the fit to a
constant (cst). We obtain the following predictions for $m_s$:
\begin{align}
\begin{aligned}
& m_s(\mu=2\text{ GeV})^\text{\nlo,cst} = 91.24(42)_\text{GF}~\text{MeV}, \\
& m_s(\mu=2\text{ GeV})^\text{\nlo,lin} = 88.0(1.3)_\text{GF}~\text{MeV}, \\
& m_s(\mu=2\text{ GeV})^\text{\nnlo,cst} = 90.45(64)_\text{GF}~\text{MeV}, \\
& m_s(\mu=2\text{ GeV})^\text{\nnlo,lin} = 88.0(1.1)_\text{GF}~\text{MeV}, 
\end{aligned}
\end{align}
Taking the full spread between both continuum limits at the full MeV level, we find 
for both \nlo{} and \nnlo{} the same value. Hence, we quote as our final result for 
$m_s$ in the \MSbar{} scheme
\begin{align}
m_s(\mu=2\text{ GeV}) &= 89(3)_\text{GF+CL}(0)_\text{PT}~\text{MeV}.
\label{Eq.Resultetas}
\end{align}
Adding the uncertainties of our result in \cref{Eq.FinalResultDs} in quadrature and
combining it with the value for $m_s$ of
\cref{Eq.Resultetas} run up to $\mu=3$ GeV, we obtain the more precise value 
\begin{align}
m_c(\mu=3\text{ GeV}) = 972(16)~\text{MeV},
\end{align} 
which we also utilize to deduce the ratio
\begin{align}
\frac{m_c}{m_s} = 12.1(4).
\end{align}
A comparison of our values for the strange and charm quark mass to the
literature is shown in \cref{Fig.Comparison}. Also, our value for the ratio $m_c/m_s$ is in good agreement with the \abbrev{FLAG} averages~\cite{FlavourLatticeAveragingGroupFLAG:2024oxs} for 2+1~\cite{Yang:2014sea,Davies:2009ih} and 2+1+1~\cite{ExtendedTwistedMass:2021gbo,FermilabLattice:2018est,Chakraborty:2014aca,EuropeanTwistedMass:2014osg} flavors.

\section{Summary}

Gradient-flowed quark propagators provide a simple, non-perturbative method to
renormalize fermionic quantities.  Matching to the perturbative \MSbar{}
scheme can be performed using \sftx{} with improved convergence by \rg{}
  running in the \gf\ scheme.  We implement and test this idea to obtain
renormalized strange- and charm-quark masses at percent-level accuracy.  The
accuracy can easily be improved by using state-of-the-art physical-point gauge
field configurations and finer lattice spacings.  The \gf\ is a mass-dependent
renormalization scheme and the matching to the mass-independent \MSbar{}
scheme requires taking the $\tau\to 0$ limit. Determination of a
non-perturbative flowed anomalous dimension $\gamma_m^\text{\gf}(\tau)$ for
the \rg\ improvement in \cref{eq:impconversion} is an interesting task for the
future.

Since the \gf\ directly renormalizes the fermionic operator, also ``mixed
action'' setups using different fermion discretizations for light and heavy
quarks (as we do for the $D_s$ meson) can be non-perturbatively
renormalized. \pt\ only enters when matching to the continuum \MSbar{} scheme
using \sftx{}.

\section*{Acknowledgments}
We thank the \abbrev{RBC/UKQCD} Collaboration for generating and making their
gauge field ensembles publicly available. The data analyzed here were generated as part of the project to calculate heavy meson lifetimes~\cite{Lifetimes:2025}. 

We thank  Jonas Kohnen, Fabian Lange, Christopher Monahan, Matthew Rizik, Andrea Shindler, Rainer Sommer, and Tobias Tsang for fruitful discussions.
A.H.~and O.W.~are grateful for the hospitality at the Kavli Institute of Theoretical Physics at UC Santa Barbara, where part of this work was carried out during the program “What is Particle Theory?”.  A.H.~thanks the 
University of Siegen for the hospitality she received during the final stages of this work.

These computations used resources provided by the OMNI cluster at the University of Siegen, the HAWK cluster at the High-Performance Computing Center Stuttgart, and LUMI-G at the CSC data center Finland (DeiC National HPC g.a.~DEIC-SDU-L5-13 and DEIC-SDU-N5-2024053).

M.B., R.V.H., and O.W.~received support from the Deutsche Forschungsgemeinschaft (DFG, German Research Foundation) through grant 396021762 --- TRR 257 ``Particle Physics Phenomenology after the Higgs Discovery''.
M.B.~was additionally funded in part by UK STFC grant ST/X000494/1.
A.H.~acknowledges support from DOE grant DE-SC001000.
This research was supported in part by grant NSF PHY-2309135 to the Kavli Institute for Theoretical Physics (KITP).

\section*{Data availability}
The \texttt{c++} lattice \qcd{} software libraries \texttt{Grid}~\cite{Boyle:2015tjk} and \texttt{Hadrons}~\cite{Hadrons,Hadrons22} are open source and publicly available. Matthew Black implemented fermionic gradient flow~\cite{Black:2023vju,Black:2024iwb,Lifetimes:2025}
in \texttt{Hadrons}: \url{https://github.com/aportelli/Hadrons/pull/137} with examples given at \url{https://github.com/mbr-phys/HeavyMesonLifetimes}. Data for the two-point correlation functions used in this project will be made available as part of the data release for Ref.\,\cite{Lifetimes:2025}.  

\bibliography{draft.bib}

\begin{thebibliography}{73}%
\makeatletter
\providecommand \@ifxundefined [1]{%
 \@ifx{#1\undefined}
}%
\providecommand \@ifnum [1]{%
 \ifnum #1\expandafter \@firstoftwo
 \else \expandafter \@secondoftwo
 \fi
}%
\providecommand \@ifx [1]{%
 \ifx #1\expandafter \@firstoftwo
 \else \expandafter \@secondoftwo
 \fi
}%
\providecommand \natexlab [1]{#1}%
\providecommand \enquote  [1]{``#1''}%
\providecommand \bibnamefont  [1]{#1}%
\providecommand \bibfnamefont [1]{#1}%
\providecommand \citenamefont [1]{#1}%
\providecommand \href@noop [0]{\@secondoftwo}%
\providecommand \href [0]{\begingroup \@sanitize@url \@href}%
\providecommand \@href[1]{\@@startlink{#1}\@@href}%
\providecommand \@@href[1]{\endgroup#1\@@endlink}%
\providecommand \@sanitize@url [0]{\catcode `\\12\catcode `\$12\catcode
  `\&12\catcode `\#12\catcode `\^12\catcode `\_12\catcode `\%12\relax}%
\providecommand \@@startlink[1]{}%
\providecommand \@@endlink[0]{}%
\providecommand \url  [0]{\begingroup\@sanitize@url \@url }%
\providecommand \@url [1]{\endgroup\@href {#1}{\urlprefix }}%
\providecommand \urlprefix  [0]{URL }%
\providecommand \Eprint [0]{\href }%
\providecommand \doibase [0]{https://doi.org/}%
\providecommand \selectlanguage [0]{\@gobble}%
\providecommand \bibinfo  [0]{\@secondoftwo}%
\providecommand \bibfield  [0]{\@secondoftwo}%
\providecommand \translation [1]{[#1]}%
\providecommand \BibitemOpen [0]{}%
\providecommand \bibitemStop [0]{}%
\providecommand \bibitemNoStop [0]{.\EOS\space}%
\providecommand \EOS [0]{\spacefactor3000\relax}%
\providecommand \BibitemShut  [1]{\csname bibitem#1\endcsname}%
\let\auto@bib@innerbib\@empty
\bibitem [{\citenamefont {Lüscher}\ \emph {et~al.}(1992)\citenamefont
  {Lüscher}, \citenamefont {Narayanan}, \citenamefont {Weisz},\ and\
  \citenamefont {Wolff}}]{Luscher:1992an}%
  \BibitemOpen
  \bibfield  {author} {\bibinfo {author} {\bibfnamefont {M.}~\bibnamefont
  {Lüscher}}, \bibinfo {author} {\bibfnamefont {R.}~\bibnamefont {Narayanan}},
  \bibinfo {author} {\bibfnamefont {P.}~\bibnamefont {Weisz}},\ and\ \bibinfo
  {author} {\bibfnamefont {U.}~\bibnamefont {Wolff}},\ }\bibfield  {title}
  {\bibinfo {title} {{The Schr\"odinger functional: A Renormalizable probe for
  nonAbelian gauge theories}},\ }\href
  {https://doi.org/10.1016/0550-3213(92)90466-O} {\bibfield  {journal}
  {\bibinfo  {journal} {Nucl. Phys. B}\ }\textbf {\bibinfo {volume} {384}},\
  \bibinfo {pages} {168} (\bibinfo {year} {1992})},\ \Eprint
  {https://arxiv.org/abs/hep-lat/9207009} {arXiv:hep-lat/9207009} \BibitemShut
  {NoStop}%
\bibitem [{\citenamefont {Lüscher}\ \emph {et~al.}(1994)\citenamefont
  {Lüscher}, \citenamefont {Sommer}, \citenamefont {Weisz},\ and\
  \citenamefont {Wolff}}]{Luscher:1993gh}%
  \BibitemOpen
  \bibfield  {author} {\bibinfo {author} {\bibfnamefont {M.}~\bibnamefont
  {Lüscher}}, \bibinfo {author} {\bibfnamefont {R.}~\bibnamefont {Sommer}},
  \bibinfo {author} {\bibfnamefont {P.}~\bibnamefont {Weisz}},\ and\ \bibinfo
  {author} {\bibfnamefont {U.}~\bibnamefont {Wolff}},\ }\bibfield  {title}
  {\bibinfo {title} {{A Precise determination of the running coupling in the
  SU(3) Yang-Mills theory}},\ }\href
  {https://doi.org/10.1016/0550-3213(94)90629-7} {\bibfield  {journal}
  {\bibinfo  {journal} {Nucl. Phys. B}\ }\textbf {\bibinfo {volume} {413}},\
  \bibinfo {pages} {481} (\bibinfo {year} {1994})},\ \Eprint
  {https://arxiv.org/abs/hep-lat/9309005} {arXiv:hep-lat/9309005} \BibitemShut
  {NoStop}%
\bibitem [{\citenamefont {Sint}(1994)}]{Sint:1993un}%
  \BibitemOpen
  \bibfield  {author} {\bibinfo {author} {\bibfnamefont {S.}~\bibnamefont
  {Sint}},\ }\bibfield  {title} {\bibinfo {title} {{On the Schrödinger
  functional in QCD}},\ }\href {https://doi.org/10.1016/0550-3213(94)90228-3}
  {\bibfield  {journal} {\bibinfo  {journal} {Nucl. Phys. B}\ }\textbf
  {\bibinfo {volume} {421}},\ \bibinfo {pages} {135} (\bibinfo {year}
  {1994})},\ \Eprint {https://arxiv.org/abs/hep-lat/9312079}
  {arXiv:hep-lat/9312079} \BibitemShut {NoStop}%
\bibitem [{\citenamefont {Capitani}\ \emph {et~al.}(1999)\citenamefont
  {Capitani}, \citenamefont {L\"uscher}, \citenamefont {Sommer},\ and\
  \citenamefont {Wittig}}]{Capitani:1998mq}%
  \BibitemOpen
  \bibfield  {author} {\bibinfo {author} {\bibfnamefont {S.}~\bibnamefont
  {Capitani}}, \bibinfo {author} {\bibfnamefont {M.}~\bibnamefont {L\"uscher}},
  \bibinfo {author} {\bibfnamefont {R.}~\bibnamefont {Sommer}},\ and\ \bibinfo
  {author} {\bibfnamefont {H.}~\bibnamefont {Wittig}},\ }\bibfield  {title}
  {\bibinfo {title} {{Non-perturbative quark mass renormalization in quenched
  lattice QCD}},\ }\href {https://doi.org/10.1016/S0550-3213(98)00857-8}
  {\bibfield  {journal} {\bibinfo  {journal} {Nucl. Phys. B}\ }\textbf
  {\bibinfo {volume} {544}},\ \bibinfo {pages} {669} (\bibinfo {year}
  {1999})},\ \bibinfo {note} {[Erratum: Nucl.Phys.B 582, 762--762 (2000)]},\
  \Eprint {https://arxiv.org/abs/hep-lat/9810063} {arXiv:hep-lat/9810063}
  \BibitemShut {NoStop}%
\bibitem [{\citenamefont {Martinelli}\ \emph {et~al.}(1995)\citenamefont
  {Martinelli}, \citenamefont {Pittori}, \citenamefont {Sachrajda},
  \citenamefont {Testa},\ and\ \citenamefont {Vladikas}}]{Martinelli:1994ty}%
  \BibitemOpen
  \bibfield  {author} {\bibinfo {author} {\bibfnamefont {G.}~\bibnamefont
  {Martinelli}}, \bibinfo {author} {\bibfnamefont {C.}~\bibnamefont {Pittori}},
  \bibinfo {author} {\bibfnamefont {C.~T.}\ \bibnamefont {Sachrajda}}, \bibinfo
  {author} {\bibfnamefont {M.}~\bibnamefont {Testa}},\ and\ \bibinfo {author}
  {\bibfnamefont {A.}~\bibnamefont {Vladikas}},\ }\bibfield  {title} {\bibinfo
  {title} {{A General method for nonperturbative renormalization of lattice
  operators}},\ }\href {https://doi.org/10.1016/0550-3213(95)00126-D}
  {\bibfield  {journal} {\bibinfo  {journal} {Nucl. Phys. B}\ }\textbf
  {\bibinfo {volume} {445}},\ \bibinfo {pages} {81} (\bibinfo {year} {1995})},\
  \Eprint {https://arxiv.org/abs/hep-lat/9411010} {arXiv:hep-lat/9411010}
  \BibitemShut {NoStop}%
\bibitem [{\citenamefont {Martinelli}\ \emph {et~al.}(1997)\citenamefont
  {Martinelli}, \citenamefont {Rossi}, \citenamefont {Sachrajda}, \citenamefont
  {Sharpe}, \citenamefont {Talevi},\ and\ \citenamefont
  {Testa}}]{Martinelli:1997zc}%
  \BibitemOpen
  \bibfield  {author} {\bibinfo {author} {\bibfnamefont {G.}~\bibnamefont
  {Martinelli}}, \bibinfo {author} {\bibfnamefont {G.~C.}\ \bibnamefont
  {Rossi}}, \bibinfo {author} {\bibfnamefont {C.~T.}\ \bibnamefont
  {Sachrajda}}, \bibinfo {author} {\bibfnamefont {S.~R.}\ \bibnamefont
  {Sharpe}}, \bibinfo {author} {\bibfnamefont {M.}~\bibnamefont {Talevi}},\
  and\ \bibinfo {author} {\bibfnamefont {M.}~\bibnamefont {Testa}},\ }\bibfield
   {title} {\bibinfo {title} {{Nonperturbative improvement of composite
  operators with Wilson fermions}},\ }\href
  {https://doi.org/10.1016/S0370-2693(97)01007-1} {\bibfield  {journal}
  {\bibinfo  {journal} {Phys. Lett. B}\ }\textbf {\bibinfo {volume} {411}},\
  \bibinfo {pages} {141} (\bibinfo {year} {1997})},\ \Eprint
  {https://arxiv.org/abs/hep-lat/9705018} {arXiv:hep-lat/9705018} \BibitemShut
  {NoStop}%
\bibitem [{\citenamefont {Gimenez}\ \emph {et~al.}(2004)\citenamefont
  {Gimenez}, \citenamefont {Giusti}, \citenamefont {Guerriero}, \citenamefont
  {Lubicz}, \citenamefont {Martinelli}, \citenamefont {Petrarca}, \citenamefont
  {Reyes}, \citenamefont {Taglienti},\ and\ \citenamefont
  {Trevigne}}]{Gimenez:2004me}%
  \BibitemOpen
  \bibfield  {author} {\bibinfo {author} {\bibfnamefont {V.}~\bibnamefont
  {Gimenez}}, \bibinfo {author} {\bibfnamefont {L.}~\bibnamefont {Giusti}},
  \bibinfo {author} {\bibfnamefont {S.}~\bibnamefont {Guerriero}}, \bibinfo
  {author} {\bibfnamefont {V.}~\bibnamefont {Lubicz}}, \bibinfo {author}
  {\bibfnamefont {G.}~\bibnamefont {Martinelli}}, \bibinfo {author}
  {\bibfnamefont {S.}~\bibnamefont {Petrarca}}, \bibinfo {author}
  {\bibfnamefont {J.}~\bibnamefont {Reyes}}, \bibinfo {author} {\bibfnamefont
  {B.}~\bibnamefont {Taglienti}},\ and\ \bibinfo {author} {\bibfnamefont
  {E.}~\bibnamefont {Trevigne}},\ }\bibfield  {title} {\bibinfo {title}
  {{Non-perturbative renormalization of lattice operators in coordinate
  space}},\ }\href {https://doi.org/10.1016/j.physletb.2004.07.053} {\bibfield
  {journal} {\bibinfo  {journal} {Phys. Lett. B}\ }\textbf {\bibinfo {volume}
  {598}},\ \bibinfo {pages} {227} (\bibinfo {year} {2004})},\ \Eprint
  {https://arxiv.org/abs/hep-lat/0406019} {arXiv:hep-lat/0406019} \BibitemShut
  {NoStop}%
\bibitem [{\citenamefont {Suzuki}(2013)}]{Suzuki:2013gza}%
  \BibitemOpen
  \bibfield  {author} {\bibinfo {author} {\bibfnamefont {H.}~\bibnamefont
  {Suzuki}},\ }\bibfield  {title} {\bibinfo {title}
  {{Energy\textendash{}momentum tensor from the Yang\textendash{}Mills gradient
  flow}},\ }\href {https://doi.org/10.1093/ptep/ptt059} {\bibfield  {journal}
  {\bibinfo  {journal} {PTEP}\ }\textbf {\bibinfo {volume} {2013}},\ \bibinfo
  {pages} {083B03} (\bibinfo {year} {2013})},\ \bibinfo {note} {[Erratum: PTEP
  2015, 079201 (2015)]},\ \Eprint {https://arxiv.org/abs/1304.0533}
  {arXiv:1304.0533 [hep-lat]} \BibitemShut {NoStop}%
\bibitem [{\citenamefont {Black}\ \emph {et~al.}(2024)\citenamefont {Black},
  \citenamefont {Harlander}, \citenamefont {Lange}, \citenamefont {Rago},
  \citenamefont {Shindler},\ and\ \citenamefont {Witzel}}]{Black:2023vju}%
  \BibitemOpen
  \bibfield  {author} {\bibinfo {author} {\bibfnamefont {M.}~\bibnamefont
  {Black}}, \bibinfo {author} {\bibfnamefont {R.}~\bibnamefont {Harlander}},
  \bibinfo {author} {\bibfnamefont {F.}~\bibnamefont {Lange}}, \bibinfo
  {author} {\bibfnamefont {A.}~\bibnamefont {Rago}}, \bibinfo {author}
  {\bibfnamefont {A.}~\bibnamefont {Shindler}},\ and\ \bibinfo {author}
  {\bibfnamefont {O.}~\bibnamefont {Witzel}},\ }\bibfield  {title} {\bibinfo
  {title} {{Using Gradient Flow to Renormalise Matrix Elements for Meson Mixing
  and Lifetimes}},\ }\href {https://doi.org/10.22323/1.453.0263} {\bibfield
  {journal} {\bibinfo  {journal} {PoS}\ }\textbf {\bibinfo {volume}
  {LATTICE2023}},\ \bibinfo {pages} {263} (\bibinfo {year} {2024})},\ \Eprint
  {https://arxiv.org/abs/2310.18059} {arXiv:2310.18059 [hep-lat]} \BibitemShut
  {NoStop}%
\bibitem [{\citenamefont {Black}\ \emph
  {et~al.}(2025{\natexlab{a}})\citenamefont {Black}, \citenamefont {Harlander},
  \citenamefont {Lange}, \citenamefont {Rago}, \citenamefont {Shindler},\ and\
  \citenamefont {Witzel}}]{Black:2024iwb}%
  \BibitemOpen
  \bibfield  {author} {\bibinfo {author} {\bibfnamefont {M.}~\bibnamefont
  {Black}}, \bibinfo {author} {\bibfnamefont {R.}~\bibnamefont {Harlander}},
  \bibinfo {author} {\bibfnamefont {F.}~\bibnamefont {Lange}}, \bibinfo
  {author} {\bibfnamefont {A.}~\bibnamefont {Rago}}, \bibinfo {author}
  {\bibfnamefont {A.}~\bibnamefont {Shindler}},\ and\ \bibinfo {author}
  {\bibfnamefont {O.}~\bibnamefont {Witzel}},\ }\bibfield  {title} {\bibinfo
  {title} {{Gradient Flow Renormalisation for Meson Mixing and Lifetimes}},\
  }\href {https://doi.org/10.22323/1.466.0243} {\bibfield  {journal} {\bibinfo
  {journal} {PoS}\ }\textbf {\bibinfo {volume} {LATTICE2024}},\ \bibinfo
  {pages} {243} (\bibinfo {year} {2025}{\natexlab{a}})},\ \Eprint
  {https://arxiv.org/abs/2409.18891} {arXiv:2409.18891 [hep-lat]} \BibitemShut
  {NoStop}%
\bibitem [{\citenamefont {Lüscher}\ and\ \citenamefont
  {Weisz}(2011)}]{Luscher:2011bx}%
  \BibitemOpen
  \bibfield  {author} {\bibinfo {author} {\bibfnamefont {M.}~\bibnamefont
  {Lüscher}}\ and\ \bibinfo {author} {\bibfnamefont {P.}~\bibnamefont
  {Weisz}},\ }\bibfield  {title} {\bibinfo {title} {{Perturbative analysis of
  the gradient flow in non-abelian gauge theories}},\ }\href
  {https://doi.org/10.1007/JHEP02(2011)051} {\bibfield  {journal} {\bibinfo
  {journal} {JHEP}\ }\textbf {\bibinfo {volume} {1102}},\ \bibinfo {pages}
  {051}},\ \Eprint {https://arxiv.org/abs/1101.0963} {arXiv:1101.0963 [hep-th]}
  \BibitemShut {NoStop}%
\bibitem [{\citenamefont {Lüscher}(2014)}]{Luscher:2013vga}%
  \BibitemOpen
  \bibfield  {author} {\bibinfo {author} {\bibfnamefont {M.}~\bibnamefont
  {Lüscher}},\ }\bibfield  {title} {\bibinfo {title} {{Future applications of
  the Yang-Mills gradient flow in lattice QCD}},\ }\href
  {https://doi.org/10.22323/1.187.0016} {\bibfield  {journal} {\bibinfo
  {journal} {PoS}\ }\textbf {\bibinfo {volume} {LATTICE2013}},\ \bibinfo
  {pages} {016} (\bibinfo {year} {2014})},\ \Eprint
  {https://arxiv.org/abs/1308.5598} {arXiv:1308.5598 [hep-lat]} \BibitemShut
  {NoStop}%
\bibitem [{\citenamefont {Narayanan}\ and\ \citenamefont
  {Neuberger}(2006)}]{Narayanan:2006rf}%
  \BibitemOpen
  \bibfield  {author} {\bibinfo {author} {\bibfnamefont {R.}~\bibnamefont
  {Narayanan}}\ and\ \bibinfo {author} {\bibfnamefont {H.}~\bibnamefont
  {Neuberger}},\ }\bibfield  {title} {\bibinfo {title} {{Infinite N phase
  transitions in continuum Wilson loop operators}},\ }\href
  {https://doi.org/10.1088/1126-6708/2006/03/064} {\bibfield  {journal}
  {\bibinfo  {journal} {JHEP}\ }\textbf {\bibinfo {volume} {0603}},\ \bibinfo
  {pages} {064}},\ \Eprint {https://arxiv.org/abs/hep-th/0601210}
  {arXiv:hep-th/0601210 [hep-th]} \BibitemShut {NoStop}%
\bibitem [{\citenamefont {Lüscher}(2010{\natexlab{a}})}]{Luscher:2010iy}%
  \BibitemOpen
  \bibfield  {author} {\bibinfo {author} {\bibfnamefont {M.}~\bibnamefont
  {Lüscher}},\ }\bibfield  {title} {\bibinfo {title} {{Properties and uses of
  the Wilson flow in lattice QCD}},\ }\href
  {https://doi.org/10.1007/JHEP08(2010)071} {\bibfield  {journal} {\bibinfo
  {journal} {JHEP}\ }\textbf {\bibinfo {volume} {1008}},\ \bibinfo {pages}
  {071}},\ \Eprint {https://arxiv.org/abs/1006.4518} {arXiv:1006.4518
  [hep-lat]} \BibitemShut {NoStop}%
\bibitem [{\citenamefont {Lüscher}(2013)}]{Luscher:2013cpa}%
  \BibitemOpen
  \bibfield  {author} {\bibinfo {author} {\bibfnamefont {M.}~\bibnamefont
  {Lüscher}},\ }\bibfield  {title} {\bibinfo {title} {{Chiral symmetry and the
  Yang--Mills gradient flow}},\ }\href
  {https://doi.org/10.1007/JHEP04(2013)123} {\bibfield  {journal} {\bibinfo
  {journal} {JHEP}\ }\textbf {\bibinfo {volume} {1304}},\ \bibinfo {pages}
  {123}},\ \Eprint {https://arxiv.org/abs/1302.5246} {arXiv:1302.5246
  [hep-lat]} \BibitemShut {NoStop}%
\bibitem [{\citenamefont {Takaura}\ \emph {et~al.}(2025)\citenamefont
  {Takaura}, \citenamefont {Harlander},\ and\ \citenamefont
  {Lange}}]{Takaura:2025pao}%
  \BibitemOpen
  \bibfield  {author} {\bibinfo {author} {\bibfnamefont {H.}~\bibnamefont
  {Takaura}}, \bibinfo {author} {\bibfnamefont {R.~V.}\ \bibnamefont
  {Harlander}},\ and\ \bibinfo {author} {\bibfnamefont {F.}~\bibnamefont
  {Lange}},\ }\bibfield  {title} {\bibinfo {title} {{A new approach to quark
  mass determination using the gradient flow}},\ }\href@noop {} {\  (\bibinfo
  {year} {2025})},\ \Eprint {https://arxiv.org/abs/2506.09537}
  {arXiv:2506.09537 [hep-lat]} \BibitemShut {NoStop}%
\bibitem [{\citenamefont {Harlander}\ \emph {et~al.}(2020)\citenamefont
  {Harlander}, \citenamefont {Lange},\ and\ \citenamefont
  {Neumann}}]{Harlander:2020duo}%
  \BibitemOpen
  \bibfield  {author} {\bibinfo {author} {\bibfnamefont {R.~V.}\ \bibnamefont
  {Harlander}}, \bibinfo {author} {\bibfnamefont {F.}~\bibnamefont {Lange}},\
  and\ \bibinfo {author} {\bibfnamefont {T.}~\bibnamefont {Neumann}},\
  }\bibfield  {title} {\bibinfo {title} {{Hadronic vacuum polarization using
  gradient flow}},\ }\href {https://doi.org/10.1007/JHEP08(2020)109} {\bibfield
   {journal} {\bibinfo  {journal} {JHEP}\ }\textbf {\bibinfo {volume} {08}},\
  \bibinfo {pages} {109}},\ \Eprint {https://arxiv.org/abs/2007.01057}
  {arXiv:2007.01057 [hep-lat]} \BibitemShut {NoStop}%
\bibitem [{\citenamefont {Hasenfratz}\ \emph {et~al.}(2022)\citenamefont
  {Hasenfratz}, \citenamefont {Monahan}, \citenamefont {Rizik}, \citenamefont
  {Shindler},\ and\ \citenamefont {Witzel}}]{Hasenfratz:2022wll}%
  \BibitemOpen
  \bibfield  {author} {\bibinfo {author} {\bibfnamefont {A.}~\bibnamefont
  {Hasenfratz}}, \bibinfo {author} {\bibfnamefont {C.~J.}\ \bibnamefont
  {Monahan}}, \bibinfo {author} {\bibfnamefont {M.~D.}\ \bibnamefont {Rizik}},
  \bibinfo {author} {\bibfnamefont {A.}~\bibnamefont {Shindler}},\ and\
  \bibinfo {author} {\bibfnamefont {O.}~\bibnamefont {Witzel}},\ }\bibfield
  {title} {\bibinfo {title} {{A novel nonperturbative renormalization scheme
  for local operators}},\ }\href {https://doi.org/10.22323/1.396.0155}
  {\bibfield  {journal} {\bibinfo  {journal} {PoS}\ }\textbf {\bibinfo {volume}
  {LATTICE2021}},\ \bibinfo {pages} {155} (\bibinfo {year} {2022})},\ \Eprint
  {https://arxiv.org/abs/2201.09740} {arXiv:2201.09740 [hep-lat]} \BibitemShut
  {NoStop}%
\bibitem [{\citenamefont {Carosso}\ \emph {et~al.}(2018)\citenamefont
  {Carosso}, \citenamefont {Hasenfratz},\ and\ \citenamefont
  {Neil}}]{Carosso:2018bmz}%
  \BibitemOpen
  \bibfield  {author} {\bibinfo {author} {\bibfnamefont {A.}~\bibnamefont
  {Carosso}}, \bibinfo {author} {\bibfnamefont {A.}~\bibnamefont
  {Hasenfratz}},\ and\ \bibinfo {author} {\bibfnamefont {E.~T.}\ \bibnamefont
  {Neil}},\ }\bibfield  {title} {\bibinfo {title} {{Nonperturbative
  Renormalization of Operators in Near-Conformal Systems Using Gradient
  Flows}},\ }\href {https://doi.org/10.1103/PhysRevLett.121.201601} {\bibfield
  {journal} {\bibinfo  {journal} {Phys. Rev. Lett.}\ }\textbf {\bibinfo
  {volume} {121}},\ \bibinfo {pages} {201601} (\bibinfo {year} {2018})},\
  \Eprint {https://arxiv.org/abs/1806.01385} {arXiv:1806.01385 [hep-lat]}
  \BibitemShut {NoStop}%
\bibitem [{\citenamefont {Artz}\ \emph {et~al.}(2019)\citenamefont {Artz},
  \citenamefont {Harlander}, \citenamefont {Lange}, \citenamefont {Neumann},\
  and\ \citenamefont {Prausa}}]{Artz:2019bpr}%
  \BibitemOpen
  \bibfield  {author} {\bibinfo {author} {\bibfnamefont {J.}~\bibnamefont
  {Artz}}, \bibinfo {author} {\bibfnamefont {R.~V.}\ \bibnamefont {Harlander}},
  \bibinfo {author} {\bibfnamefont {F.}~\bibnamefont {Lange}}, \bibinfo
  {author} {\bibfnamefont {T.}~\bibnamefont {Neumann}},\ and\ \bibinfo {author}
  {\bibfnamefont {M.}~\bibnamefont {Prausa}},\ }\bibfield  {title} {\bibinfo
  {title} {{Results and techniques for higher order calculations within the
  gradient-flow formalism}},\ }\href {https://doi.org/10.1007/JHEP06(2019)121,
  10.1007/JHEP10(2019)032} {\bibfield  {journal} {\bibinfo  {journal} {JHEP}\
  }\textbf {\bibinfo {volume} {06}},\ \bibinfo {pages} {121}},\ \bibinfo {note}
  {[erratum: JHEP10,032(2019)]},\ \Eprint {https://arxiv.org/abs/1905.00882}
  {arXiv:1905.00882 [hep-lat]} \BibitemShut {NoStop}%
\bibitem [{\citenamefont {Lüscher}(2010{\natexlab{b}})}]{Luscher:2009eq}%
  \BibitemOpen
  \bibfield  {author} {\bibinfo {author} {\bibfnamefont {M.}~\bibnamefont
  {Lüscher}},\ }\bibfield  {title} {\bibinfo {title} {{Trivializing maps, the
  Wilson flow and the HMC algorithm}},\ }\href
  {https://doi.org/10.1007/s00220-009-0953-7} {\bibfield  {journal} {\bibinfo
  {journal} {Commun. Math. Phys.}\ }\textbf {\bibinfo {volume} {293}},\
  \bibinfo {pages} {899} (\bibinfo {year} {2010}{\natexlab{b}})},\ \Eprint
  {https://arxiv.org/abs/0907.5491} {arXiv:0907.5491 [hep-lat]} \BibitemShut
  {NoStop}%
\bibitem [{\citenamefont {Endo}\ \emph {et~al.}(2015)\citenamefont {Endo},
  \citenamefont {Hieda}, \citenamefont {Miura},\ and\ \citenamefont
  {Suzuki}}]{Endo:2015iea}%
  \BibitemOpen
  \bibfield  {author} {\bibinfo {author} {\bibfnamefont {T.}~\bibnamefont
  {Endo}}, \bibinfo {author} {\bibfnamefont {K.}~\bibnamefont {Hieda}},
  \bibinfo {author} {\bibfnamefont {D.}~\bibnamefont {Miura}},\ and\ \bibinfo
  {author} {\bibfnamefont {H.}~\bibnamefont {Suzuki}},\ }\bibfield  {title}
  {\bibinfo {title} {{Universal formula for the flavor non-singlet axial-vector
  current from the gradient flow}},\ }\href
  {https://doi.org/10.1093/ptep/ptv058} {\bibfield  {journal} {\bibinfo
  {journal} {PTEP}\ }\textbf {\bibinfo {volume} {2015}},\ \bibinfo {pages}
  {053B03} (\bibinfo {year} {2015})},\ \Eprint
  {https://arxiv.org/abs/1502.01809} {arXiv:1502.01809 [hep-lat]} \BibitemShut
  {NoStop}%
\bibitem [{\citenamefont {Borgulat}\ \emph {et~al.}(2024)\citenamefont
  {Borgulat}, \citenamefont {Harlander}, \citenamefont {Kohnen},\ and\
  \citenamefont {Lange}}]{Borgulat:2023xml}%
  \BibitemOpen
  \bibfield  {author} {\bibinfo {author} {\bibfnamefont {J.}~\bibnamefont
  {Borgulat}}, \bibinfo {author} {\bibfnamefont {R.~V.}\ \bibnamefont
  {Harlander}}, \bibinfo {author} {\bibfnamefont {J.~T.}\ \bibnamefont
  {Kohnen}},\ and\ \bibinfo {author} {\bibfnamefont {F.}~\bibnamefont
  {Lange}},\ }\bibfield  {title} {\bibinfo {title} {{Short-flow-time expansion
  of quark bilinears through next-to-next-to-leading order QCD}},\ }\href
  {https://doi.org/10.1007/JHEP05(2024)179} {\bibfield  {journal} {\bibinfo
  {journal} {JHEP}\ }\textbf {\bibinfo {volume} {05}},\ \bibinfo {pages}
  {179}},\ \Eprint {https://arxiv.org/abs/2311.16799} {arXiv:2311.16799
  [hep-lat]} \BibitemShut {NoStop}%
\bibitem [{\citenamefont {Borgulat}\ \emph {et~al.}(2025)\citenamefont
  {Borgulat}, \citenamefont {Felten}, \citenamefont {Harlander},\ and\
  \citenamefont {Kohnen}}]{Borgulat:2025gys}%
  \BibitemOpen
  \bibfield  {author} {\bibinfo {author} {\bibfnamefont {J.}~\bibnamefont
  {Borgulat}}, \bibinfo {author} {\bibfnamefont {N.}~\bibnamefont {Felten}},
  \bibinfo {author} {\bibfnamefont {R.}~\bibnamefont {Harlander}},\ and\
  \bibinfo {author} {\bibfnamefont {J.~T.}\ \bibnamefont {Kohnen}},\ }\bibfield
   {title} {\bibinfo {title} {{Two-loop gradient-flow renormalization of scalar
  QCD}},\ }\href {https://doi.org/10.21468/SciPostPhysCore.8.1.032} {\bibfield
  {journal} {\bibinfo  {journal} {SciPost Phys. Core}\ }\textbf {\bibinfo
  {volume} {8}},\ \bibinfo {pages} {032} (\bibinfo {year} {2025})},\ \Eprint
  {https://arxiv.org/abs/2501.07150} {arXiv:2501.07150 [hep-ph]} \BibitemShut
  {NoStop}%
\bibitem [{\citenamefont {Harlander}\ \emph {et~al.}(2018)\citenamefont
  {Harlander}, \citenamefont {Kluth},\ and\ \citenamefont
  {Lange}}]{Harlander:2018zpi}%
  \BibitemOpen
  \bibfield  {author} {\bibinfo {author} {\bibfnamefont {R.~V.}\ \bibnamefont
  {Harlander}}, \bibinfo {author} {\bibfnamefont {Y.}~\bibnamefont {Kluth}},\
  and\ \bibinfo {author} {\bibfnamefont {F.}~\bibnamefont {Lange}},\ }\bibfield
   {title} {\bibinfo {title} {{The two-loop energy–momentum tensor within the
  gradient-flow formalism}},\ }\href
  {https://doi.org/10.1140/epjc/s10052-019-7327-x,
  10.1140/epjc/s10052-018-6415-7} {\bibfield  {journal} {\bibinfo  {journal}
  {Eur. Phys. J.}\ }\textbf {\bibinfo {volume} {C78}},\ \bibinfo {pages} {944}
  (\bibinfo {year} {2018})},\ \bibinfo {note} {[Erratum: Eur. Phys.
  J.C79,no.10,858(2019)]},\ \Eprint {https://arxiv.org/abs/1808.09837}
  {arXiv:1808.09837 [hep-lat]} \BibitemShut {NoStop}%
\bibitem [{\citenamefont {Hasenfratz}\ \emph
  {et~al.}(2023{\natexlab{a}})\citenamefont {Hasenfratz}, \citenamefont {Neil},
  \citenamefont {Shamir}, \citenamefont {Svetitsky},\ and\ \citenamefont
  {Witzel}}]{Hasenfratz:2023sqa}%
  \BibitemOpen
  \bibfield  {author} {\bibinfo {author} {\bibfnamefont {A.}~\bibnamefont
  {Hasenfratz}}, \bibinfo {author} {\bibfnamefont {E.~T.}\ \bibnamefont
  {Neil}}, \bibinfo {author} {\bibfnamefont {Y.}~\bibnamefont {Shamir}},
  \bibinfo {author} {\bibfnamefont {B.}~\bibnamefont {Svetitsky}},\ and\
  \bibinfo {author} {\bibfnamefont {O.}~\bibnamefont {Witzel}},\ }\bibfield
  {title} {\bibinfo {title} {{Infrared fixed point and anomalous dimensions in
  a composite Higgs model}},\ }\href
  {https://doi.org/10.1103/PhysRevD.107.114504} {\bibfield  {journal} {\bibinfo
   {journal} {Phys. Rev. D}\ }\textbf {\bibinfo {volume} {107}},\ \bibinfo
  {pages} {114504} (\bibinfo {year} {2023}{\natexlab{a}})},\ \Eprint
  {https://arxiv.org/abs/2304.11729} {arXiv:2304.11729 [hep-lat]} \BibitemShut
  {NoStop}%
\bibitem [{\citenamefont {Hasenfratz}\ \emph
  {et~al.}(2023{\natexlab{b}})\citenamefont {Hasenfratz}, \citenamefont {Neil},
  \citenamefont {Shamir}, \citenamefont {Svetitsky},\ and\ \citenamefont
  {Witzel}}]{Hasenfratz:2023wbr}%
  \BibitemOpen
  \bibfield  {author} {\bibinfo {author} {\bibfnamefont {A.}~\bibnamefont
  {Hasenfratz}}, \bibinfo {author} {\bibfnamefont {E.~T.}\ \bibnamefont
  {Neil}}, \bibinfo {author} {\bibfnamefont {Y.}~\bibnamefont {Shamir}},
  \bibinfo {author} {\bibfnamefont {B.}~\bibnamefont {Svetitsky}},\ and\
  \bibinfo {author} {\bibfnamefont {O.}~\bibnamefont {Witzel}},\ }\bibfield
  {title} {\bibinfo {title} {{Infrared fixed point of the SU(3) gauge theory
  with Nf=10 flavors}},\ }\href {https://doi.org/10.1103/PhysRevD.108.L071503}
  {\bibfield  {journal} {\bibinfo  {journal} {Phys. Rev. D}\ }\textbf {\bibinfo
  {volume} {108}},\ \bibinfo {pages} {L071503} (\bibinfo {year}
  {2023}{\natexlab{b}})},\ \Eprint {https://arxiv.org/abs/2306.07236}
  {arXiv:2306.07236 [hep-lat]} \BibitemShut {NoStop}%
\bibitem [{\citenamefont {Allton}\ \emph {et~al.}(2008)\citenamefont {Allton}
  \emph {et~al.}}]{Allton:2008pn}%
  \BibitemOpen
  \bibfield  {author} {\bibinfo {author} {\bibfnamefont {C.}~\bibnamefont
  {Allton}} \emph {et~al.} (\bibinfo {collaboration} {RBC/UKQCD}),\ }\bibfield
  {title} {\bibinfo {title} {{Physical Results from 2+1 Flavor Domain Wall QCD
  and SU(2) Chiral Perturbation Theory}},\ }\href
  {https://doi.org/10.1103/PhysRevD.78.114509} {\bibfield  {journal} {\bibinfo
  {journal} {Phys. Rev.}\ }\textbf {\bibinfo {volume} {D78}},\ \bibinfo {pages}
  {114509} (\bibinfo {year} {2008})},\ \Eprint
  {https://arxiv.org/abs/0804.0473} {arXiv:0804.0473 [hep-lat]} \BibitemShut
  {NoStop}%
\bibitem [{\citenamefont {Aoki}\ \emph {et~al.}(2011)\citenamefont {Aoki} \emph
  {et~al.}}]{Aoki:2010dy}%
  \BibitemOpen
  \bibfield  {author} {\bibinfo {author} {\bibfnamefont {Y.}~\bibnamefont
  {Aoki}} \emph {et~al.} (\bibinfo {collaboration} {RBC/UKQCD}),\ }\bibfield
  {title} {\bibinfo {title} {{Continuum Limit Physics from 2+1 Flavor Domain
  Wall QCD}},\ }\href {https://doi.org/10.1103/PhysRevD.83.074508} {\bibfield
  {journal} {\bibinfo  {journal} {Phys.Rev.}\ }\textbf {\bibinfo {volume}
  {D83}},\ \bibinfo {pages} {074508} (\bibinfo {year} {2011})},\ \Eprint
  {https://arxiv.org/abs/1011.0892} {arXiv:1011.0892 [hep-lat]} \BibitemShut
  {NoStop}%
\bibitem [{\citenamefont {Blum}\ \emph
  {et~al.}(2016{\natexlab{a}})\citenamefont {Blum} \emph
  {et~al.}}]{Blum:2014tka}%
  \BibitemOpen
  \bibfield  {author} {\bibinfo {author} {\bibfnamefont {T.}~\bibnamefont
  {Blum}} \emph {et~al.} (\bibinfo {collaboration} {RBC/UKQCD}),\ }\bibfield
  {title} {\bibinfo {title} {{Domain wall QCD with physical quark masses}},\
  }\href {https://doi.org/10.1103/PhysRevD.93.074505} {\bibfield  {journal}
  {\bibinfo  {journal} {Phys. Rev.}\ }\textbf {\bibinfo {volume} {D93}},\
  \bibinfo {pages} {074505} (\bibinfo {year} {2016}{\natexlab{a}})},\ \Eprint
  {https://arxiv.org/abs/1411.7017} {arXiv:1411.7017 [hep-lat]} \BibitemShut
  {NoStop}%
\bibitem [{\citenamefont {Boyle}\ \emph {et~al.}(2017)\citenamefont {Boyle}
  \emph {et~al.}}]{Boyle:2017jwu}%
  \BibitemOpen
  \bibfield  {author} {\bibinfo {author} {\bibfnamefont {P.~A.}\ \bibnamefont
  {Boyle}} \emph {et~al.} (\bibinfo {collaboration} {RBC/UKQCD}),\ }\bibfield
  {title} {\bibinfo {title} {The decay constants {$f_D$} and {$f_{D_s}$} in the
  continuum limit of {$N_f=2+1$} domain wall lattice {QCD}},\ }\href
  {https://doi.org/10.1007/JHEP12(2017)008} {\bibfield  {journal} {\bibinfo
  {journal} {JHEP}\ }\textbf {\bibinfo {volume} {12}},\ \bibinfo {pages}
  {008}},\ \Eprint {https://arxiv.org/abs/1701.02644} {arXiv:1701.02644
  [hep-lat]} \BibitemShut {NoStop}%
\bibitem [{\citenamefont {Boyle}\ \emph {et~al.}(2018)\citenamefont {Boyle}
  \emph {et~al.}}]{Boyle:2018knm}%
  \BibitemOpen
  \bibfield  {author} {\bibinfo {author} {\bibfnamefont {P.~A.}\ \bibnamefont
  {Boyle}} \emph {et~al.} (\bibinfo {collaboration} {RBC/UKQCD}),\ }\bibfield
  {title} {\bibinfo {title} {{SU(3)}-breaking ratios for {$D_{(s)}$} and
  {$B_{(s)}$} mesons},\ }\href@noop {} {\  (\bibinfo {year} {2018})},\ \Eprint
  {https://arxiv.org/abs/1812.08791} {arXiv:1812.08791 [hep-lat]} \BibitemShut
  {NoStop}%
\bibitem [{\citenamefont {Boyle}\ \emph {et~al.}(2016)\citenamefont {Boyle},
  \citenamefont {Yamaguchi}, \citenamefont {Cossu},\ and\ \citenamefont
  {Portelli}}]{Boyle:2015tjk}%
  \BibitemOpen
  \bibfield  {author} {\bibinfo {author} {\bibfnamefont {P.}~\bibnamefont
  {Boyle}}, \bibinfo {author} {\bibfnamefont {A.}~\bibnamefont {Yamaguchi}},
  \bibinfo {author} {\bibfnamefont {G.}~\bibnamefont {Cossu}},\ and\ \bibinfo
  {author} {\bibfnamefont {A.}~\bibnamefont {Portelli}},\ }\bibfield  {title}
  {\bibinfo {title} {{Grid: A next generation data parallel C++ QCD library}},\
  }\href {https://doi.org/10.22323/1.251.0023} {\bibfield  {journal} {\bibinfo
  {journal} {PoS}\ }\textbf {\bibinfo {volume} {LATTICE2015}},\ \bibinfo
  {pages} {023} (\bibinfo {year} {2016})},\ \Eprint
  {https://arxiv.org/abs/1512.03487} {arXiv:1512.03487 [hep-lat]} \BibitemShut
  {NoStop}%
\bibitem [{\citenamefont {Portelli}\ \emph {et~al.}()\citenamefont {Portelli}
  \emph {et~al.}}]{Hadrons}%
  \BibitemOpen
  \bibfield  {author} {\bibinfo {author} {\bibfnamefont {A.}~\bibnamefont
  {Portelli}} \emph {et~al.},\ }\href {https://doi.org/10.5281/zenodo.4293902}
  {\bibinfo {title} {Hadrons}}\BibitemShut {NoStop}%
\bibitem [{\citenamefont {Portelli}\ \emph {et~al.}(2022)\citenamefont
  {Portelli} \emph {et~al.}}]{Hadrons22}%
  \BibitemOpen
  \bibfield  {author} {\bibinfo {author} {\bibfnamefont {A.}~\bibnamefont
  {Portelli}} \emph {et~al.},\ }\href {https://doi.org/10.5281/zenodo.6382460}
  {\bibinfo {title} {github.com/aportelli/hadrons: Hadrons v1.3}} (\bibinfo
  {year} {2022})\BibitemShut {NoStop}%
\bibitem [{\citenamefont {Black}\ \emph
  {et~al.}(2025{\natexlab{b}})\citenamefont {Black}, \citenamefont {Harlander},
  \citenamefont {Kohnen}, \citenamefont {Lange}, \citenamefont {Rago},
  \citenamefont {Shindler},\ and\ \citenamefont {Witzel}}]{Lifetimes:2025}%
  \BibitemOpen
  \bibfield  {author} {\bibinfo {author} {\bibfnamefont {M.}~\bibnamefont
  {Black}}, \bibinfo {author} {\bibfnamefont {R.}~\bibnamefont {Harlander}},
  \bibinfo {author} {\bibfnamefont {J.}~\bibnamefont {Kohnen}}, \bibinfo
  {author} {\bibfnamefont {F.}~\bibnamefont {Lange}}, \bibinfo {author}
  {\bibfnamefont {A.}~\bibnamefont {Rago}}, \bibinfo {author} {\bibfnamefont
  {A.}~\bibnamefont {Shindler}},\ and\ \bibinfo {author} {\bibfnamefont
  {O.}~\bibnamefont {Witzel}},\ }\href@noop {} {\bibinfo {title} {Lattice
  determination of heavy meson lifetimes}} (\bibinfo {year}
  {2025}{\natexlab{b}}),\ \bibinfo {note} {(in preparation)}\BibitemShut
  {NoStop}%
\bibitem [{\citenamefont {McNeile}\ and\ \citenamefont
  {Michael}(2006)}]{McNeile:2006bz}%
  \BibitemOpen
  \bibfield  {author} {\bibinfo {author} {\bibfnamefont {C.}~\bibnamefont
  {McNeile}}\ and\ \bibinfo {author} {\bibfnamefont {C.}~\bibnamefont
  {Michael}} (\bibinfo {collaboration} {UKQCD}),\ }\bibfield  {title} {\bibinfo
  {title} {{Decay width of light quark hybrid meson from the lattice}},\ }\href
  {https://doi.org/10.1103/PhysRevD.73.074506} {\bibfield  {journal} {\bibinfo
  {journal} {Phys. Rev. D}\ }\textbf {\bibinfo {volume} {73}},\ \bibinfo
  {pages} {074506} (\bibinfo {year} {2006})},\ \Eprint
  {https://arxiv.org/abs/hep-lat/0603007} {arXiv:hep-lat/0603007} \BibitemShut
  {NoStop}%
\bibitem [{\citenamefont {Boyle}\ \emph {et~al.}(2008)\citenamefont {Boyle},
  \citenamefont {Jüttner}, \citenamefont {Kelly},\ and\ \citenamefont
  {Kenway}}]{Boyle:2008rh}%
  \BibitemOpen
  \bibfield  {author} {\bibinfo {author} {\bibfnamefont {P.~A.}\ \bibnamefont
  {Boyle}}, \bibinfo {author} {\bibfnamefont {A.}~\bibnamefont {Jüttner}},
  \bibinfo {author} {\bibfnamefont {C.}~\bibnamefont {Kelly}},\ and\ \bibinfo
  {author} {\bibfnamefont {R.~D.}\ \bibnamefont {Kenway}},\ }\bibfield  {title}
  {\bibinfo {title} {{Use of stochastic sources for the lattice determination
  of light quark physics}},\ }\href
  {https://doi.org/10.1088/1126-6708/2008/08/086} {\bibfield  {journal}
  {\bibinfo  {journal} {JHEP}\ }\textbf {\bibinfo {volume} {08}},\ \bibinfo
  {pages} {086}},\ \Eprint {https://arxiv.org/abs/0804.1501} {arXiv:0804.1501
  [hep-lat]} \BibitemShut {NoStop}%
\bibitem [{\citenamefont {Güsken}\ \emph {et~al.}(1989)\citenamefont
  {Güsken}, \citenamefont {Löw}, \citenamefont {Mütter}, \citenamefont
  {Sommer}, \citenamefont {Patel},\ and\ \citenamefont
  {Schilling}}]{Gusken:1989ad}%
  \BibitemOpen
  \bibfield  {author} {\bibinfo {author} {\bibfnamefont {S.}~\bibnamefont
  {Güsken}}, \bibinfo {author} {\bibfnamefont {U.}~\bibnamefont {Löw}},
  \bibinfo {author} {\bibfnamefont {K.~H.}\ \bibnamefont {Mütter}}, \bibinfo
  {author} {\bibfnamefont {R.}~\bibnamefont {Sommer}}, \bibinfo {author}
  {\bibfnamefont {A.}~\bibnamefont {Patel}},\ and\ \bibinfo {author}
  {\bibfnamefont {K.}~\bibnamefont {Schilling}},\ }\bibfield  {title} {\bibinfo
  {title} {{Nonsinglet Axial Vector Couplings of the Baryon Octet in Lattice
  {QCD}}},\ }\href {https://doi.org/10.1016/S0370-2693(89)80034-6} {\bibfield
  {journal} {\bibinfo  {journal} {Phys. Lett. B}\ }\textbf {\bibinfo {volume}
  {227}},\ \bibinfo {pages} {266} (\bibinfo {year} {1989})}\BibitemShut
  {NoStop}%
\bibitem [{\citenamefont {Kaplan}(1992)}]{Kaplan:1992bt}%
  \BibitemOpen
  \bibfield  {author} {\bibinfo {author} {\bibfnamefont {D.~B.}\ \bibnamefont
  {Kaplan}},\ }\bibfield  {title} {\bibinfo {title} {{A Method for simulating
  chiral fermions on the lattice}},\ }\href
  {https://doi.org/10.1016/0370-2693(92)91112-M} {\bibfield  {journal}
  {\bibinfo  {journal} {Phys. Lett. B}\ }\textbf {\bibinfo {volume} {288}},\
  \bibinfo {pages} {342} (\bibinfo {year} {1992})},\ \Eprint
  {https://arxiv.org/abs/hep-lat/9206013} {arXiv:hep-lat/9206013} \BibitemShut
  {NoStop}%
\bibitem [{\citenamefont {Shamir}(1993)}]{Shamir:1993zy}%
  \BibitemOpen
  \bibfield  {author} {\bibinfo {author} {\bibfnamefont {Y.}~\bibnamefont
  {Shamir}},\ }\bibfield  {title} {\bibinfo {title} {{Chiral fermions from
  lattice boundaries}},\ }\href {https://doi.org/10.1016/0550-3213(93)90162-I}
  {\bibfield  {journal} {\bibinfo  {journal} {Nucl. Phys. B}\ }\textbf
  {\bibinfo {volume} {406}},\ \bibinfo {pages} {90} (\bibinfo {year} {1993})},\
  \Eprint {https://arxiv.org/abs/hep-lat/9303005} {arXiv:hep-lat/9303005}
  \BibitemShut {NoStop}%
\bibitem [{\citenamefont {Furman}\ and\ \citenamefont
  {Shamir}(1995)}]{Furman:1994ky}%
  \BibitemOpen
  \bibfield  {author} {\bibinfo {author} {\bibfnamefont {V.}~\bibnamefont
  {Furman}}\ and\ \bibinfo {author} {\bibfnamefont {Y.}~\bibnamefont
  {Shamir}},\ }\bibfield  {title} {\bibinfo {title} {{Axial symmetries in
  lattice QCD with Kaplan fermions}},\ }\href
  {https://doi.org/10.1016/0550-3213(95)00031-M} {\bibfield  {journal}
  {\bibinfo  {journal} {Nucl. Phys. B}\ }\textbf {\bibinfo {volume} {439}},\
  \bibinfo {pages} {54} (\bibinfo {year} {1995})},\ \Eprint
  {https://arxiv.org/abs/hep-lat/9405004} {arXiv:hep-lat/9405004} \BibitemShut
  {NoStop}%
\bibitem [{\citenamefont {Morningstar}\ and\ \citenamefont
  {Peardon}(2004)}]{Morningstar:2003gk}%
  \BibitemOpen
  \bibfield  {author} {\bibinfo {author} {\bibfnamefont {C.}~\bibnamefont
  {Morningstar}}\ and\ \bibinfo {author} {\bibfnamefont {M.~J.}\ \bibnamefont
  {Peardon}},\ }\bibfield  {title} {\bibinfo {title} {{Analytic smearing of
  SU(3) link variables in lattice QCD}},\ }\href
  {https://doi.org/10.1103/PhysRevD.69.054501} {\bibfield  {journal} {\bibinfo
  {journal} {Phys. Rev. D}\ }\textbf {\bibinfo {volume} {69}},\ \bibinfo
  {pages} {054501} (\bibinfo {year} {2004})},\ \Eprint
  {https://arxiv.org/abs/hep-lat/0311018} {arXiv:hep-lat/0311018} \BibitemShut
  {NoStop}%
\bibitem [{\citenamefont {Brower}\ \emph {et~al.}(2017)\citenamefont {Brower},
  \citenamefont {Neff},\ and\ \citenamefont {Orginos}}]{Brower:2012vk}%
  \BibitemOpen
  \bibfield  {author} {\bibinfo {author} {\bibfnamefont {R.~C.}\ \bibnamefont
  {Brower}}, \bibinfo {author} {\bibfnamefont {H.}~\bibnamefont {Neff}},\ and\
  \bibinfo {author} {\bibfnamefont {K.}~\bibnamefont {Orginos}},\ }\bibfield
  {title} {\bibinfo {title} {{The M\"obius domain wall fermion algorithm}},\
  }\href {https://doi.org/10.1016/j.cpc.2017.01.024} {\bibfield  {journal}
  {\bibinfo  {journal} {Comput. Phys. Commun.}\ }\textbf {\bibinfo {volume}
  {220}},\ \bibinfo {pages} {1} (\bibinfo {year} {2017})},\ \Eprint
  {https://arxiv.org/abs/1206.5214} {arXiv:1206.5214 [hep-lat]} \BibitemShut
  {NoStop}%
\bibitem [{\citenamefont {Cho}\ \emph {et~al.}(2015)\citenamefont {Cho},
  \citenamefont {Hashimoto}, \citenamefont {J\"uttner}, \citenamefont {Kaneko},
  \citenamefont {Marinkovic}, \citenamefont {Noaki},\ and\ \citenamefont
  {Tsang}}]{Cho:2015ffa}%
  \BibitemOpen
  \bibfield  {author} {\bibinfo {author} {\bibfnamefont {Y.-G.}\ \bibnamefont
  {Cho}}, \bibinfo {author} {\bibfnamefont {S.}~\bibnamefont {Hashimoto}},
  \bibinfo {author} {\bibfnamefont {A.}~\bibnamefont {J\"uttner}}, \bibinfo
  {author} {\bibfnamefont {T.}~\bibnamefont {Kaneko}}, \bibinfo {author}
  {\bibfnamefont {M.}~\bibnamefont {Marinkovic}}, \bibinfo {author}
  {\bibfnamefont {J.-I.}\ \bibnamefont {Noaki}},\ and\ \bibinfo {author}
  {\bibfnamefont {J.~T.}\ \bibnamefont {Tsang}},\ }\bibfield  {title} {\bibinfo
  {title} {{Improved lattice fermion action for heavy quarks}},\ }\href
  {https://doi.org/10.1007/JHEP05(2015)072} {\bibfield  {journal} {\bibinfo
  {journal} {JHEP}\ }\textbf {\bibinfo {volume} {05}},\ \bibinfo {pages}
  {072}},\ \Eprint {https://arxiv.org/abs/1504.01630} {arXiv:1504.01630
  [hep-lat]} \BibitemShut {NoStop}%
\bibitem [{\citenamefont {Chetyrkin}\ \emph {et~al.}(2000)\citenamefont
  {Chetyrkin}, \citenamefont {K\"uhn},\ and\ \citenamefont
  {Steinhauser}}]{Chetyrkin:2000yt}%
  \BibitemOpen
  \bibfield  {author} {\bibinfo {author} {\bibfnamefont {K.~G.}\ \bibnamefont
  {Chetyrkin}}, \bibinfo {author} {\bibfnamefont {J.~H.}\ \bibnamefont
  {K\"uhn}},\ and\ \bibinfo {author} {\bibfnamefont {M.}~\bibnamefont
  {Steinhauser}},\ }\bibfield  {title} {\bibinfo {title} {{RunDec: A
  Mathematica package for running and decoupling of the strong coupling and
  quark masses}},\ }\href {https://doi.org/10.1016/S0010-4655(00)00155-7}
  {\bibfield  {journal} {\bibinfo  {journal} {Comput. Phys. Commun.}\ }\textbf
  {\bibinfo {volume} {133}},\ \bibinfo {pages} {43} (\bibinfo {year} {2000})},\
  \Eprint {https://arxiv.org/abs/hep-ph/0004189} {arXiv:hep-ph/0004189}
  \BibitemShut {NoStop}%
\bibitem [{\citenamefont {Herren}\ and\ \citenamefont
  {Steinhauser}(2018)}]{Herren:2017osy}%
  \BibitemOpen
  \bibfield  {author} {\bibinfo {author} {\bibfnamefont {F.}~\bibnamefont
  {Herren}}\ and\ \bibinfo {author} {\bibfnamefont {M.}~\bibnamefont
  {Steinhauser}},\ }\bibfield  {title} {\bibinfo {title} {{Version 3 of RunDec
  and CRunDec}},\ }\href {https://doi.org/10.1016/j.cpc.2017.11.014} {\bibfield
   {journal} {\bibinfo  {journal} {Comput. Phys. Commun.}\ }\textbf {\bibinfo
  {volume} {224}},\ \bibinfo {pages} {333} (\bibinfo {year} {2018})},\ \Eprint
  {https://arxiv.org/abs/1703.03751} {arXiv:1703.03751 [hep-ph]} \BibitemShut
  {NoStop}%
\bibitem [{\citenamefont {Suzuki}\ and\ \citenamefont
  {Takaura}(2021)}]{Suzuki:2021tlr}%
  \BibitemOpen
  \bibfield  {author} {\bibinfo {author} {\bibfnamefont {H.}~\bibnamefont
  {Suzuki}}\ and\ \bibinfo {author} {\bibfnamefont {H.}~\bibnamefont
  {Takaura}},\ }\bibfield  {title} {\bibinfo {title} {{$t \to 0$ extrapolation
  function in the small flow time expansion method for the
  energy\textendash{}momentum tensor}},\ }\href
  {https://doi.org/10.1093/ptep/ptab068} {\bibfield  {journal} {\bibinfo
  {journal} {PTEP}\ }\textbf {\bibinfo {volume} {2021}},\ \bibinfo {pages}
  {073B02} (\bibinfo {year} {2021})},\ \Eprint
  {https://arxiv.org/abs/2102.02174} {arXiv:2102.02174 [hep-lat]} \BibitemShut
  {NoStop}%
\bibitem [{\citenamefont {Schmelling}(1995)}]{Schmelling:1994pz}%
  \BibitemOpen
  \bibfield  {author} {\bibinfo {author} {\bibfnamefont {M.}~\bibnamefont
  {Schmelling}},\ }\bibfield  {title} {\bibinfo {title} {{Averaging correlated
  data}},\ }\href {https://doi.org/10.1088/0031-8949/51/6/002} {\bibfield
  {journal} {\bibinfo  {journal} {Phys. Scripta}\ }\textbf {\bibinfo {volume}
  {51}},\ \bibinfo {pages} {676} (\bibinfo {year} {1995})}\BibitemShut
  {NoStop}%
\bibitem [{\citenamefont {Aoki}\ \emph
  {et~al.}(2024{\natexlab{a}})\citenamefont {Aoki}, \citenamefont {Colquhoun},
  \citenamefont {Fukaya}, \citenamefont {Hashimoto}, \citenamefont {Kaneko},
  \citenamefont {Kellermann}, \citenamefont {Koponen},\ and\ \citenamefont
  {Kou}}]{Aoki:2023qpa}%
  \BibitemOpen
  \bibfield  {author} {\bibinfo {author} {\bibfnamefont {Y.}~\bibnamefont
  {Aoki}}, \bibinfo {author} {\bibfnamefont {B.}~\bibnamefont {Colquhoun}},
  \bibinfo {author} {\bibfnamefont {H.}~\bibnamefont {Fukaya}}, \bibinfo
  {author} {\bibfnamefont {S.}~\bibnamefont {Hashimoto}}, \bibinfo {author}
  {\bibfnamefont {T.}~\bibnamefont {Kaneko}}, \bibinfo {author} {\bibfnamefont
  {R.}~\bibnamefont {Kellermann}}, \bibinfo {author} {\bibfnamefont
  {J.}~\bibnamefont {Koponen}},\ and\ \bibinfo {author} {\bibfnamefont
  {E.}~\bibnamefont {Kou}} (\bibinfo {collaboration} {JLQCD}),\ }\bibfield
  {title} {\bibinfo {title}
  {{B{\textrightarrow}D*{\ensuremath{\ell}}{\ensuremath{\nu}}{\ensuremath{\ell}}
  semileptonic form factors from lattice QCD with M{\"o}bius domain-wall
  quarks}},\ }\href {https://doi.org/10.1103/PhysRevD.109.074503} {\bibfield
  {journal} {\bibinfo  {journal} {Phys. Rev. D}\ }\textbf {\bibinfo {volume}
  {109}},\ \bibinfo {pages} {074503} (\bibinfo {year} {2024}{\natexlab{a}})},\
  \Eprint {https://arxiv.org/abs/2306.05657} {arXiv:2306.05657 [hep-lat]}
  \BibitemShut {NoStop}%
\bibitem [{\citenamefont {Colquhoun}\ \emph {et~al.}(2022)\citenamefont
  {Colquhoun}, \citenamefont {Hashimoto}, \citenamefont {Kaneko},\ and\
  \citenamefont {Koponen}}]{Colquhoun:2022atw}%
  \BibitemOpen
  \bibfield  {author} {\bibinfo {author} {\bibfnamefont {B.}~\bibnamefont
  {Colquhoun}}, \bibinfo {author} {\bibfnamefont {S.}~\bibnamefont
  {Hashimoto}}, \bibinfo {author} {\bibfnamefont {T.}~\bibnamefont {Kaneko}},\
  and\ \bibinfo {author} {\bibfnamefont {J.}~\bibnamefont {Koponen}} (\bibinfo
  {collaboration} {JLQCD}),\ }\bibfield  {title} {\bibinfo {title} {{Form
  factors of
  B{\textrightarrow}{\ensuremath{\pi}}{\ensuremath{\ell}}{\ensuremath{\nu}} and
  a determination of |Vub| with M{\"o}bius domain-wall fermions}},\ }\href
  {https://doi.org/10.1103/PhysRevD.106.054502} {\bibfield  {journal} {\bibinfo
   {journal} {Phys. Rev. D}\ }\textbf {\bibinfo {volume} {106}},\ \bibinfo
  {pages} {054502} (\bibinfo {year} {2022})},\ \Eprint
  {https://arxiv.org/abs/2203.04938} {arXiv:2203.04938 [hep-lat]} \BibitemShut
  {NoStop}%
\bibitem [{\citenamefont {Aoki}\ \emph
  {et~al.}(2024{\natexlab{b}})\citenamefont {Aoki} \emph
  {et~al.}}]{FlavourLatticeAveragingGroupFLAG:2024oxs}%
  \BibitemOpen
  \bibfield  {author} {\bibinfo {author} {\bibfnamefont {Y.}~\bibnamefont
  {Aoki}} \emph {et~al.} (\bibinfo {collaboration} {Flavour Lattice Averaging
  Group (FLAG)}),\ }\bibfield  {title} {\bibinfo {title} {{FLAG Review 2024}},\
  }\href@noop {} {\  (\bibinfo {year} {2024}{\natexlab{b}})},\ \Eprint
  {https://arxiv.org/abs/2411.04268} {arXiv:2411.04268 [hep-lat]} \BibitemShut
  {NoStop}%
\bibitem [{\citenamefont {Yang}\ \emph {et~al.}(2015)\citenamefont {Yang} \emph
  {et~al.}}]{Yang:2014sea}%
  \BibitemOpen
  \bibfield  {author} {\bibinfo {author} {\bibfnamefont {Y.-B.}\ \bibnamefont
  {Yang}} \emph {et~al.},\ }\bibfield  {title} {\bibinfo {title} {{Charm and
  strange quark masses and $f_{D_s}$ from overlap fermions}},\ }\href
  {https://doi.org/10.1103/PhysRevD.92.034517} {\bibfield  {journal} {\bibinfo
  {journal} {Phys. Rev. D}\ }\textbf {\bibinfo {volume} {92}},\ \bibinfo
  {pages} {034517} (\bibinfo {year} {2015})},\ \Eprint
  {https://arxiv.org/abs/1410.3343} {arXiv:1410.3343 [hep-lat]} \BibitemShut
  {NoStop}%
\bibitem [{\citenamefont {Davies}\ \emph {et~al.}(2010)\citenamefont {Davies},
  \citenamefont {McNeile}, \citenamefont {Wong}, \citenamefont {Follana},
  \citenamefont {Horgan}, \citenamefont {Hornbostel}, \citenamefont {Lepage},
  \citenamefont {Shigemitsu},\ and\ \citenamefont {Trottier}}]{Davies:2009ih}%
  \BibitemOpen
  \bibfield  {author} {\bibinfo {author} {\bibfnamefont {C.~T.~H.}\
  \bibnamefont {Davies}}, \bibinfo {author} {\bibfnamefont {C.}~\bibnamefont
  {McNeile}}, \bibinfo {author} {\bibfnamefont {K.~Y.}\ \bibnamefont {Wong}},
  \bibinfo {author} {\bibfnamefont {E.}~\bibnamefont {Follana}}, \bibinfo
  {author} {\bibfnamefont {R.}~\bibnamefont {Horgan}}, \bibinfo {author}
  {\bibfnamefont {K.}~\bibnamefont {Hornbostel}}, \bibinfo {author}
  {\bibfnamefont {G.~P.}\ \bibnamefont {Lepage}}, \bibinfo {author}
  {\bibfnamefont {J.}~\bibnamefont {Shigemitsu}},\ and\ \bibinfo {author}
  {\bibfnamefont {H.}~\bibnamefont {Trottier}},\ }\bibfield  {title} {\bibinfo
  {title} {{Precise Charm to Strange Mass Ratio and Light Quark Masses from
  Full Lattice QCD}},\ }\href {https://doi.org/10.1103/PhysRevLett.104.132003}
  {\bibfield  {journal} {\bibinfo  {journal} {Phys. Rev. Lett.}\ }\textbf
  {\bibinfo {volume} {104}},\ \bibinfo {pages} {132003} (\bibinfo {year}
  {2010})},\ \Eprint {https://arxiv.org/abs/0910.3102} {arXiv:0910.3102
  [hep-ph]} \BibitemShut {NoStop}%
\bibitem [{\citenamefont {Alexandrou}\ \emph {et~al.}(2021)\citenamefont
  {Alexandrou} \emph {et~al.}}]{ExtendedTwistedMass:2021gbo}%
  \BibitemOpen
  \bibfield  {author} {\bibinfo {author} {\bibfnamefont {C.}~\bibnamefont
  {Alexandrou}} \emph {et~al.} (\bibinfo {collaboration} {Extended Twisted
  Mass}),\ }\bibfield  {title} {\bibinfo {title} {{Quark masses using
  twisted-mass fermion gauge ensembles}},\ }\href
  {https://doi.org/10.1103/PhysRevD.104.074515} {\bibfield  {journal} {\bibinfo
   {journal} {Phys. Rev. D}\ }\textbf {\bibinfo {volume} {104}},\ \bibinfo
  {pages} {074515} (\bibinfo {year} {2021})},\ \Eprint
  {https://arxiv.org/abs/2104.13408} {arXiv:2104.13408 [hep-lat]} \BibitemShut
  {NoStop}%
\bibitem [{\citenamefont {Bazavov}\ \emph {et~al.}(2018)\citenamefont {Bazavov}
  \emph {et~al.}}]{FermilabLattice:2018est}%
  \BibitemOpen
  \bibfield  {author} {\bibinfo {author} {\bibfnamefont {A.}~\bibnamefont
  {Bazavov}} \emph {et~al.} (\bibinfo {collaboration} {Fermilab Lattice, MILC,
  TUMQCD}),\ }\bibfield  {title} {\bibinfo {title} {{Up-, down-, strange-,
  charm-, and bottom-quark masses from four-flavor lattice QCD}},\ }\href
  {https://doi.org/10.1103/PhysRevD.98.054517} {\bibfield  {journal} {\bibinfo
  {journal} {Phys. Rev. D}\ }\textbf {\bibinfo {volume} {98}},\ \bibinfo
  {pages} {054517} (\bibinfo {year} {2018})},\ \Eprint
  {https://arxiv.org/abs/1802.04248} {arXiv:1802.04248 [hep-lat]} \BibitemShut
  {NoStop}%
\bibitem [{\citenamefont {Chakraborty}\ \emph {et~al.}(2015)\citenamefont
  {Chakraborty}, \citenamefont {Davies}, \citenamefont {Galloway},
  \citenamefont {Knecht}, \citenamefont {Koponen}, \citenamefont {Donald},
  \citenamefont {Dowdall}, \citenamefont {Lepage},\ and\ \citenamefont
  {McNeile}}]{Chakraborty:2014aca}%
  \BibitemOpen
  \bibfield  {author} {\bibinfo {author} {\bibfnamefont {B.}~\bibnamefont
  {Chakraborty}}, \bibinfo {author} {\bibfnamefont {C.~T.~H.}\ \bibnamefont
  {Davies}}, \bibinfo {author} {\bibfnamefont {B.}~\bibnamefont {Galloway}},
  \bibinfo {author} {\bibfnamefont {P.}~\bibnamefont {Knecht}}, \bibinfo
  {author} {\bibfnamefont {J.}~\bibnamefont {Koponen}}, \bibinfo {author}
  {\bibfnamefont {G.~C.}\ \bibnamefont {Donald}}, \bibinfo {author}
  {\bibfnamefont {R.~J.}\ \bibnamefont {Dowdall}}, \bibinfo {author}
  {\bibfnamefont {G.~P.}\ \bibnamefont {Lepage}},\ and\ \bibinfo {author}
  {\bibfnamefont {C.}~\bibnamefont {McNeile}},\ }\bibfield  {title} {\bibinfo
  {title} {{High-precision quark masses and QCD coupling from $n_f=4$ lattice
  QCD}},\ }\href {https://doi.org/10.1103/PhysRevD.91.054508} {\bibfield
  {journal} {\bibinfo  {journal} {Phys. Rev. D}\ }\textbf {\bibinfo {volume}
  {91}},\ \bibinfo {pages} {054508} (\bibinfo {year} {2015})},\ \Eprint
  {https://arxiv.org/abs/1408.4169} {arXiv:1408.4169 [hep-lat]} \BibitemShut
  {NoStop}%
\bibitem [{\citenamefont {Carrasco}\ \emph {et~al.}(2014)\citenamefont
  {Carrasco} \emph {et~al.}}]{EuropeanTwistedMass:2014osg}%
  \BibitemOpen
  \bibfield  {author} {\bibinfo {author} {\bibfnamefont {N.}~\bibnamefont
  {Carrasco}} \emph {et~al.} (\bibinfo {collaboration} {European Twisted
  Mass}),\ }\bibfield  {title} {\bibinfo {title} {{Up, down, strange and charm
  quark masses with N$_f$ = 2+1+1 twisted mass lattice QCD}},\ }\href
  {https://doi.org/10.1016/j.nuclphysb.2014.07.025} {\bibfield  {journal}
  {\bibinfo  {journal} {Nucl. Phys. B}\ }\textbf {\bibinfo {volume} {887}},\
  \bibinfo {pages} {19} (\bibinfo {year} {2014})},\ \Eprint
  {https://arxiv.org/abs/1403.4504} {arXiv:1403.4504 [hep-lat]} \BibitemShut
  {NoStop}%
\bibitem [{\citenamefont {Hu}\ \emph {et~al.}(2024)\citenamefont {Hu} \emph
  {et~al.}}]{CLQCD:2023sdb}%
  \BibitemOpen
  \bibfield  {author} {\bibinfo {author} {\bibfnamefont {Z.-C.}\ \bibnamefont
  {Hu}} \emph {et~al.} (\bibinfo {collaboration} {CLQCD}),\ }\bibfield  {title}
  {\bibinfo {title} {{Quark masses and low-energy constants in the continuum
  from the tadpole-improved clover ensembles}},\ }\href
  {https://doi.org/10.1103/PhysRevD.109.054507} {\bibfield  {journal} {\bibinfo
   {journal} {Phys. Rev. D}\ }\textbf {\bibinfo {volume} {109}},\ \bibinfo
  {pages} {054507} (\bibinfo {year} {2024})},\ \Eprint
  {https://arxiv.org/abs/2310.00814} {arXiv:2310.00814 [hep-lat]} \BibitemShut
  {NoStop}%
\bibitem [{\citenamefont {Bruno}\ \emph {et~al.}(2020)\citenamefont {Bruno},
  \citenamefont {Campos}, \citenamefont {Fritzsch}, \citenamefont {Koponen},
  \citenamefont {Pena}, \citenamefont {Preti}, \citenamefont {Ramos},\ and\
  \citenamefont {Vladikas}}]{Bruno:2019vup}%
  \BibitemOpen
  \bibfield  {author} {\bibinfo {author} {\bibfnamefont {M.}~\bibnamefont
  {Bruno}}, \bibinfo {author} {\bibfnamefont {I.}~\bibnamefont {Campos}},
  \bibinfo {author} {\bibfnamefont {P.}~\bibnamefont {Fritzsch}}, \bibinfo
  {author} {\bibfnamefont {J.}~\bibnamefont {Koponen}}, \bibinfo {author}
  {\bibfnamefont {C.}~\bibnamefont {Pena}}, \bibinfo {author} {\bibfnamefont
  {D.}~\bibnamefont {Preti}}, \bibinfo {author} {\bibfnamefont
  {A.}~\bibnamefont {Ramos}},\ and\ \bibinfo {author} {\bibfnamefont
  {A.}~\bibnamefont {Vladikas}} (\bibinfo {collaboration} {ALPHA}),\ }\bibfield
   {title} {\bibinfo {title} {{Light quark masses in ${N_\mathrm{f}=2+1}$
  lattice QCD with Wilson fermions}},\ }\href
  {https://doi.org/10.1140/epjc/s10052-020-7698-z} {\bibfield  {journal}
  {\bibinfo  {journal} {Eur. Phys. J. C}\ }\textbf {\bibinfo {volume} {80}},\
  \bibinfo {pages} {169} (\bibinfo {year} {2020})},\ \Eprint
  {https://arxiv.org/abs/1911.08025} {arXiv:1911.08025 [hep-lat]} \BibitemShut
  {NoStop}%
\bibitem [{\citenamefont {Blum}\ \emph
  {et~al.}(2016{\natexlab{b}})\citenamefont {Blum} \emph
  {et~al.}}]{RBC:2014ntl}%
  \BibitemOpen
  \bibfield  {author} {\bibinfo {author} {\bibfnamefont {T.}~\bibnamefont
  {Blum}} \emph {et~al.} (\bibinfo {collaboration} {RBC, UKQCD}),\ }\bibfield
  {title} {\bibinfo {title} {{Domain wall QCD with physical quark masses}},\
  }\href {https://doi.org/10.1103/PhysRevD.93.074505} {\bibfield  {journal}
  {\bibinfo  {journal} {Phys. Rev. D}\ }\textbf {\bibinfo {volume} {93}},\
  \bibinfo {pages} {074505} (\bibinfo {year} {2016}{\natexlab{b}})},\ \Eprint
  {https://arxiv.org/abs/1411.7017} {arXiv:1411.7017 [hep-lat]} \BibitemShut
  {NoStop}%
\bibitem [{\citenamefont {McNeile}\ \emph {et~al.}(2010)\citenamefont
  {McNeile}, \citenamefont {Davies}, \citenamefont {Follana}, \citenamefont
  {Hornbostel},\ and\ \citenamefont {Lepage}}]{McNeile:2010ji}%
  \BibitemOpen
  \bibfield  {author} {\bibinfo {author} {\bibfnamefont {C.}~\bibnamefont
  {McNeile}}, \bibinfo {author} {\bibfnamefont {C.~T.~H.}\ \bibnamefont
  {Davies}}, \bibinfo {author} {\bibfnamefont {E.}~\bibnamefont {Follana}},
  \bibinfo {author} {\bibfnamefont {K.}~\bibnamefont {Hornbostel}},\ and\
  \bibinfo {author} {\bibfnamefont {G.~P.}\ \bibnamefont {Lepage}},\ }\bibfield
   {title} {\bibinfo {title} {{High-Precision c and b Masses, and QCD Coupling
  from Current-Current Correlators in Lattice and Continuum QCD}},\ }\href
  {https://doi.org/10.1103/PhysRevD.82.034512} {\bibfield  {journal} {\bibinfo
  {journal} {Phys. Rev. D}\ }\textbf {\bibinfo {volume} {82}},\ \bibinfo
  {pages} {034512} (\bibinfo {year} {2010})},\ \Eprint
  {https://arxiv.org/abs/1004.4285} {arXiv:1004.4285 [hep-lat]} \BibitemShut
  {NoStop}%
\bibitem [{\citenamefont {Dürr}\ \emph
  {et~al.}(2011{\natexlab{a}})\citenamefont {Dürr}, \citenamefont {Fodor},
  \citenamefont {Hoelbling}, \citenamefont {Katz}, \citenamefont {Krieg},
  \citenamefont {Kurth}, \citenamefont {Lellouch}, \citenamefont {Lippert},
  \citenamefont {Szabo},\ and\ \citenamefont {Vulvert}}]{BMW:2010ucx}%
  \BibitemOpen
  \bibfield  {author} {\bibinfo {author} {\bibfnamefont {S.}~\bibnamefont
  {Dürr}}, \bibinfo {author} {\bibfnamefont {Z.}~\bibnamefont {Fodor}},
  \bibinfo {author} {\bibfnamefont {C.}~\bibnamefont {Hoelbling}}, \bibinfo
  {author} {\bibfnamefont {S.~D.}\ \bibnamefont {Katz}}, \bibinfo {author}
  {\bibfnamefont {S.}~\bibnamefont {Krieg}}, \bibinfo {author} {\bibfnamefont
  {T.}~\bibnamefont {Kurth}}, \bibinfo {author} {\bibfnamefont
  {L.}~\bibnamefont {Lellouch}}, \bibinfo {author} {\bibfnamefont
  {T.}~\bibnamefont {Lippert}}, \bibinfo {author} {\bibfnamefont {K.~K.}\
  \bibnamefont {Szabo}},\ and\ \bibinfo {author} {\bibfnamefont
  {G.}~\bibnamefont {Vulvert}} (\bibinfo {collaboration} {BMW}),\ }\bibfield
  {title} {\bibinfo {title} {{Lattice QCD at the physical point: light quark
  masses}},\ }\href {https://doi.org/10.1016/j.physletb.2011.05.053} {\bibfield
   {journal} {\bibinfo  {journal} {Phys. Lett. B}\ }\textbf {\bibinfo {volume}
  {701}},\ \bibinfo {pages} {265} (\bibinfo {year} {2011}{\natexlab{a}})},\
  \Eprint {https://arxiv.org/abs/1011.2403} {arXiv:1011.2403 [hep-lat]}
  \BibitemShut {NoStop}%
\bibitem [{\citenamefont {Dürr}\ \emph
  {et~al.}(2011{\natexlab{b}})\citenamefont {Dürr}, \citenamefont {Fodor},
  \citenamefont {Hoelbling}, \citenamefont {Katz}, \citenamefont {Krieg},
  \citenamefont {Kurth}, \citenamefont {Lellouch}, \citenamefont {Lippert},
  \citenamefont {Szabo},\ and\ \citenamefont {Vulvert}}]{BMW:2010skj}%
  \BibitemOpen
  \bibfield  {author} {\bibinfo {author} {\bibfnamefont {S.}~\bibnamefont
  {Dürr}}, \bibinfo {author} {\bibfnamefont {Z.}~\bibnamefont {Fodor}},
  \bibinfo {author} {\bibfnamefont {C.}~\bibnamefont {Hoelbling}}, \bibinfo
  {author} {\bibfnamefont {S.~D.}\ \bibnamefont {Katz}}, \bibinfo {author}
  {\bibfnamefont {S.}~\bibnamefont {Krieg}}, \bibinfo {author} {\bibfnamefont
  {T.}~\bibnamefont {Kurth}}, \bibinfo {author} {\bibfnamefont
  {L.}~\bibnamefont {Lellouch}}, \bibinfo {author} {\bibfnamefont
  {T.}~\bibnamefont {Lippert}}, \bibinfo {author} {\bibfnamefont {K.~K.}\
  \bibnamefont {Szabo}},\ and\ \bibinfo {author} {\bibfnamefont
  {G.}~\bibnamefont {Vulvert}} (\bibinfo {collaboration} {BMW}),\ }\bibfield
  {title} {\bibinfo {title} {{Lattice QCD at the physical point: Simulation and
  analysis details}},\ }\href {https://doi.org/10.1007/JHEP08(2011)148}
  {\bibfield  {journal} {\bibinfo  {journal} {JHEP}\ }\textbf {\bibinfo
  {volume} {08}},\ \bibinfo {pages} {148}},\ \Eprint
  {https://arxiv.org/abs/1011.2711} {arXiv:1011.2711 [hep-lat]} \BibitemShut
  {NoStop}%
\bibitem [{\citenamefont {Bazavov}\ \emph {et~al.}(2009)\citenamefont {Bazavov}
  \emph {et~al.}}]{MILC:2009ltw}%
  \BibitemOpen
  \bibfield  {author} {\bibinfo {author} {\bibfnamefont {A.}~\bibnamefont
  {Bazavov}} \emph {et~al.} (\bibinfo {collaboration} {MILC}),\ }\bibfield
  {title} {\bibinfo {title} {{MILC results for light pseudoscalars}},\ }\href
  {https://doi.org/10.22323/1.086.0007} {\bibfield  {journal} {\bibinfo
  {journal} {PoS}\ }\textbf {\bibinfo {volume} {CD09}},\ \bibinfo {pages} {007}
  (\bibinfo {year} {2009})},\ \Eprint {https://arxiv.org/abs/0910.2966}
  {arXiv:0910.2966 [hep-ph]} \BibitemShut {NoStop}%
\bibitem [{\citenamefont {Lytle}\ \emph {et~al.}(2018)\citenamefont {Lytle},
  \citenamefont {Davies}, \citenamefont {Hatton}, \citenamefont {Lepage},\ and\
  \citenamefont {Sturm}}]{Lytle:2018evc}%
  \BibitemOpen
  \bibfield  {author} {\bibinfo {author} {\bibfnamefont {A.~T.}\ \bibnamefont
  {Lytle}}, \bibinfo {author} {\bibfnamefont {C.~T.~H.}\ \bibnamefont
  {Davies}}, \bibinfo {author} {\bibfnamefont {D.}~\bibnamefont {Hatton}},
  \bibinfo {author} {\bibfnamefont {G.~P.}\ \bibnamefont {Lepage}},\ and\
  \bibinfo {author} {\bibfnamefont {C.}~\bibnamefont {Sturm}} (\bibinfo
  {collaboration} {HPQCD}),\ }\bibfield  {title} {\bibinfo {title}
  {{Determination of quark masses from $\mathbf{n_f=4}$ lattice QCD and the
  RI-SMOM intermediate scheme}},\ }\href
  {https://doi.org/10.1103/PhysRevD.98.014513} {\bibfield  {journal} {\bibinfo
  {journal} {Phys. Rev. D}\ }\textbf {\bibinfo {volume} {98}},\ \bibinfo
  {pages} {014513} (\bibinfo {year} {2018})},\ \Eprint
  {https://arxiv.org/abs/1805.06225} {arXiv:1805.06225 [hep-lat]} \BibitemShut
  {NoStop}%
\bibitem [{\citenamefont {Bussone}\ \emph {et~al.}(2024)\citenamefont
  {Bussone}, \citenamefont {Conigli}, \citenamefont {Frison}, \citenamefont
  {Herdo\'\i{}za}, \citenamefont {Pena}, \citenamefont {Preti}, \citenamefont
  {S\'aez},\ and\ \citenamefont {Ugarrio}}]{Bussone:2023kag}%
  \BibitemOpen
  \bibfield  {author} {\bibinfo {author} {\bibfnamefont {A.}~\bibnamefont
  {Bussone}}, \bibinfo {author} {\bibfnamefont {A.}~\bibnamefont {Conigli}},
  \bibinfo {author} {\bibfnamefont {J.}~\bibnamefont {Frison}}, \bibinfo
  {author} {\bibfnamefont {G.}~\bibnamefont {Herdo\'\i{}za}}, \bibinfo {author}
  {\bibfnamefont {C.}~\bibnamefont {Pena}}, \bibinfo {author} {\bibfnamefont
  {D.}~\bibnamefont {Preti}}, \bibinfo {author} {\bibfnamefont
  {A.}~\bibnamefont {S\'aez}},\ and\ \bibinfo {author} {\bibfnamefont
  {J.}~\bibnamefont {Ugarrio}} (\bibinfo {collaboration} {Alpha}),\ }\bibfield
  {title} {\bibinfo {title} {{Hadronic physics from a Wilson fermion
  mixed-action approach: charm quark mass and $D_{(s)}$ meson decay
  constants}},\ }\href {https://doi.org/10.1140/epjc/s10052-024-12816-4}
  {\bibfield  {journal} {\bibinfo  {journal} {Eur. Phys. J. C}\ }\textbf
  {\bibinfo {volume} {84}},\ \bibinfo {pages} {506} (\bibinfo {year} {2024})},\
  \Eprint {https://arxiv.org/abs/2309.14154} {arXiv:2309.14154 [hep-lat]}
  \BibitemShut {NoStop}%
\bibitem [{\citenamefont {Heitger}\ \emph {et~al.}(2021)\citenamefont
  {Heitger}, \citenamefont {Joswig},\ and\ \citenamefont
  {Kuberski}}]{Heitger:2021apz}%
  \BibitemOpen
  \bibfield  {author} {\bibinfo {author} {\bibfnamefont {J.}~\bibnamefont
  {Heitger}}, \bibinfo {author} {\bibfnamefont {F.}~\bibnamefont {Joswig}},\
  and\ \bibinfo {author} {\bibfnamefont {S.}~\bibnamefont {Kuberski}} (\bibinfo
  {collaboration} {ALPHA}),\ }\bibfield  {title} {\bibinfo {title}
  {{Determination of the charm quark mass in lattice QCD with $2+1$ flavours on
  fine lattices}},\ }\href {https://doi.org/10.1007/JHEP05(2021)288} {\bibfield
   {journal} {\bibinfo  {journal} {JHEP}\ }\textbf {\bibinfo {volume} {05}},\
  \bibinfo {pages} {288}},\ \Eprint {https://arxiv.org/abs/2101.02694}
  {arXiv:2101.02694 [hep-lat]} \BibitemShut {NoStop}%
\bibitem [{\citenamefont {Petreczky}\ and\ \citenamefont
  {Weber}(2019)}]{Petreczky:2019ozv}%
  \BibitemOpen
  \bibfield  {author} {\bibinfo {author} {\bibfnamefont {P.}~\bibnamefont
  {Petreczky}}\ and\ \bibinfo {author} {\bibfnamefont {J.~H.}\ \bibnamefont
  {Weber}},\ }\bibfield  {title} {\bibinfo {title} {{Strong coupling constant
  and heavy quark masses in ( 2+1 )-flavor QCD}},\ }\href
  {https://doi.org/10.1103/PhysRevD.100.034519} {\bibfield  {journal} {\bibinfo
   {journal} {Phys. Rev. D}\ }\textbf {\bibinfo {volume} {100}},\ \bibinfo
  {pages} {034519} (\bibinfo {year} {2019})},\ \Eprint
  {https://arxiv.org/abs/1901.06424} {arXiv:1901.06424 [hep-lat]} \BibitemShut
  {NoStop}%
\bibitem [{\citenamefont {Nakayama}\ \emph {et~al.}(2016)\citenamefont
  {Nakayama}, \citenamefont {Fahy},\ and\ \citenamefont
  {Hashimoto}}]{Nakayama:2016atf}%
  \BibitemOpen
  \bibfield  {author} {\bibinfo {author} {\bibfnamefont {K.}~\bibnamefont
  {Nakayama}}, \bibinfo {author} {\bibfnamefont {B.}~\bibnamefont {Fahy}},\
  and\ \bibinfo {author} {\bibfnamefont {S.}~\bibnamefont {Hashimoto}},\
  }\bibfield  {title} {\bibinfo {title} {{Short-distance charmonium correlator
  on the lattice with M\"obius domain-wall fermion and a determination of charm
  quark mass}},\ }\href {https://doi.org/10.1103/PhysRevD.94.054507} {\bibfield
   {journal} {\bibinfo  {journal} {Phys. Rev. D}\ }\textbf {\bibinfo {volume}
  {94}},\ \bibinfo {pages} {054507} (\bibinfo {year} {2016})},\ \Eprint
  {https://arxiv.org/abs/1606.01002} {arXiv:1606.01002 [hep-lat]} \BibitemShut
  {NoStop}%
\bibitem [{\citenamefont {Hatton}\ \emph {et~al.}(2020)\citenamefont {Hatton},
  \citenamefont {Davies}, \citenamefont {Galloway}, \citenamefont {Koponen},
  \citenamefont {Lepage},\ and\ \citenamefont {Lytle}}]{Hatton:2020qhk}%
  \BibitemOpen
  \bibfield  {author} {\bibinfo {author} {\bibfnamefont {D.}~\bibnamefont
  {Hatton}}, \bibinfo {author} {\bibfnamefont {C.~T.~H.}\ \bibnamefont
  {Davies}}, \bibinfo {author} {\bibfnamefont {B.}~\bibnamefont {Galloway}},
  \bibinfo {author} {\bibfnamefont {J.}~\bibnamefont {Koponen}}, \bibinfo
  {author} {\bibfnamefont {G.~P.}\ \bibnamefont {Lepage}},\ and\ \bibinfo
  {author} {\bibfnamefont {A.~T.}\ \bibnamefont {Lytle}} (\bibinfo
  {collaboration} {HPQCD}),\ }\bibfield  {title} {\bibinfo {title} {{Charmonium
  properties from lattice $QCD$+QED : Hyperfine splitting, $J/\psi$ leptonic
  width, charm quark mass, and $a^c_\mu$}},\ }\href
  {https://doi.org/10.1103/PhysRevD.102.054511} {\bibfield  {journal} {\bibinfo
   {journal} {Phys. Rev. D}\ }\textbf {\bibinfo {volume} {102}},\ \bibinfo
  {pages} {054511} (\bibinfo {year} {2020})},\ \Eprint
  {https://arxiv.org/abs/2005.01845} {arXiv:2005.01845 [hep-lat]} \BibitemShut
  {NoStop}%
\bibitem [{\citenamefont {Alexandrou}\ \emph {et~al.}(2014)\citenamefont
  {Alexandrou}, \citenamefont {Drach}, \citenamefont {Jansen}, \citenamefont
  {Kallidonis},\ and\ \citenamefont {Koutsou}}]{Alexandrou:2014sha}%
  \BibitemOpen
  \bibfield  {author} {\bibinfo {author} {\bibfnamefont {C.}~\bibnamefont
  {Alexandrou}}, \bibinfo {author} {\bibfnamefont {V.}~\bibnamefont {Drach}},
  \bibinfo {author} {\bibfnamefont {K.}~\bibnamefont {Jansen}}, \bibinfo
  {author} {\bibfnamefont {C.}~\bibnamefont {Kallidonis}},\ and\ \bibinfo
  {author} {\bibfnamefont {G.}~\bibnamefont {Koutsou}},\ }\bibfield  {title}
  {\bibinfo {title} {{Baryon spectrum with $N_f=2+1+1$ twisted mass
  fermions}},\ }\href {https://doi.org/10.1103/PhysRevD.90.074501} {\bibfield
  {journal} {\bibinfo  {journal} {Phys. Rev. D}\ }\textbf {\bibinfo {volume}
  {90}},\ \bibinfo {pages} {074501} (\bibinfo {year} {2014})},\ \Eprint
  {https://arxiv.org/abs/1406.4310} {arXiv:1406.4310 [hep-lat]} \BibitemShut
  {NoStop}%
\bibitem [{\citenamefont {Del~Debbio}\ \emph {et~al.}(2024)\citenamefont
  {Del~Debbio}, \citenamefont {Erben}, \citenamefont {Flynn}, \citenamefont
  {Mukherjee},\ and\ \citenamefont {Tsang}}]{DelDebbio:2024hca}%
  \BibitemOpen
  \bibfield  {author} {\bibinfo {author} {\bibfnamefont {L.}~\bibnamefont
  {Del~Debbio}}, \bibinfo {author} {\bibfnamefont {F.}~\bibnamefont {Erben}},
  \bibinfo {author} {\bibfnamefont {J.~M.}\ \bibnamefont {Flynn}}, \bibinfo
  {author} {\bibfnamefont {R.}~\bibnamefont {Mukherjee}},\ and\ \bibinfo
  {author} {\bibfnamefont {J.~T.}\ \bibnamefont {Tsang}} (\bibinfo
  {collaboration} {RBC, UKQCD}),\ }\bibfield  {title} {\bibinfo {title}
  {{Absorbing discretization effects with a massive renormalization scheme: The
  charm-quark mass}},\ }\href {https://doi.org/10.1103/PhysRevD.110.054512}
  {\bibfield  {journal} {\bibinfo  {journal} {Phys. Rev. D}\ }\textbf {\bibinfo
  {volume} {110}},\ \bibinfo {pages} {054512} (\bibinfo {year} {2024})},\
  \Eprint {https://arxiv.org/abs/2407.18700} {arXiv:2407.18700 [hep-lat]}
  \BibitemShut {NoStop}%
\end{thebibliography}%
\onecolumngrid
\clearpage
\appendix
\onecolumngrid
\section{Magnified plateau regions for ratio fits}
\begin{figure}[!h]
\includegraphics[width=0.46\columnwidth]{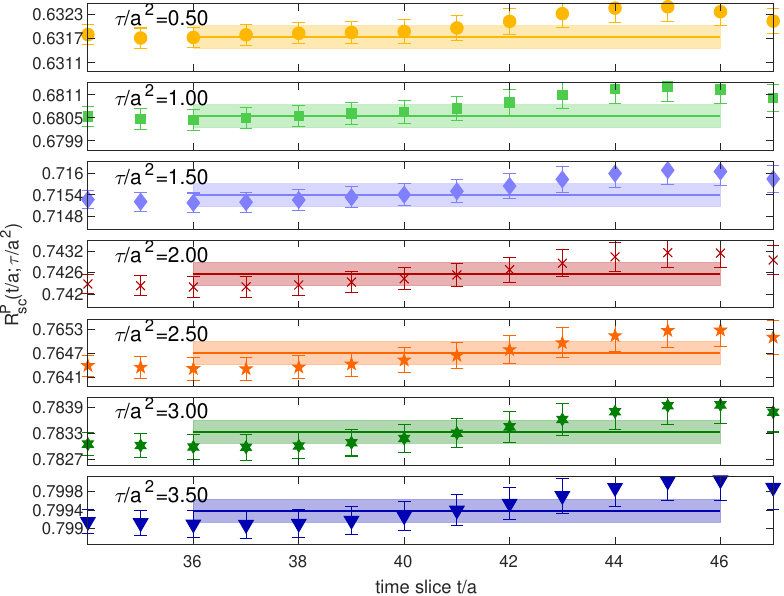}\hfill
\includegraphics[width=0.46\columnwidth]{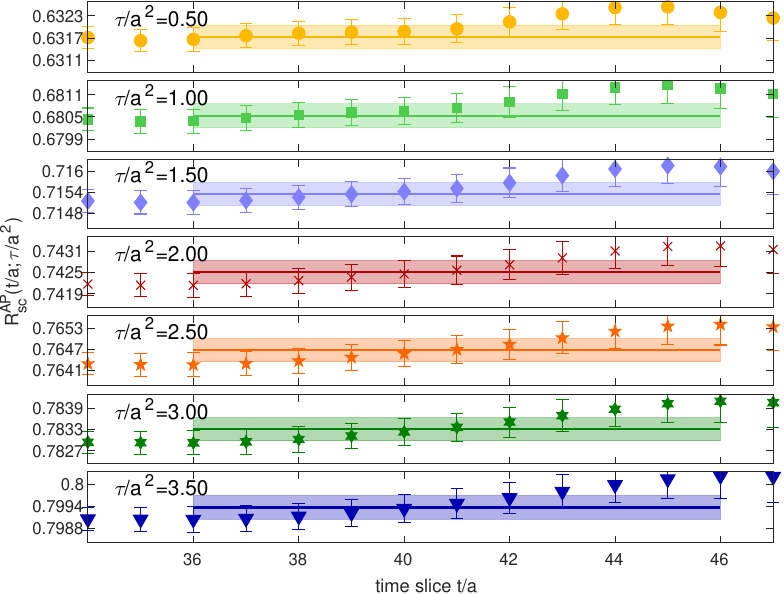}\\
\includegraphics[width=0.46\columnwidth]{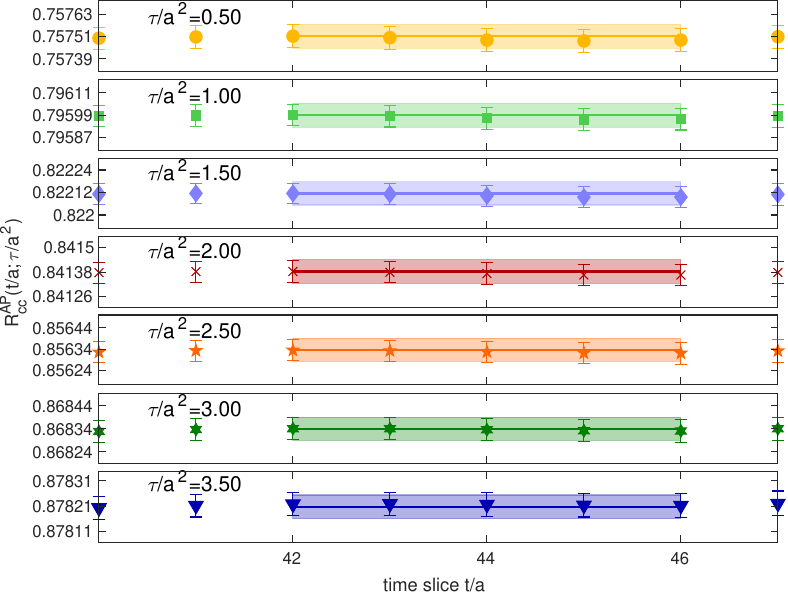}\hfill
\includegraphics[width=0.46\columnwidth]{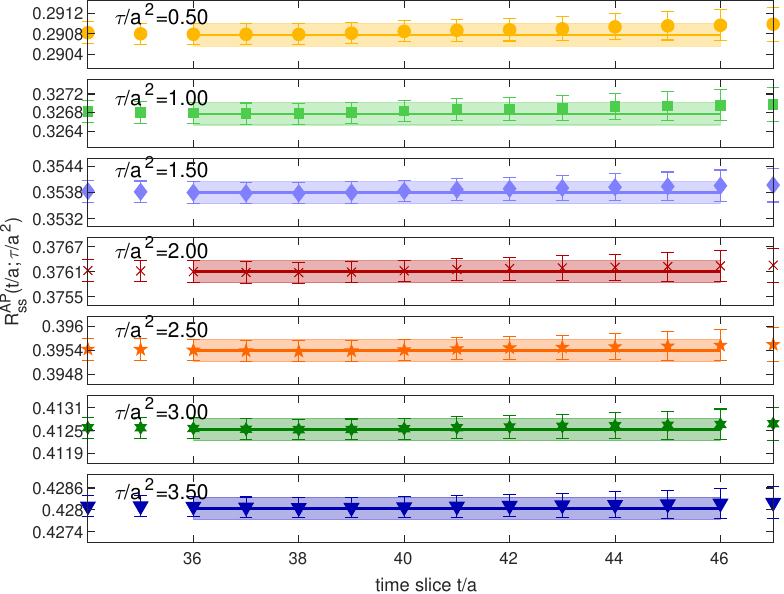}
\caption{Magnified plateau region demonstrating the extraction using ratios of correlators for the F1S ensemble. The top row shows the plateaus for the $D_s$ meson correlators $R_{sc}^P(t/a;\tau/a^2)$ on the left and $R_{sc}^{AP}(t/a;\tau/a^2)$ on the right. The bottom left displays $R_{cc}^{AP}(t/a;\tau/a^2)$ using $\eta_c$ correlators, and the bottom right $R_{ss}^{AP}(t/a;\tau/a^2)$ using $\eta_s$ correlators. Correlated fits are performed to extract the shown central values and uncertainties. Details including the goodness-of-fit are listed in Tab.~\ref{Tab.RP} and \ref{Tab.RAP}, respectively.}
\label{Fig.zoom}
\end{figure}

\section{Ratios \texorpdfstring{$\bar R^P(\tau/a^2)$ and $\bar R^{AP}(\tau/a^2)$}{RP and RAP} and continuum limit extrapolations}
\twocolumngrid

\begin{longtable*}{ccl@{~~~}ccccl@{~~~}ccccl@{~~~}cccc}  
  \caption{Values of the ratios $\bar R^P(\tau/a^2)$ for flow times $\tau/a^2$ extracted from correlated fits of the Euclidean times slices in the specified range $[t_1:t_2]$ with the corresponding $\chi^2/\text{d.o.f.}$ and $p$-value for all six ensembles analyzed. Each row lists our results using $D_s$-meson, $\eta_c$, and $\eta_s$ correlators. Values corresponding to small $\tau$ not entering in our $\tau\to 0$ extrapolation are set in italics.}\label{Tab.RP}\\
    \hline\hline  
    & && \multicolumn{4}{c}{$D_s$} && \multicolumn{4}{c}{$\eta_c$} && \multicolumn{4}{c}{$\eta_s$}\\
    \cline{4-7}  \cline{9-12} \cline{14-17}
    & $\tau/a^2$ && $\bar R^P_{sc}(\tau/a^2)$ & $[t_1:t_2]$ & ${\chi^2}/{\text{dof}}$& $p$-val. && $\bar R^P_{cc}(\tau/a^2)$ & $[t_1:t_2]$ & ${\chi^2}/{\text{dof}}$& $p$-val. && $\bar R^P_{ss}(\tau/a^2)$ & $[t_1:t_2]$ & ${\chi^2}/{\text{dof}}$ & $p$-val.\\
    \hline
    \endfirsthead
    \hline\hline
    & && \multicolumn{4}{c}{$D_s$} && \multicolumn{4}{c}{$\eta_c$} && \multicolumn{4}{c}{$\eta_s$}\\
    \cline{4-7}  \cline{9-12} \cline{14-17}    
    & $\tau/a^2$ && $\bar R^P_{sc}(\tau/a^2)$ & $[t_1:t_2]$ & ${\chi^2}/{\text{dof}}$& $p$-val. && $\bar R^P_{cc}(\tau/a^2)$ & $[t_1:t_2]$ & ${\chi^2}/{\text{dof}}$& $p$-val. && $\bar R^P_{ss}(\tau/a^2)$ & $[t_1:t_2]$ & ${\chi^2}/{\text{dof}}$ & $p$-val.\\
    \hline
    \endhead

    \hline\hline
    \endfoot
\it{F1S} & \it{ 0.10} && \it{0.57220(27)} & \it{[36:46]} & \it{1.67} & \it{8}\% && \it{0.709447(62)} & \it{[42:46]} & \it{0.78} & \it{54}\% && \it{0.24657(17)} & \it{[36:46]} & \it{0.90} & \it{54}\% \\ 
\it{F1S} & \it{ 0.20} && \it{0.58958(28)} & \it{[36:46]} & \it{1.44} & \it{16}\% && \it{0.723502(62)} & \it{[42:46]} & \it{0.78} & \it{54}\% && \it{0.25982(18)} & \it{[36:46]} & \it{0.87} & \it{56}\% \\ 
\it{F1S} & \it{ 0.30} && \it{0.60516(28)} & \it{[36:46]} & \it{1.26} & \it{25}\% && \it{0.736088(62)} & \it{[42:46]} & \it{0.78} & \it{54}\% && \it{0.27142(19)} & \it{[36:46]} & \it{0.85} & \it{58}\% \\ 
\it{F1S} & \it{ 0.40} && \it{0.61914(28)} & \it{[36:46]} & \it{1.13} & \it{34}\% && \it{0.747363(62)} & \it{[42:46]} & \it{0.78} & \it{54}\% && \it{0.28168(19)} & \it{[36:46]} & \it{0.83} & \it{60}\% \\ 
\it{F1S} & \it{ 0.50} && \it{0.63174(28)} & \it{[36:46]} & \it{1.03} & \it{41}\% && \it{0.757495(62)} & \it{[42:46]} & \it{0.79} & \it{53}\% && \it{0.29088(20)} & \it{[36:46]} & \it{0.82} & \it{61}\% \\ 
\it{F1S} & \it{ 0.60} && \it{0.64318(28)} & \it{[36:46]} & \it{0.97} & \it{47}\% && \it{0.766647(61)} & \it{[42:46]} & \it{0.79} & \it{53}\% && \it{0.29922(20)} & \it{[36:46]} & \it{0.80} & \it{62}\% \\ 
F1S &  0.70 && 0.65364(29) & [36:46] & 0.95 & 48\% && 0.774962(61) & [42:46] & 0.79 & 53\% && 0.30689(20) & [36:46] & 0.80 & 63\% \\ 
F1S &  0.80 && 0.66328(29) & [36:46] & 0.98 & 46\% && 0.782559(60) & [42:46] & 0.80 & 53\% && 0.31400(21) & [36:46] & 0.79 & 64\% \\ 
F1S &  0.90 && 0.67222(29) & [36:46] & 1.04 & 41\% && 0.789537(60) & [42:46] & 0.80 & 52\% && 0.32065(21) & [36:46] & 0.78 & 65\% \\ 
F1S &  1.00 && 0.68054(30) & [36:46] & 1.13 & 34\% && 0.795978(59) & [42:46] & 0.81 & 52\% && 0.32691(22) & [36:46] & 0.77 & 65\% \\ 
F1S &  1.10 && 0.68834(30) & [36:46] & 1.23 & 26\% && 0.801947(59) & [42:46] & 0.81 & 52\% && 0.33283(22) & [36:46] & 0.77 & 66\% \\ 
F1S &  1.20 && 0.69567(31) & [36:46] & 1.34 & 20\% && 0.807501(58) & [42:46] & 0.80 & 52\% && 0.33847(22) & [36:46] & 0.76 & 67\% \\ 
F1S &  1.30 && 0.70260(31) & [36:46] & 1.44 & 15\% && 0.812687(58) & [42:46] & 0.80 & 53\% && 0.34385(22) & [36:46] & 0.75 & 68\% \\ 
F1S &  1.40 && 0.70916(32) & [36:46] & 1.53 & 12\% && 0.817544(57) & [42:46] & 0.78 & 54\% && 0.34901(23) & [36:46] & 0.74 & 69\% \\ 
F1S &  1.50 && 0.71539(32) & [36:46] & 1.60 & 10\% && 0.822106(57) & [42:46] & 0.77 & 55\% && 0.35396(23) & [36:46] & 0.73 & 70\% \\ 
F1S &  1.60 && 0.72133(32) & [36:46] & 1.66 & 8\% && 0.826401(56) & [42:46] & 0.75 & 56\% && 0.35873(23) & [36:46] & 0.72 & 71\% \\ 
F1S &  1.70 && 0.72699(32) & [36:46] & 1.71 & 7\% && 0.830455(56) & [42:46] & 0.73 & 57\% && 0.36333(23) & [36:46] & 0.70 & 72\% \\ 
F1S &  1.80 && 0.73241(32) & [36:46] & 1.74 & 7\% && 0.834290(55) & [42:46] & 0.71 & 59\% && 0.36778(23) & [36:46] & 0.69 & 74\% \\ 
F1S &  1.90 && 0.73759(32) & [36:46] & 1.76 & 6\% && 0.837924(54) & [42:46] & 0.68 & 60\% && 0.37209(24) & [36:46] & 0.68 & 75\% \\ 
F1S &  2.00 && 0.74256(31) & [36:46] & 1.79 & 6\% && 0.841374(54) & [42:46] & 0.66 & 62\% && 0.37627(24) & [36:46] & 0.66 & 76\% \\ 
F1S &  2.10 && 0.74734(31) & [36:46] & 1.81 & 5\% && 0.844655(53) & [42:46] & 0.65 & 63\% && 0.38034(24) & [36:46] & 0.64 & 78\% \\ 
F1S &  2.20 && 0.75192(30) & [36:46] & 1.83 & 5\% && 0.847780(52) & [42:46] & 0.63 & 64\% && 0.38429(24) & [36:46] & 0.63 & 79\% \\ 
F1S &  2.30 && 0.75634(30) & [36:46] & 1.85 & 5\% && 0.850760(52) & [42:46] & 0.62 & 65\% && 0.38814(25) & [36:46] & 0.61 & 81\% \\ 
F1S &  2.40 && 0.76060(29) & [36:46] & 1.87 & 4\% && 0.853606(51) & [42:46] & 0.62 & 65\% && 0.39189(25) & [36:46] & 0.60 & 82\% \\ 
F1S &  2.50 && 0.76471(28) & [36:46] & 1.90 & 4\% && 0.856327(51) & [42:46] & 0.63 & 64\% && 0.39555(25) & [36:46] & 0.59 & 83\% \\ 
F1S &  2.60 && 0.76868(28) & [36:46] & 1.92 & 4\% && 0.858932(50) & [42:46] & 0.64 & 63\% && 0.39913(25) & [36:46] & 0.58 & 83\% \\ 
F1S &  2.70 && 0.77252(27) & [36:46] & 1.95 & 3\% && 0.861428(50) & [42:46] & 0.66 & 62\% && 0.40263(26) & [36:46] & 0.59 & 83\% \\     
\hline
\it{M1} & \it{ 0.10} && \it{0.58992(36)} & \it{[24:30]} & \it{0.60} & \it{73}\% && \it{0.72703(10)} & \it{[28:31]} & \it{2.06} & \it{10}\% && \it{0.24811(30)} & \it{[24:30]} & \it{0.62} & \it{72}\% \\ 
\it{M1} & \it{ 0.20} && \it{0.60900(36)} & \it{[24:30]} & \it{0.64} & \it{70}\% && \it{0.74227(10)} & \it{[28:31]} & \it{1.99} & \it{11}\% && \it{0.26243(32)} & \it{[24:30]} & \it{0.58} & \it{75}\% \\ 
\it{M1} & \it{ 0.30} && \it{0.62611(37)} & \it{[24:30]} & \it{0.68} & \it{67}\% && \it{0.75579(10)} & \it{[28:31]} & \it{1.98} & \it{11}\% && \it{0.27510(34)} & \it{[24:30]} & \it{0.58} & \it{75}\% \\ 
\it{M1} & \it{ 0.40} && \it{0.64144(37)} & \it{[24:30]} & \it{0.69} & \it{66}\% && \it{0.76780(10)} & \it{[28:31]} & \it{1.97} & \it{12}\% && \it{0.28638(36)} & \it{[24:30]} & \it{0.60} & \it{73}\% \\ 
M1 &  0.50 && 0.65522(37) & [24:30] & 0.65 & 69\% && 0.778515(100) & [28:31] & 1.96 & 12\% && 0.29654(37) & [24:30] & 0.63 & 71\% \\ 
M1 &  0.60 && 0.66769(37) & [24:30] & 0.57 & 75\% && 0.788120(99) & [28:31] & 1.95 & 12\% && 0.30579(38) & [24:30] & 0.67 & 67\% \\ 
M1 &  0.70 && 0.67906(37) & [24:30] & 0.46 & 84\% && 0.796785(97) & [28:31] & 1.92 & 12\% && 0.31432(39) & [24:30] & 0.71 & 64\% \\ 
M1 &  0.80 && 0.68948(37) & [24:30] & 0.35 & 91\% && 0.804649(96) & [28:31] & 1.90 & 13\% && 0.32225(40) & [24:30] & 0.74 & 62\% \\ 
M1 &  0.90 && 0.69911(37) & [24:30] & 0.26 & 95\% && 0.811825(94) & [28:31] & 1.88 & 13\% && 0.32968(40) & [24:30] & 0.76 & 60\% \\ 
M1 &  1.00 && 0.70805(36) & [24:30] & 0.20 & 98\% && 0.818407(92) & [28:31] & 1.86 & 13\% && 0.33668(40) & [24:30] & 0.76 & 60\% \\ 
M1 &  1.10 && 0.71638(36) & [24:30] & 0.19 & 98\% && 0.824471(90) & [28:31] & 1.84 & 14\% && 0.34332(41) & [24:30] & 0.76 & 60\% \\ 
M1 &  1.20 && 0.72420(35) & [24:30] & 0.21 & 98\% && 0.830080(88) & [28:31] & 1.81 & 14\% && 0.34963(41) & [24:30] & 0.74 & 62\% \\ 
M1 &  1.30 && 0.73155(35) & [24:30] & 0.25 & 96\% && 0.835288(86) & [28:31] & 1.78 & 15\% && 0.35567(41) & [24:30] & 0.71 & 64\% \\ 
M1 &  1.40 && 0.73849(34) & [24:30] & 0.32 & 93\% && 0.840138(83) & [28:31] & 1.75 & 15\% && 0.36146(41) & [24:30] & 0.66 & 68\% \\ 
M1 &  1.50 && 0.74507(34) & [24:30] & 0.40 & 88\% && 0.844669(81) & [28:31] & 1.72 & 16\% && 0.36702(41) & [24:30] & 0.61 & 73\% \\ 
M1 &  1.60 && 0.75132(33) & [24:30] & 0.47 & 83\% && 0.848912(79) & [28:31] & 1.69 & 17\% && 0.37239(41) & [24:30] & 0.55 & 77\% \\ 
M1 &  1.70 && 0.75728(33) & [24:30] & 0.53 & 78\% && 0.852896(77) & [28:31] & 1.66 & 17\% && 0.37756(41) & [24:30] & 0.49 & 81\% \\ 
M1 &  1.80 && 0.76297(33) & [24:30] & 0.59 & 74\% && 0.856645(75) & [28:31] & 1.65 & 18\% && 0.38258(41) & [24:30] & 0.44 & 85\% \\ 
M1 &  1.90 && 0.76841(33) & [24:30] & 0.63 & 71\% && 0.860178(73) & [28:31] & 1.65 & 17\% && 0.38743(41) & [24:30] & 0.39 & 89\% \\ 
\hline
\it{M2} & \it{ 0.10} && \it{0.58959(37)} & \it{[24:30]} & \it{1.76} & \it{10}\% && \it{0.726823(100)} & \it{[28:31]} & \it{0.48} & \it{70}\% && \it{0.24798(30)} & \it{[24:30]} & \it{1.44} & \it{19}\% \\ 
\it{M2} & \it{ 0.20} && \it{0.60866(37)} & \it{[24:30]} & \it{1.55} & \it{16}\% && \it{0.742078(99)} & \it{[28:31]} & \it{0.54} & \it{66}\% && \it{0.26229(31)} & \it{[24:30]} & \it{1.49} & \it{18}\% \\ 
\it{M2} & \it{ 0.30} && \it{0.62575(37)} & \it{[24:30]} & \it{1.43} & \it{20}\% && \it{0.755607(99)} & \it{[28:31]} & \it{0.73} & \it{54}\% && \it{0.27496(33)} & \it{[24:30]} & \it{1.54} & \it{16}\% \\ 
\it{M2} & \it{ 0.40} && \it{0.64106(37)} & \it{[24:30]} & \it{1.38} & \it{22}\% && \it{0.767622(99)} & \it{[28:31]} & \it{1.09} & \it{35}\% && \it{0.28625(34)} & \it{[24:30]} & \it{1.58} & \it{15}\% \\ 
M2 &  0.50 && 0.65484(37) & [24:30] & 1.36 & 23\% && 0.778333(100) & [28:31] & 1.59 & 19\% && 0.29642(35) & [24:30] & 1.62 & 14\% \\ 
M2 &  0.60 && 0.66732(37) & [24:30] & 1.36 & 23\% && 0.787935(99) & [28:31] & 2.14 & 9\% && 0.30569(36) & [24:30] & 1.67 & 12\% \\ 
M2 &  0.70 && 0.67912(52) & [29:30] & 0.50 & 48\% && 0.796642(90) & [29:31] & 1.00 & 37\% && 0.31397(35) & [29:30] & 0.86 & 35\% \\ 
M2 &  0.80 && 0.68962(52) & [29:30] & 0.33 & 56\% && 0.804509(89) & [29:31] & 1.12 & 33\% && 0.32190(35) & [29:30] & 0.89 & 34\% \\ 
M2 &  0.90 && 0.69933(53) & [29:30] & 0.20 & 65\% && 0.811689(87) & [29:31] & 1.25 & 29\% && 0.32934(36) & [29:30] & 0.93 & 33\% \\ 
M2 &  1.00 && 0.70836(53) & [29:30] & 0.11 & 74\% && 0.818274(86) & [29:31] & 1.41 & 24\% && 0.33635(37) & [29:30] & 0.98 & 32\% \\ 
M2 &  1.10 && 0.71679(53) & [29:30] & 0.04 & 83\% && 0.824342(85) & [29:31] & 1.59 & 20\% && 0.34299(37) & [29:30] & 1.03 & 31\% \\ 
M2 &  1.20 && 0.72471(53) & [29:30] & 0.01 & 92\% && 0.829954(84) & [29:31] & 1.79 & 17\% && 0.34932(38) & [29:30] & 1.10 & 29\% \\ 
M2 &  1.30 && 0.73216(54) & [29:30] & 0.00 & 99\% && 0.835165(83) & [29:31] & 2.02 & 13\% && 0.35537(38) & [29:30] & 1.16 & 28\% \\ 
M2 &  1.40 && 0.73919(54) & [29:30] & 0.01 & 90\% && 0.840019(82) & [29:31] & 2.26 & 10\% && 0.36117(38) & [29:30] & 1.24 & 27\% \\ 
M2 &  1.50 && 0.74586(54) & [29:30] & 0.05 & 82\% && 0.844553(81) & [29:31] & 2.52 & 8\% && 0.36674(39) & [29:30] & 1.31 & 25\% \\ 
M2 &  1.60 && 0.75219(54) & [29:30] & 0.11 & 74\% && 0.848800(80) & [29:31] & 2.79 & 6\% && 0.37212(39) & [29:30] & 1.38 & 24\% \\ 
M2 &  1.70 && 0.75821(54) & [29:30] & 0.18 & 67\% && 0.852788(79) & [29:31] & 3.06 & 5\% && 0.37730(39) & [29:30] & 1.46 & 23\% \\ 
M2 &  1.80 && 0.76395(54) & [29:30] & 0.27 & 61\% && 0.856540(78) & [29:31] & 3.33 & 4\% && 0.38232(40) & [29:30] & 1.53 & 22\% \\ 
M2 &  1.90 && 0.76944(54) & [29:30] & 0.37 & 55\% && 0.860078(77) & [29:31] & 3.60 & 3\% && 0.38719(40) & [29:30] & 1.61 & 21\% \\ 
\hline
\it{M3} & \it{ 0.10} && \it{0.58910(44)} & \it{[24:30]} & \it{1.08} & \it{37}\% && \it{0.72658(12)} & \it{[28:31]} & \it{2.70} & \it{4}\% && \it{0.24752(26)} & \it{[24:30]} & \it{1.59} & \it{15}\% \\ 
\it{M3} & \it{ 0.20} && \it{0.60818(44)} & \it{[24:30]} & \it{1.08} & \it{37}\% && \it{0.74185(12)} & \it{[28:31]} & \it{2.74} & \it{4}\% && \it{0.26182(28)} & \it{[24:30]} & \it{1.50} & \it{17}\% \\ 
\it{M3} & \it{ 0.30} && \it{0.62529(44)} & \it{[24:30]} & \it{1.08} & \it{37}\% && \it{0.75539(12)} & \it{[28:31]} & \it{2.78} & \it{4}\% && \it{0.27448(29)} & \it{[24:30]} & \it{1.42} & \it{20}\% \\ 
\it{M3} & \it{ 0.40} && \it{0.64063(44)} & \it{[24:30]} & \it{1.10} & \it{36}\% && \it{0.76742(12)} & \it{[28:31]} & \it{2.81} & \it{4}\% && \it{0.28576(30)} & \it{[24:30]} & \it{1.36} & \it{23}\% \\ 
M3 &  0.50 && 0.65445(44) & [24:30] & 1.13 & 34\% && 0.77815(11) & [28:31] & 2.83 & 4\% && 0.29593(31) & [24:30] & 1.31 & 25\% \\ 
M3 &  0.60 && 0.66699(45) & [24:30] & 1.18 & 31\% && 0.78777(11) & [28:31] & 2.85 & 4\% && 0.30520(33) & [24:30] & 1.29 & 26\% \\ 
M3 &  0.70 && 0.67843(45) & [24:30] & 1.24 & 28\% && 0.79645(11) & [28:31] & 2.89 & 3\% && 0.31375(34) & [24:30] & 1.28 & 26\% \\ 
M3 &  0.80 && 0.68895(45) & [24:30] & 1.31 & 25\% && 0.80432(11) & [28:31] & 2.94 & 3\% && 0.32171(36) & [24:30] & 1.26 & 27\% \\ 
M3 &  0.90 && 0.69867(45) & [24:30] & 1.39 & 21\% && 0.81151(11) & [28:31] & 2.99 & 3\% && 0.32917(37) & [24:30] & 1.24 & 28\% \\ 
M3 &  1.00 && 0.70771(45) & [24:30] & 1.47 & 18\% && 0.81810(10) & [28:31] & 3.04 & 3\% && 0.33620(38) & [24:30] & 1.20 & 30\% \\ 
M3 &  1.10 && 0.71614(46) & [24:30] & 1.56 & 15\% && 0.82418(10) & [28:31] & 3.08 & 3\% && 0.34288(39) & [24:30] & 1.16 & 32\% \\ 
M3 &  1.20 && 0.72405(46) & [24:30] & 1.65 & 13\% && 0.829795(99) & [28:31] & 3.12 & 2\% && 0.34923(40) & [24:30] & 1.13 & 34\% \\ 
M3 &  1.30 && 0.73149(46) & [24:30] & 1.72 & 11\% && 0.835011(97) & [28:31] & 3.14 & 2\% && 0.35531(41) & [24:30] & 1.11 & 35\% \\ 
M3 &  1.40 && 0.73851(46) & [24:30] & 1.78 & 10\% && 0.839871(95) & [28:31] & 3.16 & 2\% && 0.36113(41) & [24:30] & 1.11 & 35\% \\ 
M3 &  1.50 && 0.74515(46) & [24:30] & 1.82 & 9\% && 0.844411(93) & [28:31] & 3.17 & 2\% && 0.36673(42) & [24:30] & 1.14 & 34\% \\ 
M3 &  1.60 && 0.75146(45) & [24:30] & 1.85 & 9\% && 0.848663(91) & [28:31] & 3.16 & 2\% && 0.37213(42) & [24:30] & 1.18 & 32\% \\ 
M3 &  1.70 && 0.75746(45) & [24:30] & 1.87 & 8\% && 0.852656(89) & [28:31] & 3.15 & 2\% && 0.37734(42) & [24:30] & 1.22 & 29\% \\ 
M3 &  1.80 && 0.76319(45) & [24:30] & 1.88 & 8\% && 0.856414(87) & [28:31] & 3.12 & 2\% && 0.38238(42) & [24:30] & 1.28 & 26\% \\ 
M3 &  1.90 && 0.76865(44) & [24:30] & 1.89 & 8\% && 0.859957(85) & [28:31] & 3.09 & 3\% && 0.38727(43) & [24:30] & 1.33 & 24\% \\ 
\hline
\it{C1} & \it{ 0.10} && \it{0.62485(41)} & \it{[24:30]} & \it{0.31} & \it{93}\% && \it{0.75068(12)} & \it{[28:31]} & \it{5.42} & \it{0}\% && \it{0.26723(28)} & \it{[24:30]} & \it{0.60} & \it{73}\% \\ 
\it{C1} & \it{ 0.20} && \it{0.64784(41)} & \it{[24:30]} & \it{0.28} & \it{95}\% && \it{0.77005(11)} & \it{[28:31]} & \it{3.97} & \it{1}\% && \it{0.28476(29)} & \it{[24:30]} & \it{0.47} & \it{83}\% \\ 
C1 &  0.30 && 0.66832(42) & [24:30] & 0.28 & 95\% && 0.78659(10) & [28:31] & 3.11 & 3\% && 0.30054(31) & [24:30] & 0.38 & 89\% \\ 
C1 &  0.40 && 0.68654(42) & [24:30] & 0.28 & 95\% && 0.800809(96) & [28:31] & 2.58 & 5\% && 0.31477(32) & [24:30] & 0.31 & 93\% \\ 
C1 &  0.50 && 0.70282(43) & [24:30] & 0.28 & 95\% && 0.813127(91) & [28:31] & 2.25 & 8\% && 0.32769(33) & [24:30] & 0.25 & 96\% \\ 
C1 &  0.60 && 0.71747(43) & [24:30] & 0.28 & 95\% && 0.823888(87) & [28:31] & 2.03 & 11\% && 0.33954(34) & [24:30] & 0.21 & 97\% \\ 
C1 &  0.70 && 0.73073(43) & [24:30] & 0.29 & 94\% && 0.833371(84) & [28:31] & 1.88 & 13\% && 0.35048(35) & [24:30] & 0.18 & 98\% \\ 
C1 &  0.80 && 0.74283(43) & [24:30] & 0.32 & 92\% && 0.841793(81) & [28:31] & 1.77 & 15\% && 0.36068(35) & [24:30] & 0.16 & 99\% \\ 
C1 &  0.90 && 0.75392(44) & [24:30] & 0.39 & 88\% && 0.849326(79) & [28:31] & 1.70 & 17\% && 0.37025(36) & [24:30] & 0.13 & 99\% \\ 
C1 &  1.00 && 0.76415(44) & [24:30] & 0.50 & 81\% && 0.856106(76) & [28:31] & 1.63 & 18\% && 0.37926(37) & [24:30] & 0.12 & 99\% \\ 
C1 &  1.10 && 0.77363(44) & [24:30] & 0.63 & 70\% && 0.862241(75) & [28:31] & 1.58 & 19\% && 0.38781(37) & [24:30] & 0.10 & 100\% \\ 
\hline
\it{C2} & \it{ 0.10} && \it{0.62354(35)} & \it{[24:30]} & \it{1.71} & \it{11}\% && \it{0.74987(12)} & \it{[28:31]} & \it{4.52} & \it{0}\% && \it{0.26633(25)} & \it{[24:30]} & \it{0.89} & \it{50}\% \\ 
\it{C2} & \it{ 0.20} && \it{0.64654(36)} & \it{[24:30]} & \it{1.67} & \it{12}\% && \it{0.769353(98)} & \it{[28:31]} & \it{1.46} & \it{22}\% && \it{0.28387(27)} & \it{[24:30]} & \it{1.05} & \it{39}\% \\ 
C2 &  0.30 && 0.66705(36) & [24:30] & 1.60 & 14\% && 0.785979(88) & [28:31] & 0.64 & 59\% && 0.29968(28) & [24:30] & 1.19 & 31\% \\ 
C2 &  0.40 && 0.68531(36) & [24:30] & 1.52 & 17\% && 0.800265(82) & [28:31] & 0.65 & 58\% && 0.31395(30) & [24:30] & 1.31 & 25\% \\ 
C2 &  0.50 && 0.70164(36) & [24:30] & 1.43 & 20\% && 0.812634(79) & [28:31] & 0.85 & 47\% && 0.32692(31) & [24:30] & 1.41 & 21\% \\ 
C2 &  0.60 && 0.71635(37) & [24:30] & 1.35 & 23\% && 0.823437(76) & [28:31] & 1.03 & 38\% && 0.33882(32) & [24:30] & 1.49 & 18\% \\ 
C2 &  0.70 && 0.72969(37) & [24:30] & 1.30 & 25\% && 0.832954(73) & [28:31] & 1.16 & 32\% && 0.34982(32) & [24:30] & 1.56 & 15\% \\ 
C2 &  0.80 && 0.74187(37) & [24:30] & 1.28 & 26\% && 0.841407(71) & [28:31] & 1.28 & 28\% && 0.36009(33) & [24:30] & 1.61 & 14\% \\ 
C2 &  0.90 && 0.75307(37) & [24:30] & 1.28 & 26\% && 0.848969(70) & [28:31] & 1.39 & 24\% && 0.36971(34) & [24:30] & 1.65 & 13\% \\ 
C2 &  1.00 && 0.76341(37) & [24:30] & 1.30 & 25\% && 0.855774(68) & [28:31] & 1.51 & 21\% && 0.37880(34) & [24:30] & 1.68 & 12\% \\ 
C2 &  1.10 && 0.77300(37) & [24:30] & 1.31 & 25\% && 0.861934(66) & [28:31] & 1.64 & 18\% && 0.38740(34) & [24:30] & 1.70 & 12\% \\ 
\end{longtable*}

\begin{longtable*}{ccl@{~~~}ccccl@{~~~}ccccl@{~~~}cccc}  
  \caption{Values of $\bar R^{AP}(\tau/a^2) \equiv \sqrt{\bar R^P(\tau/a^2)\bar R^A(\tau/a^2)}$ for flow times $\tau/a^2$ extracted from correlated fits of the Euclidean times slices in the specified range $[t_1:t_2]$ with the corresponding $\chi^2/\text{d.o.f.}$ and $p$-value for all six ensembles analyzed. Each row lists our results using $D_s$-meson, $\eta_c$, and $\eta_s$, correlators. Values corresponding to small $\tau$ not entering in our $\tau\to 0$ extrapolation are set in italics.}\label{Tab.RAP}\\  
    \hline\hline  
    & && \multicolumn{4}{c}{$D_s$} && \multicolumn{4}{c}{$\eta_c$} && \multicolumn{4}{c}{$\eta_s$}\\
    \cline{4-7}  \cline{9-12} \cline{14-17}
    & $\tau/a^2$ && $\bar R^{AP}_{cs}(\tau/a^2)$ & $[t_1:t_2]$ & ${\chi^2}/{\text{dof}}$& $p$-val. && $\bar R^{AP}_{cc}(\tau/a^2)$ & $[t_1:t_2]$ & ${\chi^2}/{\text{dof}}$& $p$-val. && $\bar R^{AP}_{ss}(\tau/a^2)$ & $[t_1:t_2]$ & ${\chi^2}/{\text{dof}}$ & $p$-val.\\
    \hline
    \endfirsthead
    \hline\hline
    & && \multicolumn{4}{c}{$D_s$} && \multicolumn{4}{c}{$\eta_c$} && \multicolumn{4}{c}{$\eta_s$}\\
    \cline{4-7}  \cline{9-12} \cline{14-17}    
    & $\tau/a^2$ && $\bar R^{AP}_{sc}(\tau/a^2)$ & $[t_1:t_2]$ & ${\chi^2}/{\text{dof}}$& $p$-val. && $\bar R^{AP}_{cc}(\tau/a^2)$ & $[t_1:t_2]$ & ${\chi^2}/{\text{dof}}$& $p$-val. && $\bar R^{AP}_{ss}(\tau/a^2)$ & $[t_1:t_2]$ & ${\chi^2}/{\text{dof}}$ & $p$-val.\\
    \hline
    \endhead

    \hline\hline
    \endfoot
\it{F1S} & \it{ 0.10} && \it{0.57221(30)} & \it{[36:46]} & \it{1.42} & \it{16}\% && \it{0.709463(63)} & \it{[42:46]} & \it{0.58} & \it{68}\% && \it{0.24649(19)} & \it{[36:46]} & \it{0.38} & \it{96}\% \\ 
\it{F1S} & \it{ 0.20} && \it{0.58958(30)} & \it{[36:46]} & \it{1.21} & \it{28}\% && \it{0.723519(63)} & \it{[42:46]} & \it{0.61} & \it{65}\% && \it{0.25974(20)} & \it{[36:46]} & \it{0.36} & \it{96}\% \\ 
\it{F1S} & \it{ 0.30} && \it{0.60516(31)} & \it{[36:46]} & \it{1.07} & \it{38}\% && \it{0.736105(63)} & \it{[42:46]} & \it{0.67} & \it{61}\% && \it{0.27135(21)} & \it{[36:46]} & \it{0.35} & \it{97}\% \\ 
\it{F1S} & \it{ 0.40} && \it{0.61914(31)} & \it{[36:46]} & \it{0.99} & \it{45}\% && \it{0.747379(64)} & \it{[42:46]} & \it{0.75} & \it{56}\% && \it{0.28160(22)} & \it{[36:46]} & \it{0.35} & \it{97}\% \\ 
\it{F1S} & \it{ 0.50} && \it{0.63174(32)} & \it{[36:46]} & \it{0.95} & \it{48}\% && \it{0.757511(64)} & \it{[42:46]} & \it{0.84} & \it{50}\% && \it{0.29078(22)} & \it{[36:46]} & \it{0.36} & \it{96}\% \\ 
\it{F1S} & \it{ 0.60} && \it{0.64318(32)} & \it{[36:46]} & \it{0.95} & \it{49}\% && \it{0.766662(64)} & \it{[42:46]} & \it{0.94} & \it{44}\% && \it{0.29912(23)} & \it{[36:46]} & \it{0.37} & \it{96}\% \\ 
F1S &  0.70 && 0.65364(32) & [36:46] & 0.96 & 47\% && 0.774976(64) & [42:46] & 1.05 & 38\% && 0.30678(23) & [36:46] & 0.38 & 96\% \\ 
F1S &  0.80 && 0.66328(32) & [36:46] & 0.99 & 45\% && 0.782572(64) & [42:46] & 1.15 & 33\% && 0.31388(24) & [36:46] & 0.39 & 95\% \\ 
F1S &  0.90 && 0.67221(32) & [36:46] & 1.03 & 41\% && 0.789550(64) & [42:46] & 1.25 & 29\% && 0.32052(24) & [36:46] & 0.39 & 95\% \\ 
F1S &  1.00 && 0.68053(32) & [36:46] & 1.07 & 38\% && 0.795989(64) & [42:46] & 1.33 & 26\% && 0.32678(24) & [36:46] & 0.39 & 95\% \\ 
F1S &  1.10 && 0.68832(32) & [36:46] & 1.10 & 36\% && 0.801958(63) & [42:46] & 1.41 & 23\% && 0.33269(25) & [36:46] & 0.39 & 95\% \\ 
F1S &  1.20 && 0.69565(32) & [36:46] & 1.13 & 33\% && 0.807512(63) & [42:46] & 1.47 & 21\% && 0.33832(25) & [36:46] & 0.39 & 95\% \\ 
F1S &  1.30 && 0.70257(33) & [36:46] & 1.15 & 32\% && 0.812697(63) & [42:46] & 1.51 & 19\% && 0.34370(25) & [36:46] & 0.39 & 95\% \\ 
F1S &  1.40 && 0.70913(33) & [36:46] & 1.17 & 30\% && 0.817554(62) & [42:46] & 1.55 & 19\% && 0.34885(26) & [36:46] & 0.38 & 96\% \\ 
F1S &  1.50 && 0.71536(33) & [36:46] & 1.19 & 29\% && 0.822115(62) & [42:46] & 1.56 & 18\% && 0.35380(26) & [36:46] & 0.37 & 96\% \\ 
F1S &  1.60 && 0.72129(33) & [36:46] & 1.20 & 28\% && 0.826410(61) & [42:46] & 1.57 & 18\% && 0.35857(26) & [36:46] & 0.37 & 96\% \\ 
F1S &  1.70 && 0.72695(33) & [36:46] & 1.21 & 28\% && 0.830464(60) & [42:46] & 1.55 & 18\% && 0.36317(26) & [36:46] & 0.36 & 96\% \\ 
F1S &  1.80 && 0.73236(33) & [36:46] & 1.22 & 27\% && 0.834298(60) & [42:46] & 1.53 & 19\% && 0.36762(27) & [36:46] & 0.36 & 96\% \\ 
F1S &  1.90 && 0.73754(33) & [36:46] & 1.24 & 26\% && 0.837932(59) & [42:46] & 1.50 & 20\% && 0.37192(27) & [36:46] & 0.35 & 97\% \\ 
F1S &  2.00 && 0.74251(33) & [36:46] & 1.25 & 26\% && 0.841382(58) & [42:46] & 1.46 & 21\% && 0.37611(27) & [36:46] & 0.35 & 97\% \\ 
F1S &  2.10 && 0.74729(33) & [36:46] & 1.26 & 25\% && 0.844663(57) & [42:46] & 1.42 & 22\% && 0.38017(27) & [36:46] & 0.34 & 97\% \\ 
F1S &  2.20 && 0.75188(33) & [36:46] & 1.27 & 24\% && 0.847788(56) & [42:46] & 1.39 & 24\% && 0.38412(28) & [36:46] & 0.34 & 97\% \\ 
F1S &  2.30 && 0.75630(33) & [36:46] & 1.28 & 23\% && 0.850769(56) & [42:46] & 1.35 & 25\% && 0.38797(28) & [36:46] & 0.34 & 97\% \\ 
F1S &  2.40 && 0.76056(33) & [36:46] & 1.30 & 22\% && 0.853615(55) & [42:46] & 1.32 & 26\% && 0.39173(28) & [36:46] & 0.34 & 97\% \\ 
F1S &  2.50 && 0.76467(33) & [36:46] & 1.32 & 22\% && 0.856337(54) & [42:46] & 1.30 & 27\% && 0.39539(28) & [36:46] & 0.34 & 97\% \\ 
F1S &  2.60 && 0.76864(33) & [36:46] & 1.33 & 21\% && 0.858942(53) & [42:46] & 1.29 & 27\% && 0.39897(29) & [36:46] & 0.34 & 97\% \\ 
F1S &  2.70 && 0.77249(33) & [36:46] & 1.35 & 20\% && 0.861438(53) & [42:46] & 1.28 & 27\% && 0.40247(29) & [36:46] & 0.35 & 97\% \\ 
\hline
\it{M1} & \it{ 0.10} && \it{0.58969(36)} & \it{[24:30]} & \it{0.27} & \it{95}\% && \it{0.72706(10)} & \it{[28:31]} & \it{1.04} & \it{37}\% && \it{0.24816(33)} & \it{[24:30]} & \it{0.63} & \it{71}\% \\ 
\it{M1} & \it{ 0.20} && \it{0.60876(36)} & \it{[24:30]} & \it{0.35} & \it{91}\% && \it{0.742299(99)} & \it{[28:31]} & \it{0.80} & \it{49}\% && \it{0.26248(35)} & \it{[24:30]} & \it{0.59} & \it{74}\% \\ 
\it{M1} & \it{ 0.30} && \it{0.62587(36)} & \it{[24:30]} & \it{0.41} & \it{87}\% && \it{0.755820(97)} & \it{[28:31]} & \it{0.69} & \it{56}\% && \it{0.27515(37)} & \it{[24:30]} & \it{0.57} & \it{76}\% \\ 
\it{M1} & \it{ 0.40} && \it{0.64120(37)} & \it{[24:30]} & \it{0.44} & \it{85}\% && \it{0.767831(96)} & \it{[28:31]} & \it{0.64} & \it{59}\% && \it{0.28644(39)} & \it{[24:30]} & \it{0.57} & \it{75}\% \\ 
M1 &  0.50 && 0.65499(38) & [24:30] & 0.42 & 86\% && 0.778540(94) & [28:31] & 0.61 & 61\% && 0.29661(40) & [24:30] & 0.59 & 74\% \\ 
M1 &  0.60 && 0.66747(39) & [24:30] & 0.36 & 90\% && 0.788142(93) & [28:31] & 0.59 & 62\% && 0.30587(41) & [24:30] & 0.62 & 71\% \\ 
M1 &  0.70 && 0.67883(39) & [24:30] & 0.28 & 95\% && 0.796805(91) & [28:31] & 0.60 & 61\% && 0.31441(42) & [24:30] & 0.66 & 68\% \\ 
M1 &  0.80 && 0.68926(39) & [24:30] & 0.19 & 98\% && 0.804666(89) & [28:31] & 0.63 & 60\% && 0.32234(42) & [24:30] & 0.69 & 66\% \\ 
M1 &  0.90 && 0.69889(39) & [24:30] & 0.12 & 99\% && 0.811841(88) & [28:31] & 0.68 & 57\% && 0.32977(43) & [24:30] & 0.72 & 63\% \\ 
M1 &  1.00 && 0.70783(38) & [24:30] & 0.07 & 100\% && 0.818423(86) & [28:31] & 0.73 & 53\% && 0.33678(44) & [24:30] & 0.75 & 61\% \\ 
M1 &  1.10 && 0.71618(38) & [24:30] & 0.05 & 100\% && 0.824487(84) & [28:31] & 0.79 & 50\% && 0.34342(44) & [24:30] & 0.76 & 60\% \\ 
M1 &  1.20 && 0.72400(37) & [24:30] & 0.07 & 100\% && 0.830097(82) & [28:31] & 0.85 & 47\% && 0.34974(45) & [24:30] & 0.77 & 60\% \\ 
M1 &  1.30 && 0.73135(36) & [24:30] & 0.10 & 100\% && 0.835306(80) & [28:31] & 0.89 & 44\% && 0.35578(45) & [24:30] & 0.77 & 59\% \\ 
M1 &  1.40 && 0.73830(36) & [24:30] & 0.15 & 99\% && 0.840158(79) & [28:31] & 0.91 & 43\% && 0.36157(46) & [24:30] & 0.77 & 59\% \\ 
M1 &  1.50 && 0.74489(36) & [24:30] & 0.21 & 97\% && 0.844691(77) & [28:31] & 0.91 & 43\% && 0.36714(46) & [24:30] & 0.77 & 59\% \\ 
M1 &  1.60 && 0.75115(36) & [24:30] & 0.27 & 95\% && 0.848936(76) & [28:31] & 0.89 & 44\% && 0.37250(47) & [24:30] & 0.78 & 59\% \\ 
M1 &  1.70 && 0.75711(36) & [24:30] & 0.32 & 93\% && 0.852922(74) & [28:31] & 0.85 & 47\% && 0.37768(47) & [24:30] & 0.78 & 58\% \\ 
M1 &  1.80 && 0.76281(35) & [24:30] & 0.36 & 91\% && 0.856672(73) & [28:31] & 0.79 & 50\% && 0.38269(47) & [24:30] & 0.80 & 57\% \\ 
M1 &  1.90 && 0.76825(35) & [24:30] & 0.39 & 89\% && 0.860207(72) & [28:31] & 0.72 & 54\% && 0.38754(48) & [24:30] & 0.81 & 56\% \\ 
\hline
\it{M2} & \it{ 0.10} && \it{0.58957(42)} & \it{[24:30]} & \it{1.69} & \it{12}\% && \it{0.72685(10)} & \it{[29:31]} & \it{0.21} & \it{81}\% && \it{0.24771(29)} & \it{[24:30]} & \it{0.64} & \it{70}\% \\ 
\it{M2} & \it{ 0.20} && \it{0.60864(42)} & \it{[24:30]} & \it{1.46} & \it{19}\% && \it{0.742111(100)} & \it{[29:31]} & \it{0.22} & \it{81}\% && \it{0.26199(30)} & \it{[24:30]} & \it{0.67} & \it{67}\% \\ 
\it{M2} & \it{ 0.30} && \it{0.62573(42)} & \it{[24:30]} & \it{1.33} & \it{24}\% && \it{0.755645(98)} & \it{[29:31]} & \it{0.23} & \it{79}\% && \it{0.27464(32)} & \it{[24:30]} & \it{0.71} & \it{64}\% \\ 
\it{M2} & \it{ 0.40} && \it{0.64105(42)} & \it{[24:30]} & \it{1.27} & \it{27}\% && \it{0.767668(96)} & \it{[29:31]} & \it{0.25} & \it{78}\% && \it{0.28591(33)} & \it{[24:30]} & \it{0.75} & \it{61}\% \\ 
M2 &  0.50 && 0.65483(41) & [24:30] & 1.25 & 28\% && 0.778386(94) & [29:31] & 0.26 & 77\% && 0.29606(34) & [24:30] & 0.79 & 58\% \\ 
M2 &  0.60 && 0.66733(41) & [24:30] & 1.27 & 27\% && 0.787996(93) & [29:31] & 0.27 & 76\% && 0.30532(35) & [24:30] & 0.83 & 54\% \\ 
M2 &  0.70 && 0.67873(40) & [24:30] & 1.31 & 25\% && 0.796664(91) & [29:31] & 0.30 & 74\% && 0.31386(35) & [24:30] & 0.88 & 51\% \\ 
M2 &  0.80 && 0.68921(39) & [24:30] & 1.34 & 24\% && 0.804530(90) & [29:31] & 0.36 & 70\% && 0.32180(36) & [24:30] & 0.92 & 48\% \\ 
M2 &  0.90 && 0.69891(39) & [24:30] & 1.35 & 23\% && 0.811709(89) & [29:31] & 0.46 & 63\% && 0.32924(37) & [24:30] & 0.97 & 45\% \\ 
M2 &  1.00 && 0.70793(38) & [24:30] & 1.35 & 23\% && 0.818293(87) & [29:31] & 0.60 & 55\% && 0.33625(38) & [24:30] & 1.01 & 42\% \\ 
M2 &  1.10 && 0.71637(37) & [24:30] & 1.32 & 24\% && 0.824360(86) & [29:31] & 0.79 & 46\% && 0.34290(38) & [24:30] & 1.05 & 39\% \\ 
M2 &  1.20 && 0.72428(36) & [24:30] & 1.28 & 26\% && 0.829972(85) & [29:31] & 1.00 & 37\% && 0.34923(39) & [24:30] & 1.09 & 37\% \\ 
M2 &  1.30 && 0.73174(36) & [24:30] & 1.24 & 28\% && 0.835183(84) & [29:31] & 1.24 & 29\% && 0.35528(40) & [24:30] & 1.12 & 35\% \\ 
M2 &  1.40 && 0.73878(35) & [24:30] & 1.21 & 30\% && 0.840036(83) & [29:31] & 1.50 & 22\% && 0.36108(40) & [24:30] & 1.15 & 33\% \\ 
M2 &  1.50 && 0.74545(34) & [24:30] & 1.18 & 31\% && 0.844571(82) & [29:31] & 1.75 & 17\% && 0.36665(41) & [24:30] & 1.17 & 32\% \\ 
M2 &  1.60 && 0.75179(34) & [24:30] & 1.17 & 32\% && 0.848818(81) & [29:31] & 2.01 & 13\% && 0.37202(42) & [24:30] & 1.20 & 30\% \\ 
M2 &  1.70 && 0.75783(33) & [24:30] & 1.16 & 33\% && 0.852805(80) & [29:31] & 2.25 & 11\% && 0.37721(42) & [24:30] & 1.22 & 29\% \\ 
M2 &  1.80 && 0.76358(33) & [24:30] & 1.15 & 33\% && 0.856558(79) & [29:31] & 2.47 & 8\% && 0.38223(42) & [24:30] & 1.23 & 28\% \\ 
M2 &  1.90 && 0.76908(33) & [24:30] & 1.15 & 33\% && 0.860096(78) & [29:31] & 2.67 & 7\% && 0.38709(42) & [24:30] & 1.25 & 28\% \\ 
\hline
\it{M3} & \it{ 0.10} && \it{0.58908(45)} & \it{[24:30]} & \it{1.28} & \it{26}\% && \it{0.72660(14)} & \it{[28:31]} & \it{1.78} & \it{15}\% && \it{0.24744(28)} & \it{[24:30]} & \it{1.19} & \it{31}\% \\ 
\it{M3} & \it{ 0.20} && \it{0.60814(46)} & \it{[24:30]} & \it{1.28} & \it{26}\% && \it{0.74187(14)} & \it{[28:31]} & \it{1.79} & \it{15}\% && \it{0.26174(29)} & \it{[24:30]} & \it{1.12} & \it{35}\% \\ 
\it{M3} & \it{ 0.30} && \it{0.62523(46)} & \it{[24:30]} & \it{1.27} & \it{27}\% && \it{0.75542(14)} & \it{[28:31]} & \it{1.82} & \it{14}\% && \it{0.27439(30)} & \it{[24:30]} & \it{1.05} & \it{39}\% \\ 
\it{M3} & \it{ 0.40} && \it{0.64055(46)} & \it{[24:30]} & \it{1.28} & \it{26}\% && \it{0.76745(13)} & \it{[28:31]} & \it{1.83} & \it{14}\% && \it{0.28567(32)} & \it{[24:30]} & \it{0.97} & \it{45}\% \\ 
M3 &  0.50 && 0.65436(45) & [24:30] & 1.30 & 25\% && 0.77818(13) & [28:31] & 1.83 & 14\% && 0.29584(33) & [24:30] & 0.89 & 50\% \\ 
M3 &  0.60 && 0.66688(45) & [24:30] & 1.32 & 24\% && 0.78780(13) & [28:31] & 1.83 & 14\% && 0.30512(34) & [24:30] & 0.81 & 56\% \\ 
M3 &  0.70 && 0.67832(45) & [24:30] & 1.35 & 23\% && 0.79648(13) & [28:31] & 1.82 & 14\% && 0.31367(35) & [24:30] & 0.75 & 61\% \\ 
M3 &  0.80 && 0.68883(45) & [24:30] & 1.37 & 22\% && 0.80435(13) & [28:31] & 1.82 & 14\% && 0.32164(37) & [24:30] & 0.69 & 66\% \\ 
M3 &  0.90 && 0.69854(45) & [24:30] & 1.37 & 22\% && 0.81154(12) & [28:31] & 1.82 & 14\% && 0.32910(38) & [24:30] & 0.65 & 69\% \\ 
M3 &  1.00 && 0.70756(45) & [24:30] & 1.35 & 23\% && 0.81813(12) & [28:31] & 1.82 & 14\% && 0.33615(39) & [24:30] & 0.63 & 70\% \\ 
M3 &  1.10 && 0.71598(45) & [24:30] & 1.32 & 24\% && 0.82420(11) & [28:31] & 1.82 & 14\% && 0.34283(40) & [24:30] & 0.64 & 70\% \\ 
M3 &  1.20 && 0.72386(45) & [24:30] & 1.27 & 27\% && 0.82982(11) & [28:31] & 1.82 & 14\% && 0.34919(41) & [24:30] & 0.66 & 68\% \\ 
M3 &  1.30 && 0.73128(45) & [24:30] & 1.22 & 29\% && 0.83504(11) & [28:31] & 1.81 & 14\% && 0.35527(42) & [24:30] & 0.70 & 65\% \\ 
M3 &  1.40 && 0.73828(45) & [24:30] & 1.17 & 32\% && 0.83990(10) & [28:31] & 1.79 & 15\% && 0.36110(43) & [24:30] & 0.74 & 62\% \\ 
M3 &  1.50 && 0.74492(46) & [24:30] & 1.12 & 35\% && 0.844436(100) & [28:31] & 1.77 & 15\% && 0.36670(43) & [24:30] & 0.79 & 58\% \\ 
M3 &  1.60 && 0.75121(46) & [24:30] & 1.08 & 37\% && 0.848688(97) & [28:31] & 1.74 & 16\% && 0.37210(44) & [24:30] & 0.83 & 54\% \\ 
M3 &  1.70 && 0.75721(46) & [24:30] & 1.05 & 39\% && 0.852682(94) & [28:31] & 1.70 & 16\% && 0.37731(44) & [24:30] & 0.88 & 51\% \\ 
M3 &  1.80 && 0.76292(46) & [24:30] & 1.02 & 41\% && 0.856440(91) & [28:31] & 1.67 & 17\% && 0.38236(45) & [24:30] & 0.93 & 47\% \\ 
M3 &  1.90 && 0.76839(46) & [24:30] & 1.00 & 42\% && 0.859983(89) & [28:31] & 1.63 & 18\% && 0.38725(45) & [24:30] & 0.97 & 44\% \\ 
\hline
\it{C1} & \it{ 0.10} && \it{0.62479(44)} & \it{[24:30]} & \it{0.76} & \it{60}\% && \it{0.75072(11)} & \it{[28:31]} & \it{3.35} & \it{2}\% && \it{0.26727(32)} & \it{[24:30]} & \it{0.60} & \it{73}\% \\ 
\it{C1} & \it{ 0.20} && \it{0.64780(45)} & \it{[24:30]} & \it{0.66} & \it{68}\% && \it{0.77009(10)} & \it{[28:31]} & \it{2.81} & \it{4}\% && \it{0.28480(34)} & \it{[24:30]} & \it{0.64} & \it{70}\% \\ 
C1 &  0.30 && 0.66829(45) & [24:30] & 0.61 & 72\% && 0.786622(98) & [28:31] & 2.52 & 6\% && 0.30058(36) & [24:30] & 0.69 & 66\% \\ 
C1 &  0.40 && 0.68653(46) & [24:30] & 0.59 & 74\% && 0.800837(93) & [28:31] & 2.37 & 7\% && 0.31481(38) & [24:30] & 0.73 & 62\% \\ 
C1 &  0.50 && 0.70282(47) & [24:30] & 0.57 & 75\% && 0.813149(88) & [28:31] & 2.27 & 8\% && 0.32774(39) & [24:30] & 0.77 & 60\% \\ 
C1 &  0.60 && 0.71748(47) & [24:30] & 0.56 & 76\% && 0.823906(84) & [28:31] & 2.18 & 9\% && 0.33958(40) & [24:30] & 0.79 & 58\% \\ 
C1 &  0.70 && 0.73075(48) & [24:30] & 0.56 & 76\% && 0.833384(81) & [28:31] & 2.08 & 10\% && 0.35053(40) & [24:30] & 0.79 & 57\% \\ 
C1 &  0.80 && 0.74285(48) & [24:30] & 0.57 & 76\% && 0.841803(77) & [28:31] & 1.98 & 11\% && 0.36073(41) & [24:30] & 0.79 & 58\% \\ 
C1 &  0.90 && 0.75394(48) & [24:30] & 0.58 & 75\% && 0.849333(74) & [28:31] & 1.90 & 13\% && 0.37029(41) & [24:30] & 0.79 & 58\% \\ 
C1 &  1.00 && 0.76418(48) & [24:30] & 0.59 & 74\% && 0.856112(72) & [28:31] & 1.83 & 14\% && 0.37930(41) & [24:30] & 0.78 & 58\% \\ 
C1 &  1.10 && 0.77366(49) & [24:30] & 0.61 & 73\% && 0.862247(70) & [28:31] & 1.78 & 15\% && 0.38784(42) & [24:30] & 0.79 & 58\% \\ 
\hline
\it{C2} & \it{ 0.10} && \it{0.62355(40)} & \it{[24:30]} & \it{1.46} & \it{19}\% && \it{0.74999(11)} & \it{[28:31]} & \it{3.78} & \it{1}\% && \it{0.26660(24)} & \it{[24:30]} & \it{0.45} & \it{85}\% \\ 
\it{C2} & \it{ 0.20} && \it{0.64655(40)} & \it{[24:30]} & \it{1.48} & \it{18}\% && \it{0.769448(99)} & \it{[28:31]} & \it{2.77} & \it{4}\% && \it{0.28416(25)} & \it{[24:30]} & \it{0.51} & \it{80}\% \\ 
C2 &  0.30 && 0.66705(41) & [24:30] & 1.47 & 18\% && 0.786060(94) & [28:31] & 2.16 & 9\% && 0.29998(27) & [24:30] & 0.55 & 77\% \\ 
C2 &  0.40 && 0.68532(41) & [24:30] & 1.44 & 20\% && 0.800334(89) & [28:31] & 1.76 & 15\% && 0.31425(28) & [24:30] & 0.59 & 74\% \\ 
C2 &  0.50 && 0.70166(42) & [24:30] & 1.40 & 21\% && 0.812695(86) & [28:31] & 1.49 & 22\% && 0.32721(29) & [24:30] & 0.62 & 71\% \\ 
C2 &  0.60 && 0.71639(43) & [24:30] & 1.36 & 23\% && 0.823491(83) & [28:31] & 1.30 & 27\% && 0.33909(30) & [24:30] & 0.65 & 69\% \\ 
C2 &  0.70 && 0.72974(43) & [24:30] & 1.34 & 23\% && 0.833004(80) & [28:31] & 1.17 & 32\% && 0.35007(31) & [24:30] & 0.67 & 67\% \\ 
C2 &  0.80 && 0.74193(44) & [24:30] & 1.34 & 23\% && 0.841451(78) & [28:31] & 1.08 & 36\% && 0.36032(32) & [24:30] & 0.69 & 66\% \\ 
C2 &  0.90 && 0.75314(44) & [24:30] & 1.36 & 23\% && 0.849008(76) & [28:31] & 1.02 & 38\% && 0.36992(33) & [24:30] & 0.69 & 66\% \\ 
C2 &  1.00 && 0.76348(44) & [24:30] & 1.37 & 22\% && 0.855809(74) & [28:31] & 0.97 & 40\% && 0.37899(34) & [24:30] & 0.69 & 66\% \\ 
C2 &  1.10 && 0.77306(44) & [24:30] & 1.36 & 22\% && 0.861964(72) & [28:31] & 0.95 & 42\% && 0.38758(34) & [24:30] & 0.68 & 66\% \\ 
\end{longtable*}

\begin{table}[tb]
  \caption{Results of the $a\to 0$ continuum limit extrapolations for quantities based on $D_s$ and $\eta_c$ correlators at fixed flow time $\tau$ using an ansatz linear in $a^2$. Listed are the flow time $\tau$ in physical units, the continuum limit value as well as the goodness of fit expressed by the corresponding $\chi^2/\text{d.o.f.}$ and the $p$-value for both quantities. Values corresponding to small $\tau$ not entering in our $\tau\to 0$ extrapolation are set in italics.} 
  \label{Tab.ContinuumExtra}
  \begin{tabular}{c@{~~}cccc@{~~}ccc}
    \hline\hline
    $\tau$ & \multicolumn{3}{c}{$D_s$} && \multicolumn{3}{c}{$\eta_c$} \\
    \cline{2-4}  \cline{6-8} 
    
     [$\text{GeV}^2$] & $R^{AP}_{sc}(\tau)M_{D_s}$ & $\chi^2/\text{dof}$ & $p$-val. &&
    $R^{AP}_{cc}(\tau)M_{\eta_c}$ & $\chi^2/\text{dof}$ & $p$-val. \\
    \hline
\it{0.039} & \it{583.5(2.7)} & \it{0.40} & \it{81\%} && 	\it{1096.8(4.9)} & \it{2.33} & \it{5\%} \\ 
\it{0.052} & \it{600.1(2.8)} & \it{0.32} & \it{86\%} && 	\it{1116.1(5)} & \it{2.02} & \it{9\%} \\ 
\it{0.064} & \it{614.2(2.8)} & \it{0.23} & \it{92\%} && 	\it{1132.3(5.1)} & \it{1.68} & \it{15\%} \\ 
\it{0.077} & \it{627.6(2.9)} & \it{0.22} & \it{93\%} && 	\it{1148.0(5.1)} & \it{1.54} & \it{19\%} \\ 
0.090 & 639.3(2.9) & 0.22 & 93\% && 	1161.5(5.2) & 1.44 & 22\% \\ 
0.103 & 650.1(3) & 0.21 & 93\% && 	1174.1(5.2) & 1.35 & 25\% \\ 
0.116 & 659.9(3) & 0.20 & 94\% && 	1185.3(5.2) & 1.26 & 28\% \\ 
0.129 & 668.8(3) & 0.20 & 94\% && 	1195.5(5.3) & 1.19 & 31\% \\ 
0.142 & 677.4(3.1) & 0.23 & 92\% && 	1205.3(5.3) & 1.18 & 32\% \\ 
0.155 & 685.0(3.1) & 0.21 & 93\% && 	1213.8(5.4) & 1.12 & 35\% \\ 
0.168 & 692.4(3.1) & 0.22 & 93\% && 	1222.1(5.4) & 1.09 & 36\% \\ 
0.181 & 699.3(3.2) & 0.22 & 92\% && 	1229.8(5.4) & 1.06 & 37\% \\ 
0.193 & 705.8(3.2) & 0.23 & 92\% && 	1236.9(5.5) & 1.05 & 38\% \\ 
0.206 & 711.9(3.2) & 0.23 & 92\% && 	1243.6(5.5) & 1.02 & 39\% \\ 
0.219 & 717.6(3.2) & 0.23 & 92\% && 	1249.7(5.5) & 0.99 & 41\% \\ 
0.232 & 723.2(3.3) & 0.24 & 92\% && 	1255.8(5.5) & 0.99 & 41\% \\ 
0.245 & 728.5(3.3) & 0.24 & 92\% && 	1261.4(5.6) & 0.98 & 42\% \\ 
0.258 & 733.5(3.3) & 0.24 & 92\% && 	1266.7(5.6) & 0.96 & 43\% \\ 
0.271 & 738.4(3.3) & 0.25 & 91\% && 	1271.7(5.6) & 0.95 & 43\% \\ 
0.284 & 742.9(3.3) & 0.25 & 91\% && 	1276.4(5.6) & 0.94 & 44\% \\ 
0.297 & 747.5(3.4) & 0.26 & 91\% && 	1281.1(5.6) & 0.94 & 44\% \\ 
0.309 & 751.7(3.4) & 0.26 & 90\% && 	1285.3(5.7) & 0.92 & 45\% \\ 
0.322 & 755.9(3.4) & 0.27 & 90\% && 	1289.5(5.7) & 0.92 & 45\% \\ 
0.335 & 759.9(3.4) & 0.28 & 89\% && 	1293.5(5.7) & 0.92 & 45\% \\ 
0.348 & 763.7(3.4) & 0.28 & 89\% && 	1297.2(5.7) & 0.90 & 46\% \\
\hline\hline
  \end{tabular}
\end{table}

\begin{table}[tb]
  \caption{Results of the $a\to 0$ continuum limit extrapolations for ratios based on $\eta_s$ correlators at fixed flow time $\tau$ using either an ansatz linear in $a^2$ or a fit to a constant. Listed are the flow time $\tau$ in physical units, the continuum limit value as well as the goodness of fit given by the corresponding $\chi^2/\text{d.o.f.}$ and the $p$-value for both continuum limits. Values corresponding to  small $\tau$ not entering in our $\tau\to 0$ extrapolation are set in italics.} 
  \label{Tab.ContinuumExtraEtas}
  \begin{tabular}{c@{~}cccc@{~~}ccc}
    \hline\hline
    $\tau$ & \multicolumn{3}{c}{$\eta_s^{a\to 0\text{ linear}}$} && \multicolumn{3}{c}{$\eta_s^{a\to 0 \text{ const}}$} \\
    \cline{2-4}  \cline{6-8} 
    
     [$\text{GeV}^2$] & $R^{AP}_{sc}(\tau)M_{\eta_s}$ & $\chi^2/\text{dof}$ & $p$-val. &&
    $R^{AP}_{cc}(\tau)M_{\eta_s}$ & $\chi^2/\text{dof}$ & $p$-val. \\
    \hline
\it{0.035} & \it{87.99(50)} & \it{0.19} & \it{91\%} && \it{92.56(14)} & \it{22.57} & \it{0\%} \\ 
\it{0.053} & \it{93.65(53)} & \it{0.21} & \it{89\%} && \it{96.50(14)} & \it{8.12} & \it{0\%} \\ 
\it{0.070} & \it{98.40(55)} & \it{0.22} & \it{88\%} && \it{100.10(15)} & \it{2.81} & \it{2\%} \\ 
0.088 & 102.51(57) & 0.24 & 87\% && 103.42(15) & 0.88 & 48\% \\ 
0.106 & 106.17(58) & 0.26 & 85\% && 106.48(16) & 0.27 & 90\% \\ 
0.123 & 109.38(60) & 0.28 & 84\% && 109.36(16) & 0.21 & 93\% \\ 
0.141 & 112.41(62) & 0.30 & 83\% && 112.02(17) & 0.33 & 86\% \\ 
0.158 & 115.10(63) & 0.32 & 81\% && 114.57(17) & 0.43 & 79\% \\ 
0.176 & 117.70(65) & 0.34 & 80\% && 116.95(17) & 0.62 & 65\% \\ 
0.194 & 120.08(66) & 0.36 & 78\% && 119.25(18) & 0.71 & 59\% \\ 
0.211 & 122.37(67) & 0.38 & 77\% && 121.42(18) & 0.83 & 51\% \\ 
0.229 & 124.53(68) & 0.40 & 76\% && 123.51(18) & 0.90 & 46\% \\ 
0.246 & 126.58(70) & 0.42 & 74\% && 125.52(19) & 0.94 & 44\% \\ 
0.264 & 128.57(71) & 0.44 & 73\% && 127.44(19) & 1.02 & 40\% \\ 
0.282 & 130.44(72) & 0.46 & 71\% && 129.31(19) & 1.01 & 40\% \\ 
0.299 & 132.29(73) & 0.48 & 69\% && 131.10(20) & 1.08 & 37\% \\ 
0.317 & 134.03(74) & 0.51 & 68\% && 132.85(20) & 1.07 & 37\% \\ 
0.335 & 135.75(75) & 0.53 & 66\% && 134.54(20) & 1.11 & 35\% \\ 
0.352 & 137.40(76) & 0.56 & 64\% && 136.18(20) & 1.12 & 35\% \\ 
\hline\hline
  \end{tabular}
\end{table}

\clearpage

\onecolumngrid
\section{Residual masses}\label{Mres}
\begin{figure}[!h]
\includegraphics[width=0.48\textwidth]{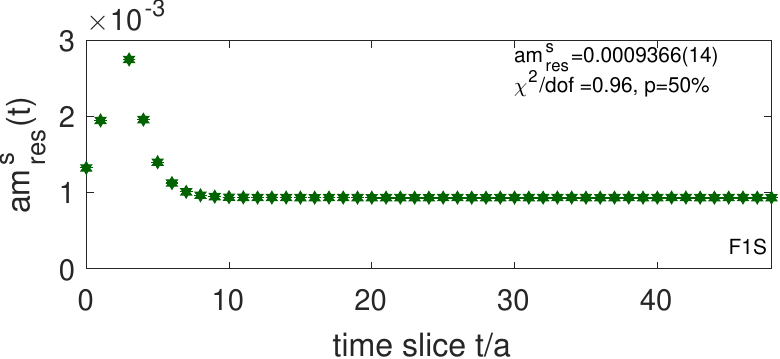}\hfill
\includegraphics[width=0.48\textwidth]{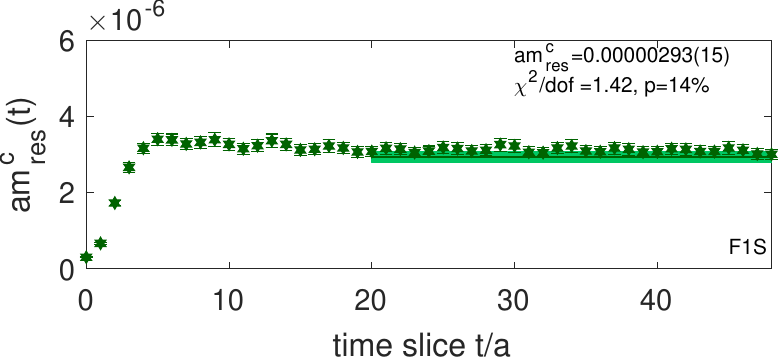}\\
\includegraphics[width=0.48\textwidth]{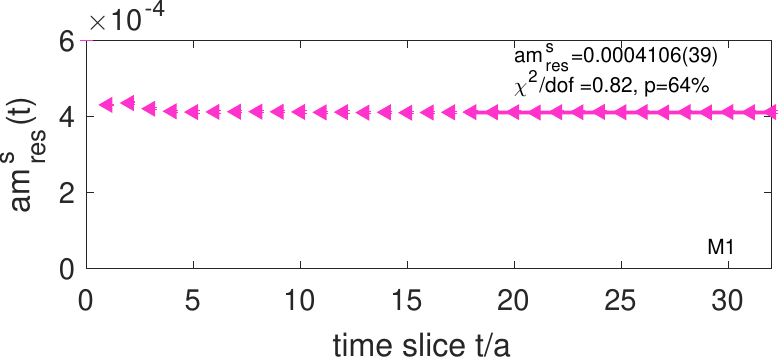}\hfill  
\includegraphics[width=0.48\textwidth]{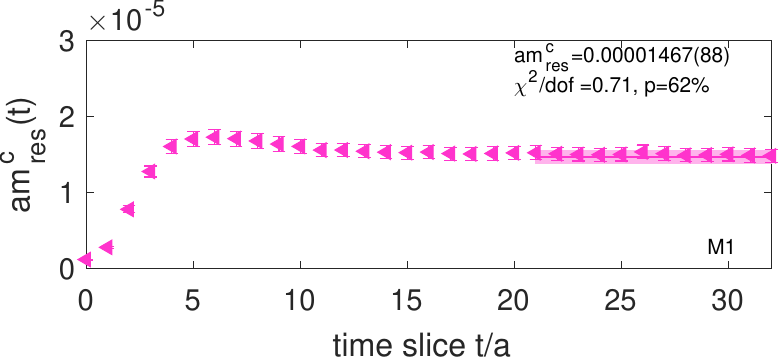}\\
\includegraphics[width=0.48\textwidth]{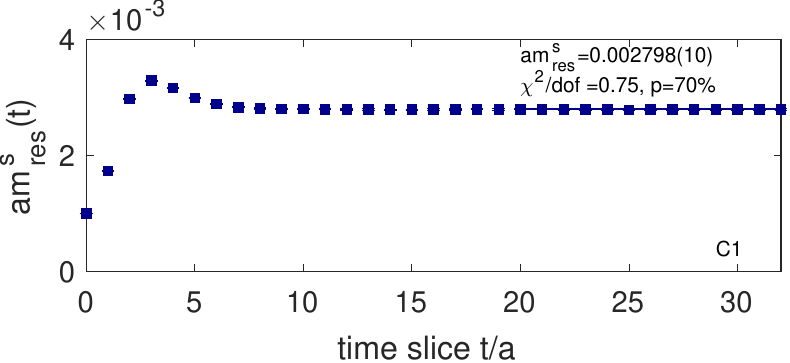 }\hfill
\includegraphics[width=0.48\textwidth]{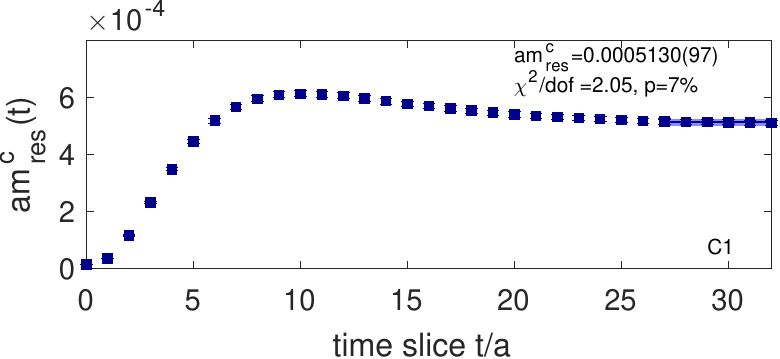}
\caption{Extraction of the residual mass using strange quark correlators
  (left) and charm quark correlators (right) on the F1S ensemble (top), M1
  (center), and C1 (bottom). In all cases we observe the expected flat curve
  indicating that \dwf\ discretization errors are well under control.}
\end{figure}

\end{document}